\documentclass[12pt,onecolumn,oneside]{IEEEtran}
\IEEEoverridecommandlockouts
\usepackage{amsmath}
\usepackage{amsfonts}
\usepackage{cases}
\usepackage{setspace}
\usepackage{fancybox}
\usepackage{subfigure}
\usepackage{epsfig}
\usepackage{graphicx}
\usepackage{epstopdf}
\usepackage{float}
\usepackage{multirow}
\newtheorem{lemma}{Lemma}
\newtheorem{theorem}{Theorem}
\newtheorem{fact}{Fact}

\newtheorem{definition}{Definition}

\usepackage{color}

\def\x{{\mathbf x}}

\def\A{{\mathbf A}}
\def\I{{\mathbf I}}
\def\v{\mathbf{v}}
\def\u{{\mathbf u}}
\def\e{{\mathbf e}}
\def\p{{\mathbf p}}
\def\z{{\mathbf z}}

\def\y{{\mathbf y}}

\def\F{\mathcal{F}}
\def\D{\mathcal{D}}
\def\P{\mathcal{P}}

\usepackage{bm}
\usepackage{amsmath}
\usepackage{multirow}

\usepackage{indentfirst}

\usepackage{times,amsmath,color,amssymb,epsfig,cite,subfigure,algorithm,algorithmic}

\begin{document}

\title{Decoding Nonbinary LDPC Codes via Proximal-ADMM Approach}
\author{\IEEEauthorblockN{Yongchao Wang, Jing Bai}
%
}

\maketitle

\begin{abstract}
In this paper, we focus on decoding nonbinary low-density parity-check (LDPC) codes in Galois fields of characteristic two via the proximal alternating direction method of multipliers (proximal-ADMM). By exploiting Flanagan/Constant-Weighting embedding techniques and the decomposition technique based on three-variables parity-check equations, two efficient proximal-ADMM decoders for nonbinary LDPC codes are proposed. We show that both of them are theoretically guaranteed convergent to some stationary point of the decoding model and either of their computational complexities in each proximal-ADMM iteration scales linearly with LDPC code's length and the size of the considered Galois field. Moreover, the decoder based on the Constant-Weight embedding technique satisfies the favorable property of codeword symmetry. Simulation results demonstrate their effectiveness in comparison with state-of-the-art LDPC decoders.
\end{abstract}

\small{Keywords: Nonbinary Low-density Parity-check (LDPC) codes, Galois Fields of Characteristic Two, Proximal Alternating Direction Method of Multipliers (Proximal-ADMM), Quadratic Programming (QP).}

\section{Introduction}\label{introduction}

Nonbinary low-density parity-check (LDPC) codes \cite{Davey-nonBP} in Galois fields of characteristic 2 ($\mathbb{F}_{2^q}$) are favorable in high-data-rate communication systems and storage systems \cite{optical-communication} since they possess many desirable merits from the viewpoints of practical applications.
For example, nonbinary LDPC codes have greater ability to eliminate short cycles (especially 4-cycles) and display better error-correction performance \cite{better-performance}.
Moreover, nonbinary LDPC codes have good ability to resist burst errors by combining multiple burst bit errors into fewer nonbinary symbol errors \cite{burst-error}.
Furthermore, a nonbinary LDPC code can provide a higher data transmission rate and spectral efficiency when it is combined with a high-order modulation scheme \cite{combined-highmodu}.

Typical decoding algorithms for nonbinary LDPC codes, such as the sum-product \cite{Davey-nonBP}\cite{FHT-nonBP}, are based on the belief propagation (BP) strategy.
In \cite{Davey-nonBP}, Davey and Mackay first investigated nonbinary LDPC codes and the corresponding nonbinary BP algorithm.
Later, Djordjevic and Vasic in \cite{FHT-nonBP} optimized nonbinary BP algorithms by reducing the computational complexity of check-node processing.
However, nonbinary BP-like decoding algorithms are heuristic from a theoretical viewpoint since their convergence performance cannot be guaranteed in theory. Meanwhile, analyzing the behavior of nonbinary BP-like decoding algorithms is often difficult and the corresponding results are very limited \cite{nonBP-converanaly}.

In recent years, mathematical programming (MP) techniques, such as linear programming (LP) and quadratic programming (QP) \cite{Bertsekas_NP}, are applied to decoding LDPC codes. These decoding techniques have attracted significant attention from researchers in the error-correction coding/decoding field due to their analyzable decoding performance, such as convergence and codeword symmetrical property.
The first MP decoding technique was proposed by Feldman et al. \cite{FeldmanLP}, who relaxed the maximum-likelihood (ML) decoding problem to a linear program for binary LDPC codes. In comparison with classical BP decoders, two issues must be considered: one is computational complexity of the general LP solving algorithms, such as the interior point method \cite{interior-point} and simplex method \cite{revised-simplex}, which are prohibitive in practical applications and the other is its inferior error-correction performance in low SNR regions\cite{FeldmanLP}. For the first issue, several improved MP decoders for binary LDPC codes are proposed.
In \cite{Barman-ADMM}, Barman et al. applied the alternating direction method of the multipliers (ADMM) technique \cite{boyd-ADMM} to solve the original LP decoding problem \cite{FeldmanLP}.
Comparable with existing LP decoders, the decoding complexity of the ADMM-based decoders is greatly reduced. However, it is still expensive from a practical viewpoint since it involves
costly projection operations onto parity polytopes in each iteration.
Later, Zhang and Siegel in \cite{efficient-projection1} optimized the parity polytope projection algorithm based on the cut search algorithm proposed in \cite{Adaptive-cut}.
In \cite{efficient-projection2}, the parity polytope projection was further simplified to an efficient simplex projection algorithm.
In \cite{look-up-SPL}, Jiao et al. proposed a cheap projection algorithm, which can be implemented. Specifically, this projection algorithm employed a cut-searching method in \cite{efficient-projection1} to identify which facet to project onto and then performed the projection operation via the simplex method \cite{efficient-projection2}.
In \cite{efficient-projection3}, Wei and Banihashemi proposed an iterative check-polytope projection algorithm to reduce the complexity of the LP decoding algorithm.
Moreover, a projection reduction technique was investigated in \cite{jiao-zhang} to reduce the number of Euclidean projections onto the check polytope.
Bai et al. in \cite{my-wcl} proposed an efficient ADMM-based LP decoding algorithm via the three-variables parity-check equations decomposition technique.
For the second issue of error-correction performance, Taagavi and Siegel in \cite{adaptive-LP} designed an adaptive LP decoder to improve LP decoding performance by adaptively adding necessary parity-check constraints. Later, different cut-generating algorithms were designed in \cite{cut-plane-algorithm,separation-cut,Adaptive-cut} to eliminate unexpected pseudo-codewords and improve the error-correction performance of LP decoding. Rosnes in \cite{Adaptive_Rosnes} considered adaptive linear programming decoding of linear codes over prime fields and efficient separation of
the underlying inequalities describing the decoding polytope which was done through dynamic programming.
In addition, Liu et al. in \cite{penalty-decoder} and Bai et al. in \cite{bai-admm-qp-binary} proposed improved ADMM-based penalized decoding algorithms to enhance the error-correction performance of LP decoding in low SNR regions.

In comparison with MP decoding techniques for binary LDPC codes, the corresponding approaches for nonbinary cases are limited.
Of particular relevance is the work \cite{Flanagan} where LP decoding was first generalized to nonbinary LDPC codes.
However, nonbinary LP decoding encounters a similar computational complexity problem as the binary case when using general LP solvers. To overcome this problem, Goldin et al. in [31] and Punekar et al. in [32] applied the coordinate ascent method to solve the dual problems of the original LP decoding problem in [33] and [34].
Punekar and Flanagan in \cite{Trellis-LP-nonbinary} adopted the binary LP decoding idea in \cite{Trellis-LP-binary} and proposed a trellis-based algorithm for check node processing to reduce the complexity of nonbinary LP decoding.
In addition, another nonbinary LP decoding scheme was introduced in \cite{Fast-LP-nonbinary} by using constant-weight binary vectors to represent elements in $\mathbb{F}_{2^q}$, but no efficient algorithm was developed to solve the resulting LP problem.
Recently, Liu and Draper in \cite{Liu-nonbinary-journal} extended the binary LP decoding idea in \cite{Barman-ADMM} and developed an LP decoding algorithm based on the ADMM technique for nonbinary LDPC codes in $\mathbb{F}_{2^q}$. Specifically, the algorithm is implemented by transforming nonbinary parity-check constraints to an equivalent binary factor graph representation and then relaxing the representation to linear constraints, and finally applying the ADMM algorithm to solve the resulting decoding problem. However, the proposed nonbinary ADMM decoder involves sorting or iteration operations to implement Euclidean projections onto high dimensional parity-check/simplex polytopes, which is time-consuming from a practical viewpoint.

In this paper, we focus on designing new nonbinary LDPC decoders with theoretically-guaranteed convergence, low complexity, and competitive error-correction performance.
Specifically, {the main contents of this paper are} summarized as follows.
\begin{itemize}
  \item Based on the parity-check equation decomposition method, we decompose a general multi-variables parity-check equation into a set of three-variables parity-check equations. Then, by exploiting the equivalent binary parity-check formulation of every three-variables parity-check equation and Flanagan embedding technique, we transform the nonbinary ML decoding problem to an equivalent linear integer program. Finally, by relaxing binary constraints to box constraints, adding a quadratic penalty term into the objective, and introducing extra linear constraints, a new quadratic programming (QP) decoding model is established for nonbinary LDPC codes in $\mathbb{F}_{2^q}$.
  \item We develop a proximal-ADMM algorithm to solve the resulting QP decoding model. By exploiting the inherent structures of the QP problem, variables in one ADMM step are blocked and the blocks can be updated in parallel. Moreover, variables in other ADMM steps are updated in parallel.
  \item The proposed proximal-ADMM decoding algorithm {can be proven to} converge to a stationary point of the formulated QP decoding problem. In addition, its complexity in each iteration
      scales linearly with block length and Galois field's size of the nonbinary LDPC codes.
  \item {Besides, we leverage the Constant-Weight embedding technique to develop a different proximal-ADMM decoder, which has a similar convergence property and computational complexity; meanwhile, it satisfies the favorable property of codeword symmetry, i.e., all the transmitted codewords have the same error probability if the noisy channel is symmetrical.}
\end{itemize}


To facilitate reading of this manuscript, notations used in this paper are summarized in \emph{Table I}. Moreover, we have the following remarks.
\begin{itemize}
  \item Throughout the paper, all the vectors are column vectors.
  \item $\rm{diag}(\cdot)$ performs different operations depending on its input.
      \begin{itemize}
       \item If the input is a vector $\mathbf{a}$, $\rm {diag}(\mathbf{a})$ denotes a diagonal matrix and $\mathbf{a}$ is its main-diagonal vector.
       \item If the input is a square matrix $\mathbf{A}$, $\rm {diag}(\mathbf{A}) = \mathbf{a}$, where $\mathbf{a}$ is the matrix's main-diagonal vector.
       \item If the input is vectors or matrices, ``$\textrm{diag}(\cdot,\ldots,\cdot)$'' builds a diagonal matrix and the inputs are located in the main-diagonal line of the matrix.
      \end{itemize}
  \item The operator ``$\otimes$'' is specifically defined as follows.
      \begin{itemize}
        \item For matrices $\mathbf{A} \in \mathbb{R}^{m\times n} $ and $\mathbf{B} \in \mathbb{R}^{p\times q}$, $\mathbf{A} \otimes \mathbf{B}=\begin{bmatrix}
    a_{11}\mathbf{B} & \cdots & a_{1n}\mathbf{B} \\
       \vdots        & \ddots & \vdots \\
    a_{m1}\mathbf{B} & \cdots & a_{mn}\mathbf{B}
  \end{bmatrix}
  \in \mathbb{R}^{mp \times nq}$.
   \item For column vectors $\mathbf{a}=[a_1,\ldots,a_m]^T$ and $\mathbf{b}=(b_1,\ldots,b_n)^T$, $\mathbf{a} \otimes \mathbf{b}=[a_1\mathbf{b};\ldots;a_m\mathbf{b}]\in \mathbb{R}^{mn}$.
  \item For column vector $\mathbf{a}=[a_1,\ldots,a_m]^T$ and matrix $\mathbf{B} \in \mathbb{R}^{p\times q}$, $\mathbf{a} \otimes \mathbf{B}=[a_1\mathbf{B},\ldots,a_m\mathbf{B}]\in \mathbb{R}^{mp\times n}$.
  \end{itemize}
\end{itemize}

\begin{table*}[t]
\caption{Notations and Descriptions.}
\label{notation-table}
\renewcommand{\arraystretch}{1.3}
\begin{center}
\begin{tabular}{|c|l|}
\hline
\hline
\textbf{Notations}  &  \hspace{3cm} \textbf{Descriptions}                                     \\\hline
$\mathbb{F}_{2^q}$  & Galois field of characteristic two                    \\\hline
$\mathbb{R}$        & {T}he set of real numbers                                            \\\hline
$\mathbf{A}$        & Matrix                                            \\\hline
$\mathbf{a}$        & {C}olumn vector
 \\\hline
$a$                 & Scalar
 \\\hline
 $\{0,1\}^a$        & a-length binary column vector
 \\\hline
 \hspace{0.2cm}$\{0,1\}^{a\times b}$         & a-by-b binary matrix
 \\\hline
$[\mathbf{a}; \mathbf{b}]$ or $[\mathbf{A}; \mathbf{B}]$             & Vectors or matrices {are} concatenated in column-wise                 \\\hline
$[\mathbf{a}, \mathbf{b}]$ or $[\mathbf{A},\mathbf{B}]$             & Vectors or matrices are concatenated in row-wise                      \\\hline
$\mathbf{1}_{a}$ or $\mathbf{1}_{a\times b}$             & Length-$a$ all-ones vector or $a$-by-$b$ all-ones matrix                   \\\hline
$\mathbf{0}_{a}$ or $\mathbf{0}_{a\times b}$             & Length-$a$ all-zeros vector or $a$-by-$b$ all-zeros matrix                      \\\hline
$\mathbf{I}_{a}$             & $a\times a$ identity matrix                   \\\hline
$(\cdot)^T$         & Transpose operator                                    \\\hline
$\|\cdot\|_{2}$     &  $\ell_2$-norm                                  \\\hline
$\preceq$           & Generalized inequality          \\\hline
$\otimes$           & Kronecker product                                     \\\hline
$\delta_{\mathbf{A}}$             & Spectral norm of matrix $\mathbf{A}$                      \\\hline
$\textrm{diag}(\cdot)$    & Vector/matrix diagonalization operator                                \\\hline
$\lambda_{\min}(\A^T\!\A)$    & Minimum eigenvalue of matrix $\A^T\A$                                 \\\hline
$\underset{\mathcal{X}}\Pi$    & Euclidean projection onto  set $\mathcal{X}$                                 \\\hline
\hline\hline
\end{tabular}
\end{center}
\end{table*}

The rest of this paper is organized as follows. In Section \ref{nonbinary-ML-model}, we briefly introduce the formulation of the ML decoding problem for nonbinary linear block codes.
In Section \ref{Problem formulation}, we establish a relaxed QP decoding model for nonbinary LDPC codes in $\mathbb{F}_{2^q}$ via decomposition and relaxation techniques of the three-variables parity-check equation.
Moreover, an efficient proximal-ADMM algorithm for solving the formulated QP problem is presented in Section \ref{admm-qp-section}.
Section \ref{Analysis-admm-qp-decoding} shows the convergence and complexity analyses of the proposed proximal-ADMM decoding algorithm.
Simulation results demonstrate the effectiveness of our proposed decoders for LDPC codes in Section \ref{simulation-result}.
Finally, Section \ref{Conclusion} concludes this paper.

\section{ML decoding problem formulation}\label{nonbinary-ML-model}

This section presents a brief review on the ML decoding problem formulation for nonbinary LDPC codes in $\mathbb{F}_{2^q}$. More details can also be found in \cite{Flanagan}\cite{Liu-nonbinary-journal}.

  Consider a nonbinary LDPC code {defined by an $m$-by-$n$ check matrix $\mathbf{H}$} in $\mathbb{F}_{2^q}$. Its feasible codeword set $\mathcal{C}$ can be denoted by
  \begin{equation}\label{C}
    \mathcal{C}\!=\!\bigg\{\!\mathbf{c}|{\mathbf{h}_j^T\mathbf{c}=0}, j\in\mathcal{J},       \mathbf{c}\in \mathbb{F}_{2^q}^n \!\bigg\},
  \end{equation}
   where $\mathbf{h}_j^T$, $j\in\mathcal{J}=\{1,2,\dotsb,m\}$ denotes the $j$th row vector of the parity-check matrix $\mathbf{H}$.

   Assume that a codeword $\mathbf{c} \in \mathcal{C}$ is transmitted through an additional white Gaussian noise (AWGN) channel and its corresponding output is denoted as $\mathbf{r} \in \mathbb{R}^n$. In the receiver, the aim of ML decoding is to determine which codeword has the largest {\it a priori} probability $p(\mathbf{r}|\mathbf{c})$ throughout the feasible codeword set $\mathcal{C}$. So, the ML decoding problem can be described as
    \begin{equation}\label{ML}
                \mathbf{c}^* = \underset{\mathbf{c}\in\mathcal{C}}{\rm argmax} \hspace{0.2cm} p(\mathbf{r}|\mathbf{c}).
    \end{equation}

  {Before processing the above ML decoding problem, we introduce two embedding techniques, which can map element $c_i$ in $\mathbb{F}_{2^q}$ to a binary vector in real space.}
  {\begin{enumerate}
      \item{\it Flanagan embedding\cite{Flanagan}:} the mapped vector $\mathbf{x}_i=[x_{i,1};\dotsb;
  x_{i,2^q-1}]\in\{0,1\}^{2^q-1}$. Specifically, for the nonzero element in $\mathbb{F}_{2^q}$, there are
    \begin{equation}\label{Mq2b}
       x_{i,\sigma}= \begin{cases}1,&\ \sigma=c_i, \\0,& \ \sigma\neq c_i, \end{cases}
    \end{equation}
    and for the zero element, it is a $(2^q-1)$-length all-zeros vector.
  \item {\it Constant-Weight embedding\cite{CR-journal}:} the mapped vector $\mathbf{x}_i=[x_{i,0};\dotsb;
  x_{i,2^q-1}]\in\{0,1\}^{2^q}$, where
    \begin{equation}\label{Mq2b}
       x_{i,\sigma}= \begin{cases}1,&\ \sigma=c_i, \\0,& \ \sigma\neq c_i. \end{cases}
    \end{equation}
 \end{enumerate}}

  {{\it Remarks:} Both of the above two embedding techniques\footnotemark \ can be applied to formulating the ML decoding problem \eqref{ML} to a decoding model in real space. Since most of their derivations are similar, in the following, we only provide the details on how to leverage the former embedding technique. In Appendix \ref{CR}, we provide a brief description and discussion on the latter. Moreover, both of their decoding performances are provided in the simulation section.}\footnotetext{{{\bf Example:} In $\mathbb{F}_4$, Flanagan embedding: $0\mapsto[0,0,0]$, $1\mapsto[1,0,0]$, $2\mapsto[0,1,0]$, and $3\mapsto[0,0,1]$ and Constant-Weight embedding: $0\mapsto[1,0,0,0]$, $1\mapsto[0,1,0,0]$, $2\mapsto[0,0,1,0]$, and $3\mapsto[0,0,0,1]$.}}

  When the Flanagan embedding technique is applied, any codeword $\mathbf{c}$ can be mapped to a binary vector $\mathbf{x}=[\mathbf{x}_1; \dotsb; \mathbf{x}_n]\in\{0,1\}^{n(2^q-1)}$. {Specifically, we call $\mathbf{x}$ as an equivalent binary codeword of the nonbinary codeword $\mathbf{c}$.} Let $\mathcal{X}$ denote the set consisting of all equivalent binary codewords. Then, the ML decoding problem \eqref{ML} can be transformed to
  \begin{equation}\label{ML_X}
                \mathbf{x}^* = \underset{\mathbf{x}\in\mathcal{X}}{\rm argmax} \hspace{0.2cm} p(\mathbf{r}|\mathbf{x}),
    \end{equation}
 which can be further derived as
 \begin{equation}\label{ML 2}
  \begin{split}
    \underset{\mathbf{x}\in\mathcal{X}}{\rm argmax} \hspace{0.2cm} p(\mathbf{r}|\mathbf{x}) & = \underset{\mathbf{x}\in\mathcal{X}}{\rm argmax} \hspace{0.1cm} \prod_{i=1}^n\prod_{\sigma=1}^{2^q-1} p(r_i|x_{i,\sigma}) \\
    & = \underset{\mathbf{x}\in\mathcal{X}}{\rm argmin} \hspace{0.1cm} \sum_{i=1}^n\sum_{\sigma=1}^{2^q-1} -\log p(r_i|x_{i,\sigma}).
    \end{split}
 \end{equation}
 By adding the constant $\displaystyle\sum_{i=1}^n\sum_{\sigma=1}^{2^q-1} \log p(r_i|x_{i,\sigma}=0)$ to \eqref{ML 2}, we obtain
   \begin{equation}\label{ML_3}
        \begin{split}
            \underset{\mathbf{x}\in\mathcal{X}}{\rm argmax}\ p(\!\mathbf{r}|\mathbf{x}\!) \!\!&= \underset{\mathbf{x}\in\mathcal{X}}{\rm argmin} \displaystyle\sum_{i=1}^n\sum_{\sigma=1}^{2^q-1} \log\frac{p(r_i|x_{i,\sigma}=0)}{p(r_{i}|x_{i,\sigma})} \\
                                    \hspace{-0.1cm} &= \underset{\mathbf{x}\in\mathcal{X}}{\rm argmin} \displaystyle\sum_{i=1}^n\sum_{\sigma=1}^{2^q-1} x_{i,\sigma}\log\frac{p(r_{i}|x_{i,\sigma}=0)}{p(r_{i}|x_{i,\sigma}=1)} \\
                                     &=  \underset{\mathbf{x}\in \mathcal{X}}{\rm argmin}\ \boldsymbol\gamma^T\mathbf{x},
        \end{split}
    \end{equation}
    where $\boldsymbol{\gamma}\in\mathbb{R}^{n(2^q-1)}$ is defined by \eqref{cost vector}.
    \begin{figure*}
    \begin{equation}\label{cost vector}
        \begin{split}
        {\boldsymbol{\gamma}}\!=\!\bigg[
                                    \log\frac{p(r_1|x_{1,1}\!=\!0)}{p(r_1|x_{1,1}\!=\!1)}\!, \!\dotsb,\! \log\frac{p(r_1|x_{1,2^q-1}\!=\!0)}{p(r_{1}|x_{1,2^q-1}\!=\!1)}\!,\!
             \dotsb,
                                   \log\frac{p(r_n|x_{n,1}\!=\!0)}{p(r_n|x_{n,1}\!=\!1)}\!,\!
                                    \dotsb,\! \log\frac{p(r_n|x_{n,2^q-1}\!=\!0)}{p(r_n|x_{n,2^q-1}\!=\!1)}\!
                                \bigg].
        \end{split}
     \end{equation}
     \end{figure*}
    Then, the ML decoding problem \eqref{ML_X} can be formulated as the following integer program
    \begin{subequations}\label{ML_x}
        \begin{align}
                &\underset{\mathbf{x}}{\rm min} \hspace{0.31cm} \boldsymbol\gamma^T\mathbf{x},
                 \label{ML_x_a} \\
                & \hspace{0.1cm} {\rm s.t.} \hspace{0.355cm}     \mathbf{x}\in\mathcal{X}. \label{ML_x_b}
        \end{align}
    \end{subequations}

    {The problem \eqref{ML_x} is difficult to solve since constraint \eqref{ML_x_b} is nonconvex. In the following section,
    we exploit techniques of decomposition, relaxation, penalty, and tightness to transform \eqref{ML_x} to a nonconvex, but tractable quadratic optimization problem.}

\section{ML Problem Relaxation}\label{Problem formulation}
{In this section, we consider how to relax the ML decoding problem \eqref{ML_x} to a tractable QP decoding model for nonbinary LDPC codes in $\mathbb{F}_{2^q}$.}

\subsection{Decomposition of the multi-variables parity-check equation}

Consider the $j$th parity-check equation in \eqref{C}. Without loss
of generality, we assume it involves $d_j\geq3$ variables, which are denoted by $c_{\sigma_1},\dotsb, c_{\sigma_{d_j}}$ and the corresponding coefficients are $h_{\sigma_1},\dotsb, h_{\sigma_{d_j}}$. In the following, we show that it can be decomposed equivalently to $d_j-2$ three-variables parity-check equations by introducing $d_j-3$ auxiliary variables. The detailed decomposing procedure consists of three steps:

Step 1: for the first two variables $c_{\sigma_1}$ and $c_{\sigma_2}$, we introduce an auxiliary variable $g_1$ in $\mathbb{F}_{2^q}$ and let them satisfy
\begin{equation}\label{decom-1}
 {h_{\sigma_1}c_{\sigma_1}+h_{\sigma_2}c_{\sigma_2}+g_1}=0.
\end{equation}

Step 2: for the variables $c_{\sigma_{3}},\dotsb,c_{\sigma_{d_j-2}}$, we introduce auxiliary variables $g_{2},\dotsb, g_{d_j-3}$ in $\mathbb{F}_{2^q}$ and let them satisfy \begin{equation}\label{decom-2}
{g_{t-1}+h_{\sigma_{t+1}}c_{\sigma_{t+1}}+g_{t}}=0,~~ t=2, \dotsb, d_j-3.
\end{equation}

Step 3: let auxiliary variable $g_{d_j-3}$ in $\mathbb{F}_{2^q}$ and the last two variables $c_{\sigma_{d_j-1}}$ and $c_{\sigma_{d_j}}$ satisfy
\begin{equation}\label{decom-3}
   {g_{d_j-3}+h_{\sigma_{d_j-1}}c_{\sigma_{d_j-1}}+h_{\sigma_{d_j}}c_{\sigma_{d_j}}}=0.
\end{equation}

For the above decomposition procedure, we have the following fact:

\begin{fact} The set of the three-variables parity-check equations \eqref{decom-1}-\eqref{decom-3} is equivalent to the $j$th parity-check equation $\mathbf{h}_j^T\mathbf{c}=0$ in \eqref{C} in the sense that their solutions are one-to-one correspondent.
\end{fact}
{\it Proof:}
See Appendix \ref{check-equ-proof}.

Applying \eqref{decom-1}-\eqref{decom-3} to all of the parity-check equations in \eqref{C}, the total numbers of the three-variables parity-check equations in $\mathbb{F}_{2^q}$ and the introduced auxiliary variables are
\begin{equation}\label{gamma_a_c}
  \begin{split}
   &\Gamma_{c} = \sum_{j=1}^{m}(d_j-2),  ~~ \Gamma_{a} = \sum_{j=1}^{m}(d_j-3),
  \end{split}
\end{equation} respectively.

In the following, we show that any three-variables parity-check equation in $\mathbb{F}_{2^q}$ has an equivalent expression in real space.

\subsection{Equivalent expression of the three-variables parity-check equation in real space}\label{embed-section}
\begin{figure}[tp]
  \centering
  \centerline{\psfig{figure=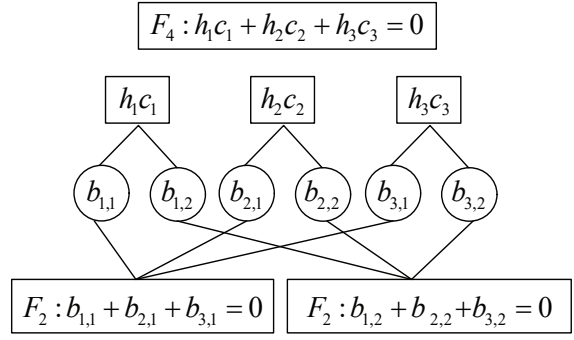,width=7.5cm,height=4.5cm}}
  \caption{The factor graph representation for equivalence between the three-variables parity-check equation in $\mathbb{F}_4$ and the two three-variables parity-check equations in $\mathbb{F}_2$, where $h_1, h_2, h_3\in\mathbb{F}_4\setminus0$, $c_1, c_2, c_3\in\mathbb{F}_4$ and $b_{1,1}, b_{1,2}, b_{2,1}, b_{2,2}, b_{3,1}, b_{3,2}\in\mathbb{F}_2$.}
  \label{factor_graph}
\end{figure}

 Consider the following three-variables parity-check equation in $\mathbb{F}_{2^q}$
\begin{equation}\label{three-variabls check euqation}
  {\displaystyle\sum_{k=1}^3 h_kc_k =0,}
\end{equation}
where {$h_k, c_k\in\mathbb{F}_{2^q}$ but $h_k$ is nonzero.} Since $h_kc_k \in \mathbb{F}_{2^q}$ can be expressed exactly as\footnotemark
\begin{equation}\label{c}
   \mathcal{F}(h_kc_k) = \displaystyle\sum_{i=1}^q b_{i,k}\zeta^{i-1},
\end{equation}
where {$b_{i,k}\in\mathbb{F}_2$}, $\zeta$ is the primitive element in $\mathbb{F}_{2^{q}}$, and function $\mathcal{F}(\cdot)$ denotes $h_kc_k$'s  polynomial representation in $\mathbb{F}_{2^{q}}$. Then,  \eqref{three-variabls check euqation} is equivalent to the following $q$ three-variables parity-check equations in $\mathbb{F}_2$
\footnotetext{
 {Different representations of nonzero elements in $\mathbb{F}_4$:}
  \[\begin{tabular}{|c|c|c|c|}
  \hline
              & polynomial &bits & integer \\ \hline
    $\zeta^0$ & 1          &01   & 1     \\ \hline
    $\zeta^1$ & $\zeta$    &10   & 2     \\ \hline
    $\zeta^2$ & $\zeta+1$  &11   & 3     \\ \hline
\end{tabular}
\]}
\begin{equation}\label{three-variable pairty-check}
  \begin{split}
    \sum_{k=1}^3 b_{i,k}=0,\ i=1,2,\dotsb,q.\footnotemark
  \end{split}
\end{equation}
 Figure \ref{factor_graph} shows an example of equivalence between a three-variables  parity-check equation in $\mathbb{F}_4$ and two three-variables parity-check equations in $\mathbb{F}_2$.
\footnotetext{{\textbf{Example}: Consider $1\cdot 2+2\cdot 2+3\cdot 2=0$ in $\mathbb{F}_4$. Since $1\cdot 2=2$, $2\cdot 2=3$, $3\cdot 2=1$, and  $\mathcal{F}(2)=\zeta$, $\mathcal{F}(3)=1+\zeta$, and $\mathcal{F}(1)=1$ respectively, we see that the corresponding coefficient vectors of the polynomials are $[b_{1,1},b_{2,1}]=[0, 1]$, $[b_{1,2},b_{2,2}]=[1, 1]$, and $[b_{1,3},b_{2,3}]=[1, 0]$ respectively. Then, one can see that $1\cdot 2+2\cdot 2+3\cdot 2=0$ in $\mathbb{F}_4$ is equivalent to two parity-check equations in $\mathbb{F}_2$: $0+1+1=0$ and $1+1+0=0$.}}

Next, we consider how to obtain equivalent expression of the parity-check equations \eqref{three-variable pairty-check} in real space. First, notice any nonzero element $h_k \in \mathbb{F}_{2^{q}}$ can be mapped to an elementary matrix  $\mathbf{D}(2^q,h_k)\in\{0,1\}^{(2^q-1)\times(2^q-1)}$ \cite{Liu-nonbinary-journal}, whose entries are defined by \footnote{\textbf{Example:} In $\mathbb{F}_4$, $\mathbf{D}(4,1) = \begin{bmatrix} 1 & 0 & 0\\ 0 & 1 & 0\\ 0 & 0& 1\end{bmatrix}$, $\mathbf{D}(4,2) = \begin{bmatrix} 0 & 0 & 1\\ 1 & 0 & 0\\ 0& 1& 0\end{bmatrix}$, and $\mathbf{D}(4,3) = \begin{bmatrix} 0 & 1 & 0\\ 0 & 0 & 1\\ 1& 0& 0\end{bmatrix}$.}
\begin{equation}\label{permu-matrix}
D(2^q,h_k)_{ij}=
\begin{cases}
1,& \textrm{if}~ i=j\cdot h_k,\\
0,& \textrm{otherwise},
\end{cases}
\end{equation}
{where the multiplication in $j\cdot h_k$ is in $\mathbb{F}_{2^q}$.} Then, we have the following lemma.
\begin{lemma}\label{hu-dD-u-define}
   Let $\mathbf{x}_k$ be $c_k$'s equivalent binary codeword according to mapping rule \eqref{Mq2b}. Then, $\mathbf{D}(2^q,h_k)\mathbf{x}_k$ is {an equivalent binary codeword to $h_kc_k$ mapped according to the same rule.}
\end{lemma}
 {\it Proof:} See Appendix \ref{hu-dD-u-define-proof}.

Let binary vector $\tilde{\mathbf{b}}(\tilde{\alpha})=[\tilde{b}_1; \dotsb; \tilde{b}_q]$ be bits expression for any nonzero element  $\tilde{\alpha}\in\mathbb{F}_{2^q}$ and formulate the following $q$-by-$(2^q-1)$ binary matrix\footnotemark
\begin{equation}\label{B}
   \mathbf{B}\!\!=\!\!\begin{bmatrix}\tilde{\mathbf{b}}(1), \tilde{\mathbf{b}}(2),\ldots,\tilde{\mathbf{b}}(\!2^q\!-\!1\!)\!\end{bmatrix}=\begin{bmatrix}
           1&       0&   \dotsb & 0&       1 \\
           0&       1&   \dotsb & 1&       1  \\
      \vdots&  \vdots&   \vdots & \vdots& \vdots \\
           0&       0&   \dotsb & 1&       1
   \end{bmatrix}.
\end{equation}
\footnotetext{\textbf{Example:} In $\mathbb{F}_{4}$,
   $\mathbf{B}=\begin{bmatrix} \tilde{\mathbf{b}}(1), \tilde{\mathbf{b}}(2), \tilde{\mathbf{b}}(3)\end{bmatrix}=\begin{bmatrix}
           1&   0&  1  \\
           0&   1&  1
   \end{bmatrix}.$}
Let $\hat{\mathbf{b}}_i^T$ denote the $i$th row vector of matrix $\mathbf{B}$. Then, we can obtain\footnotemark
\begin{equation}\label{cik}
  b_{i,k} = \hat{\mathbf{b}}_i^T\mathbf{D}(2^q,h_k)\mathbf{x}_k, \ \forall i=1,2,\ldots,q; \ k=1,2,3.
\end{equation}
{Note that the multiplications/additions in \eqref{cik} operate in real space, but the result can only be 1 or 0  since matrix $\mathbf{D}(2^q,h_k)$ is elementary and $\mathbf{x}_k$ only includes at most one nonzero element 1. It means $b_{i,k}$ lies in $\mathbb{F}_2$. Therefore, three-variables parity-check equations \eqref{three-variable pairty-check} in $\mathbb{F}_2$ can be rewritten as
\begin{equation}\label{TDX0}
       \hat{\mathbf{b}}_i^T\mathbf{D}(2^q,h_1)\mathbf{x}_1\!+\!\hat{\mathbf{b}}_i^T\mathbf{D}(2^q,h_2)\mathbf{x}_2\!+\!\hat{\mathbf{b}}_i^T\mathbf{D}(2^q,h_3)\mathbf{x}_3\!=\!0, \end{equation}
where the sum is in $\mathbb{F}_2$ and $i=1,2,\ldots,q$.}
\footnotetext{{\bf Example:} To be clear, consider multiplication $3\cdot2=1$ in $\mathbb{F}_4$: according to Lemma 1, equivalent binary codeword of $3\cdot2$ can be determined by $\begin{bmatrix} 0 & 1 & 0\\ 0 & 0 & 1\\ 1& 0& 0\end{bmatrix}\begin{bmatrix} 0 \\  1\\  0\end{bmatrix} =\begin{bmatrix} 1 \\  0\\  0\end{bmatrix}$. Moreover, $\mathbf{B}\begin{bmatrix} 0 \\  1\\  0\end{bmatrix}=\begin{bmatrix}
           1&   0&  1  \\
           0&   1&  1
   \end{bmatrix}\begin{bmatrix} 1 \\  0\\  0\end{bmatrix}=\begin{bmatrix} 1 \\  0 \end{bmatrix}$.}
Moreover, three-variables parity-check equations \eqref{three-variable pairty-check} can be equivalent to the following inequality system defined in real space
 \begin{equation}\label{four-inequa}
   \begin{split}
  & f_{i,1} \leq f_{i,2}+f_{i,3},~~ f_{i,2} \leq f_{i,1}+f_{i,3}, \\
  & f_{i,3} \leq f_{i,1}+f_{i,2},~~ f_{i,1}+f_{i,2}+f_{i,3} \leq 2,\\
  & f_{i,1},f_{i,2},f_{i,3} \in \{0,1\},~ i=1,2,\dotsb,q,
   \end{split}
 \end{equation}
in the sense that {solutions $b_{i,k}$ and $f_{i,k}$ are one-to-one correspondent.\footnotemark}
\footnotetext{The one-to-one correspondence can be seen clearly from the following example: assume that $[b_{1,1},b_{2,1}]=[1, 0]$, $[b_{1,2},b_{2,2}]=[0, 1]$, and $[b_{1,3},b_{2,3}]=[1, 1]$. It is easy to check that they satisfy \eqref{three-variable pairty-check}. Moreover, it is obvious that reals $[f_{1,1},f_{2,1}]=[1, 0]$, $[f_{1,2},f_{2,2}]=[0, 1]$, and $[f_{1,3},f_{2,3}]=[1, 1]$ satisfy the inequality system \eqref{four-inequa}. {This example shows that if $b_{i,k}=1$, then $f_{i,k}=1$, or if $b_{i,k}=0$, then $f_{i,k}=0$.} Notice $b_{i,k}\in \mathbb{F}_2$ and $f_{i,k}$ are in real space.} Let
   \begin{equation}\label{t F matrix}
      \begin{split}
         \mathbf{t}=\begin{bmatrix}\ 0\ \\ \ 0\ \\ \ 0\ \\ \ 2\ \end{bmatrix}, \ \
         \mathbf{P} = \begin{bmatrix}
                           ~~1  &-1  & -1 \\
                          -1  &~~1 & -1 \\
                          -1  &-1  &~~1 \\
                          ~~1  &~~1 &~~1
                       \end{bmatrix}.
      \end{split}
   \end{equation}
Then, \eqref{four-inequa} can be rewritten as
\begin{equation}\label{three-variable-checks-matrix-form}
  \begin{split}
   \mathbf{P}\mathbf{f}_{i} \preceq \mathbf{t},\ \mathbf{f}_i \in \{0,1\}^3,~ i=1,2,\ldots,q,
  \end{split}
\end{equation}
where $\mathbf{f}_i=[f_{i,1}; f_{i,2}; f_{i,3}]$. {Notice that $f_{i,k}$ can take the place of the $k$th term $b_{i,k}$ in \eqref{TDX0}.} Therefore, define
\begin{subequations}\label{t F matrix2}
\begin{align}
&\mathbf{T}_{i}={\rm{diag}}(\hat{\mathbf{b}}^T_{i},\hat{\mathbf{b}}^T_{i},\hat{\mathbf{b}}^T_{i}) \in \{0,1\}^{3\times3(2^q-1)},\quad \quad\quad\quad\,\,\label{Ti}\\
&\mathbf{D}\!=\!{\rm diag}(\!\mathbf{D}(\!2^q\!,\!h_1\!)\!,\!\mathbf{D}(\!2^q\!,\!h_2)\!,\!\mathbf{D}(2^q\!,\!h_3)\!)\!\!\in\! \!\{\!0\!,\!1\!\}\!^{3\!(\!2^q\!-\!1\!)\!\!\times\!3\!(\!2^q\!-\!1\!)\!}\!,\label{D}\\
&\mathbf{x}=[\mathbf{x}_{1}; \mathbf{x}_{2}; \mathbf{x}_{3}] \in \{0,1\}^{3(2^q-1)}. \label{x}
\end{align}
\end{subequations}
We can obtain
\begin{equation}\label{f}
 \mathbf{f}_i=\mathbf{T}_i\mathbf{Dx}.
\end{equation}
Then, plugging \eqref{f} into \eqref{three-variable-checks-matrix-form}, we have
\begin{equation}\label{sum T-matrix-form-2}
   \begin{split}
 &  \mathbf{P}\mathbf{T}_{i}\mathbf{D}\mathbf{x} \preceq \mathbf{t}, ~ \forall i\in\{1,2,\ldots,q\}.
  \end{split}
\end{equation}

Furthermore, let
\begin{subequations}\label{WwE}
 \begin{align}
    &\mathbf{W}=[\mathbf{P}\mathbf{T}_{1}\mathbf{D};\ldots;\mathbf{P}\mathbf{T}_{q}\mathbf{D}] \in \mathbb{R}^{4q \times 3(2^q-1)}, \label{W}\\
    &\mathbf{w}=\mathbf{1}_{q} \otimes\mathbf{t} \in \mathbb{R}^{4q}. \label{w}
 \end{align}
 \end{subequations}
{Then, the three-variables parity-check equation \eqref{three-variabls check euqation} can be equivalent to}
\begin{equation}\label{Ww}
  \mathbf{Wx}\preceq\mathbf{w}, \ \mathbf{x}\in \{0,1\}^{3(2^q-1)}
\end{equation}
{in the sense that their solutions $\mathbf{c}$ and $\mathbf{x}$ are} one-to-one correspondent {under the mapping rule \eqref{Mq2b}.}

\subsection{Equivalent ML decoding problem}
In this subsection, we consider to establish a linear integer program equivalent to the ML decoding problem \eqref{ML_x} for nonbinary LDPC codes in $\mathbb{F}_{2^q}$.

First, we define
\begin{equation}\label{v}
  \mathbf{v}=[\mathbf{x}; \mathbf{s}] \in \{0,1\}^{(2^q-1)(n+\Gamma_a)},
\end{equation}
  where $\mathbf{s}=[\mathbf{s}_1; \dotsb; \mathbf{s}_i;\dotsb; \mathbf{s}_{\Gamma_a}]\in\{0,1\}^{(2^q-1)\Gamma_a}$ and $\mathbf{s}_i\in\{0,1\}^{2^q-1}$ correspond to the $i$th auxiliary variables introduced in the decomposition procedure \eqref{decom-1}-\eqref{decom-3}. Define a variable-selecting matrix $\mathbf{Q}_{\tau}\in\{0,1\}^{3\times(n+\Gamma_a)}$ corresponding to the $\tau$th decomposed three-variables parity-check equation in $\mathbb{F}_{2^q}$ for parity-check equations in \eqref{C}, where $\tau=1,\dotsb, \Gamma_c$. Its every row vector includes only one ``1'', whose index corresponds to the variable in \eqref{C}.\footnotemark \    Therefore,  $(\mathbf{Q}_{\tau}\otimes\mathbf{I}_{2^q-1})\mathbf{v}$ are the variables involved in the $\tau$th three-variables parity-check equation in $\mathbb{F}_2$.
  \footnotetext{{\bf Example}: Consider the parity-check equation in $\mathbb{F}_4$: $c_1+c_2+2c_3+3c_4=0$. According to \eqref{decom-1}-\eqref{decom-3}, it can be decomposed to two three-variables parity-check equations: $g_1+c_1+c_2=0$, $g_1+2c_3+3c_4=0$. Then, the corresponding variable-selecting matrices  $\mathbf{Q}_1=\begin{bmatrix}
                            1& 0 & 0 & 0 & 0  \\                              0& 1 & 0 & 0 & 0   \\
                            0& 0 & 0 & 0 & 1  \end{bmatrix}$ and $\mathbf{Q}_2=\begin{bmatrix}
                            0& 0 & 1 & 0 & 0  \\                              0& 0 & 0 & 1 & 0  \\
                            0& 0 & 0 & 0 & 1 \end{bmatrix}$.}
  Moreover, we define
\begin{subequations}\label{Abq}
     \begin{align}
     &\mathbf{F}=[\mathbf{W}_1(\mathbf{Q}_1\otimes\mathbf{I}_{{2^q-1}}); \cdots;\mathbf{W}_{\tau} (\mathbf{Q}_\tau\otimes\mathbf{I}_{{2^q-1}});\cdots;\nonumber\\
     &\;\;\;\;\quad\mathbf{W}_{\Gamma_c}(\mathbf{Q}_{\Gamma_c}\otimes\mathbf{I}_{{2^q-1}})] \in \mathbb{R}^{4q\Gamma_c \times (n+\Gamma_a)(2^q-1)}, \label{Abq-b}\\
     &\mathbf{u}=\mathbf{1}_{{\Gamma_c}} \otimes {\mathbf{w}}  \in \mathbb{R}^{4q\Gamma_c}.                \label{Abq-d}
    \end{align}
\end{subequations}
Then, by \eqref{Ww}, we have
\begin{equation}\label{Ff}
  {\mathbf{F}}\mathbf{v} \preceq \mathbf{u}.
\end{equation}
Therefore, we can transform the ML decoding problem \eqref{ML_x} to the following linear integer program
\begin{subequations}\label{ML-decoding}
\begin{align}
&\underset{\mathbf{v}}{\rm min} \hspace{0.35cm} {\boldsymbol\lambda}^{T}\mathbf{v},\\
&\hspace{0.1cm} \rm{s. t.} \hspace{0.25cm} {\mathbf{F}}\mathbf{v} \preceq \mathbf{u},\  \label{ML-decoding-b}\\
& \hspace{0.9cm} \mathbf{v}\in \{0,1\}^{(2^q-1)(n+\Gamma_a)}, \label{ML-decoding-c}
\end{align}
\end{subequations}
where $\boldsymbol\lambda=[\pmb{\gamma}; \mathbf{0}_{\Gamma_a(2^q-1)}]\in \mathbb{R}^{(2^q-1)(n+\Gamma_a)}$.

Due to the binary constraints \eqref{ML-decoding-c}, problem \eqref{ML-decoding} is difficult to solve. {In the following, we relax it to a continuous, nonconvex, but tractable  model.}

{\subsection{Proposed Decoding Model}
In this subsection, we exploit a simple relaxation method for the linear integer program \eqref{ML_x} and then introduce three techniques to alleviate the {\it relaxation} effect on the optimal solution, which leads to the proposed decoding model for nonbinary LDPC codes in $\mathbb{F}_{2^q}$.}

The typical way for the binary constraint \eqref{ML-decoding-c} is to relax it to a box constraint, i.e.,  $\v \in[0,1]^{(2^q-1)(n+\Gamma_a)}$, which can simplify the nonconvex problem \eqref{ML-decoding} to a convex one. However, the resulting optimization problem's optimal solution could be fractional especially when the decoder works in low SNR regions. {In the following, we deploy three techniques to handle this problem.}

The first one is to add a quadratic penalty term into the objective, i.e., \begin{equation}\label{penalty}
  \pmb{\lambda}^{T}\mathbf{v}-\frac{\alpha}{2}\|\mathbf{v}-0.5\|_{2}^{2},
\end{equation}
where $\alpha >0$ is a preset constant. Empirically, the quadratic penalty makes the optimal integer solutions favorable and improves error-correction performance of LDPC decoders significantly \cite{bai-admm-qp-binary} \cite{penalty-decoder} \cite{Liu-nonbinary-journal}.

The second one is to introduce extra linear constraints to {tighten the relaxation}. {Specifically, let $\{\mathcal{K}_{\ell}|\ell=1,\dotsb,2^q-1\}$ denote all the subsets of the set $\{1,\dotsb,q\}$. Then, \eqref{TDX0} can be equivalent to \eqref{Bxik},
\begin{figure*}
\begin{equation}\label{Bxik}
  \begin{split}
       \bigg(\!\displaystyle\sum_{i\in\mathcal{K}_{\ell}}\hat{\mathbf{b}}_i^T\!\bigg)\mathbf{D}(2^q,h_1)\mathbf{x}_1\!+\!
       \bigg(\!\displaystyle\sum_{i\in\mathcal{K}_{\ell}}\hat{\mathbf{b}}_i^T\!\bigg)\mathbf{D}(2^q,h_2)\mathbf{x}_2\!+\!
       \bigg(\!\displaystyle\sum_{i\in\mathcal{K}_{\ell}}\hat{\mathbf{b}}_i^T\!\bigg)\mathbf{D}(2^q,h_3)\mathbf{x}_3\!=\!0,
       \ell\!=\!1,\ldots,2^q-1,
  \end{split}
\end{equation}
\end{figure*}
where the addition $\displaystyle\sum_{i\in\mathcal{K}_{\ell}}$ is in $\mathbb{F}_2$.
Define
\begin{subequations}\label{Ww_hat}
 \begin{align}
    &\hat{\mathbf{T}}_{\ell} = {\rm diag}(\displaystyle\sum_{i\in\mathcal{K}_{\ell}}\hat{\mathbf{b}}_i^T, \displaystyle\sum_{i\in\mathcal{K}_{\ell}}\hat{\mathbf{b}}_i^T, \displaystyle\sum_{i\in\mathcal{K}_{\ell}}\hat{\mathbf{b}}_i^T), \label{T_l} \\
    &\hat{\mathbf{W}}=[\mathbf{P}\hat{\mathbf{T}}_{1}\mathbf{D};\ldots;\mathbf{P}\hat{\mathbf{T}}_{2^q-1}\mathbf{D}] \in \mathbb{R}^{4(2^q-1)\times 3(2^q-1)}, \label{W_hat}\\
    &\hat{\mathbf{w}}=\mathbf{1}_{2^q-1} \otimes\mathbf{t}\in\mathbb{R}^{4(2^q-1)}, \label{w_hat}
 \end{align}
 \end{subequations}
}
{Then, according to \eqref{four-inequa}, we can obtain  \eqref{Bxik}'s equivalent inequalities system in real space as follows:}
\begin{equation}\label{ineq redundant}
\begin{split}
  & \hat{\mathbf{W}}\mathbf{x}\preceq \hat{\mathbf{w}}, \ \  \mathbf{x}\in\{0,1\}^{3(2^q-1)}.
\end{split}
\end{equation}

Since $\hat{\mathbf{w}}$ is a $4(2^q-1)$-length vector and $\hat{\mathbf{W}}$ is a $4(2^q-1)$-by-$3(2^q-1)$ matrix, \eqref{ineq redundant} consists of $4(2^q-1)$ inequalities. Besides $4q$ inequalities same as in \eqref{Ww}, other inequalities in \eqref{ineq redundant} can be seen as redundant ones since they are combined by inequalities in \eqref{Ww}. However, when the binary constraint $\x \in\{0,1\}^{3(2^q-1)}$ is relaxed  to the box constraint $\x \in[0,1]^{3(2^q-1)}$, these redundant inequalities can play a role in tightening the relaxation. By defining
\begin{subequations}\label{Fv}
\begin{align}
&{\hat{\mathbf{F}}}=[\hat{\mathbf{W}}_1(\mathbf{Q}_1\otimes\mathbf{I}_{{2^q-1}}); \cdots;\hat{\mathbf{W}}_{\tau} (\mathbf{Q}_\tau\otimes\mathbf{I}_{{2^q-1}});\cdots;\quad\quad\quad\quad\;\;\nonumber\\
&\quad\quad\hat{\mathbf{W}}_{\Gamma_c}\!(\mathbf{Q}_{\Gamma_c}\!\otimes\!\mathbf{I}_{{2^q-1}}\!)]\! \in\! \mathbb{R}^{4(2^q\!-\!1)\Gamma_{c}\!\times\! (n+\Gamma_a)(2^q\!-\!1)},\label{Fv-F} \\
&{\hat{\u}}=\mathbf{1}_{{\Gamma_c}} \otimes \hat{\mathbf{w}}\in\mathbb{R}^{4(2^q-1)\Gamma_c}. \label{Fv-v}
\end{align}
\end{subequations}
we can transform \eqref{Ff} equivalently to
\begin{equation}\label{Ff_hat}
  \hat{\mathbf{F}}\mathbf{v}\preceq\hat{\mathbf{u}}.
\end{equation}

The third one is used to exploit the structure so that the equivalent binary codeword of {the element in $\mathbb{F}_{2^q}$} has at most one 1 (see \eqref{Mq2b}). To do so, we define a binary matrix
\begin{equation}\label{S}
\mathbf{S}\!\!=\!\!{\rm diag}(\underbrace{\mathbf{1}_{2^q-1}^T\!,\dotsb,\!\mathbf{1}_{2^q-1}^T}_{n+\Gamma_a})\!\! \in \!\! \{0,1\}^{(n\!+\!\Gamma_a)\times(2^q\!-\!1)(n\!+\!\Gamma_a)},
\end{equation}
which leads to
\begin{equation}\label{Sv}
 \mathbf{Sv}\preceq \mathbf{1}_{n+\Gamma_{a}}.
\end{equation}

{Then, by combining \eqref{penalty}, \eqref{Ff_hat}, and \eqref{Sv}, we relax the ML decoding problem \eqref{ML_x} to \begin{subequations}\label{ML-decoding-all1}
\begin{align}
&\underset{\mathbf{v}}{\rm min} \hspace{0.3cm} {\boldsymbol\lambda}^{T}\mathbf{v}-\frac{\alpha}{2}\|\mathbf{v}-0.5\|_{2}^{2},\\
& \hspace{0.1cm} \rm{s.t.} \hspace{0.3cm} \hat{\mathbf{F}}\mathbf{v} \preceq \hat{\u}, \label{ML-decoding-all-b}\\
&\hspace{0.9cm}\mathbf{S}\mathbf{v}\preceq \mathbf{1}_{n+\Gamma_a}, \label{ML-decoding-all-c} \\
&\hspace{0.9cm} \mathbf{v} \in [0,1]^{(2^q-1)(n+\Gamma_a)}. \label{ML-decoding-all-d}
\end{align}
\end{subequations}}

In the next section, we present an efficient solving algorithm via the proximal-ADMM technique for the decoding problem \eqref{ML-decoding-all}, where variables in every ADMM iteration are solved analytically and in parallel.
Moreover, we show the proposed decoder is theoretically-guaranteed convergent to a stationary point of the problem \eqref{ML-decoding-all1} and its computational complexity in each iteration scales linearly with code length and size of the Galois field.

\section{Proximal-ADMM solving algorithm}\label{admm-qp-section}
{In this section, we develop a proximal-ADMM algorithm to solve the decoding problem \eqref{ML-decoding-all1}. Moreover, by exploiting its inherent structures, each subproblem in the proximal-ADMM iteration can be solved efficiently.}

\subsection{Proximal-ADMM algorithm framework}

Define
\begin{subequations}\label{Ab-construct}
\begin{align}
&\hspace{-7pt}{\mathbf{A}}=[\hat{\mathbf{W}}_1(\mathbf{Q}_1\otimes\mathbf{I}_{{2^q-1}}); \cdots;\hat{\mathbf{W}}_{\tau} (\mathbf{Q}_\tau\otimes\mathbf{I}_{{2^q-1}});\cdots;\nonumber\\
&\quad\;\;\hat{\mathbf{W}}_{\Gamma_c}(\mathbf{Q}_{\Gamma_c}\otimes\mathbf{I}_{{2^q-1}});\mathbf{S}] \in \mathbb{R}^{M\times N},\label{Ab-construct_A} \\
&\hspace{-7pt}{\boldsymbol\varrho}=[\hat{\mathbf{u}}; \mathbf{1}_{n+{\Gamma_a}}] \in \mathbb{R}^{M}, \label{Ab-construct_b}
\end{align}
\end{subequations}
where $M=4(2^q-1)\Gamma_{c}+n+\Gamma_{a}$ and $N=(n+\Gamma_a)(2^q-1)$. Then, we can transform \eqref{ML-decoding-all1} to
\begin{subequations}\label{ML-decoding-all}
\begin{align}
&\underset{\mathbf{v}}{\rm min} \hspace{0.3cm} {\boldsymbol\lambda}^{T}\mathbf{v}-\frac{\alpha}{2}\|\mathbf{v}-0.5\|_{2}^{2},\\
& \hspace{0.1cm} \rm{s.t.} \hspace{0.25cm} {\mathbf{A}}\mathbf{v} \preceq {\boldsymbol\varrho}, \label{ML-decoding-all-b}\\
&\hspace{1.0cm} \mathbf{v} \in [0,1]^{(2^q-1)(n+\Gamma_a)}. \label{ML-decoding-all-c}
\end{align}
\end{subequations}
Moreover, by introducing two auxiliary variables, $\mathbf{e}_{1}$ and $\mathbf{e}_2$, the decoding problem \eqref{ML-decoding-all} is equivalent to
\begin{subequations}\label{pADMM-frame-problem}
\begin{align}
& \hspace{0.0cm} \underset{\mathbf{v},\mathbf{e}_1,\mathbf{e}_2}{\min} \hspace{0.25cm}  \pmb{\lambda}^{T}\mathbf{v}-\frac{\alpha}{2}\|\mathbf{v}-0.5\|_{2}^{2}, \\
&\ \ {\rm s.\ t.} \hspace{0.28cm} \mathbf{A}\mathbf{v}+\mathbf{e}_1 = \boldsymbol\varrho, \label{pADMM-frame-b}\\
& \hspace{1.22cm} \mathbf{v}=\mathbf{e}_{2}, \label{pADMM-frame-c} \\
& \hspace{1.22cm} \mathbf{e}_{1} \succeq \mathbf{0}_{M}, \ \mathbf{0}_N \preceq \mathbf{e}_{2} \preceq \mathbf{1}_{N}.  \label{pADMM-frame-d}
\end{align}
\end{subequations}
Its augmented Lagrangian function  can be written as
\begin{equation}\label{aug-Lagrangian}
\begin{split}
\mathcal{L}_{\mu}(\!\mathbf{ v}\!,\!\mathbf{e}_{1}\!,\!\mathbf{e}_{2}\!,\!\mathbf{y}_{1}\!,\!\mathbf{y}_{2}\!)\!\!=\!& \pmb{\lambda}^{T}\!\mathbf{v}\!\!-\!\!\frac{\alpha}{2}\!\|\!\mathbf{v}\!\!-\!\!0.5\|_{2}^{2} \!+\!\mathbf{y}_{1}^{T}\!(\!\mathbf{A}\!\mathbf{v}\! +\!\mathbf{e}_{1}\!\!-\!\!\boldsymbol\varrho\!)\\
&\!\!\!\!\!+\!\!\mathbf{y}_{2}^{T}\!(\!\mathbf{v}\!\!-\!\!\mathbf{e}_{2}\!)\!\!+\!\! \frac{\mu}{2}\!\|\!\mathbf{A}\!\mathbf{v}\!\! +\!\!\mathbf{e}_{1}\!\!-\!\! \boldsymbol\varrho\!\|_{2}^{2} \!\!+\!\! \frac{\mu}{2}\! \|\!\mathbf{v}\!\!-\!\!\mathbf{e}_{2}\!\|\!_{2}^{2},
\end{split}
\end{equation}
where $\mathbf{y}_{1}\in\mathbb{R}^M$ and $\mathbf{y}_{2}\in\mathbb{R}^N$ are Lagrangian multipliers corresponding to equality constraints in \eqref{pADMM-frame-b} and \eqref{pADMM-frame-c} respectively and $\mu>0$ is a preset penalty parameter.
Based on the augmented Lagrangian \eqref{aug-Lagrangian}, the proximal-ADMM solving algorithm for model \eqref{pADMM-frame-problem} can be described as follows
\begin{subequations}\label{proximal-ADMM update_LP}
\begin{align}
& \v^{k+1} \!= \!\mathop{\arg \min}\limits_{\v} \mathcal{L}_{\mu}(\v,\e_{1}^{k},\e_{2}^{k},\y_{1}^{k},\y_{2}^{k})+\frac{\rho}{2}\|\v-\mathbf{p}^{k}\|_2^2, \label{proximal-x-update}  \\
& \e_{1}^{k+1} \!=\! \mathop{\arg \min}\limits_{\e_{1}\succeq\mathbf{0}_M} \mathcal{L}_{\mu}(\v^{k+1}\!,\!\e_{1}\!,\!\e_{2}^{k}\!,\!\y_{1}^{k}\!,\!\y_{2}^{k})\!+\!\frac{\rho}{2}\|\e_1\!-\!\z_1^{k}\|_2^2, \label{proximal-v1-update} \\
& \e_{2}^{k+1} \!=\!\! \mathop{\arg \min}\limits_{\mathbf{0}_N \preceq \mathbf{e}_{2} \preceq \mathbf{1}_N} \mathcal{L}_{\mu}\!(\v^{k+1}\!,\!\e_{1}^{k}\!,\!\e_{2}\!,\!\y_{1}^{k}\!,\!\y_{2}^{k})\!+\!\frac{\rho}{2}\|\e_2\!\!-\!\!\z_2^{k}\|_2^2, \label{proximal-v2-update} \\
& \mathbf{p}^{k+1}=\mathbf{p}^{k}+\beta(\v^{k+1}-\mathbf{p}^{k}),
\label{proximal-p-update}\\
& \z_{1}^{k+1}=\z_{1}^{k}+\beta(\e_{1}^{k+1}-\z_{1}^{k}), \label{proximal-z1-update}\\
& \z_{2}^{k+1}=\z_{2}^{k}+\beta(\e_{2}^{k+1}-\z_{2}^{k}),
\label{proximal-z2-update}\\
& \y_1^{k+1} = \y_1^{k} + \mu(\A\v^{k+1}+ \e_{1}^{k+1}- \boldsymbol\varrho), \label{proximal-y1-update}\\
& \y_2^{k+1} = \y_2^{k} + \mu(\v^{k+1}-\e_{2}^{k+1}), \label{proximal-y2-update}
\end{align}
\end{subequations}
where $k$ is the iteration number, $\|\v-\mathbf{p}^{k}\|_2^2$, $\|\e_1-\z_1^{k}\|_2^2$, and $\|\e_2-\z_2^{k}\|_2^2$ are the so-called  proximal terms, $\rho$ is the corresponding penalty parameter, and $\beta$ belongs to $(0,1]$. In the above iterations, variables $\mathbf{p}$, $\mathbf{z}_1$, and $\mathbf{z}_2$ are generated by an exponential averaging (or smoothing) scheme. Since extra quadratic proximal terms, centered at $\mathbf{p}$, $\mathbf{z}_1$, and $\mathbf{z}_2$, are inserted into the augmented Lagrangian function with respect to variables $\mathbf{x}$, $\mathbf{e}_1$, and $\mathbf{e}_2$ intuitively, $\mathbf{x}^{k+1}$, $\mathbf{e}_1^{k+1}$ and $\mathbf{e}_2^{k+1}$ may not deviate too much from the stabilized iterates $\mathbf{p}^k$, $\mathbf{z}_1^k$, and $\mathbf{z}_2^k$ respectively(see more details in \cite{proximal-admm}).

In the following, we show that subproblems \eqref{proximal-x-update}-\eqref{proximal-v2-update} can be solved efficiently by exploiting the inherent structures of the model.

\subsection{Solving subproblem \eqref{proximal-x-update}}

Obviously, choosing the values of parameters $\rho$, $\mu$, and $\alpha$ properly, problem \eqref{proximal-x-update} can be reduced to a strongly quadratic convex one with respect to $\mathbf{v}$.
In this case, its solving procedure can be described as follows:
by setting the gradient of the function $\mathcal{L}_{\mu}(\v,\u_{1}^{k},\u_{2}^{k},\y_{1}^{k},\y_{2}^{k})+\frac{\rho}{2}\|\v-\mathbf{p}^{k}\|_2^2$ to be zero and solving the corresponding linear equation,
we update $\mathbf{v}$ as
\begin{equation}\label{lp-x-update-solution}
\begin{split}
& \hspace{0.0cm} \mathbf{v}^{k+1} = \big(\A^{T}\A+\epsilon\I_{N})^{-1}\pmb{\varphi}^{k},\\
\end{split}
\end{equation}
where
\begin{subequations}\label{epsilon_varphi}
\begin{align}
&\epsilon=1+\frac{\rho}{\mu}-\frac{\alpha}{\mu}, \label{epsilon_varphi_a}\\
&\pmb{\varphi}^{k}\!\!=\!\! \A^{T}\!\bigg(\!\!\boldsymbol\varrho\!-\!\e_{1}^{k}\!-\!\frac{\y_1^{k}}{\mu}\bigg)\!
\!\!+\!\!\bigg(\!\e_{2}^{k}\!-\!\frac{\y_2^{k}}{\mu}\bigg)\!+\!\frac{\rho}{\mu}\mathbf{p}^{k}\!-\!\frac{\pmb{\lambda}\!+\!0.5\alpha}{\mu}. \label{epsilon_varphi_b}
\end{align}
\end{subequations}

Note that $\big(\A^{T}\A+\epsilon\I_{N}\big)^{-1}$ is fixed for a given nonbinary LDPC code. Thus, it only needs to be calculated only once throughout the proximal-ADMM iterations. Therefore, the main computational cost lies in $\big(\A^{T}\A+\epsilon\I_{N}\big)^{-1}\pmb{\varphi}^{k}$, which requires $\mathcal{O}\big((2^q-1)^2(n+\Gamma_{a})^2\big)$ multiplications. It is prohibitive for large-scale problems in practice. In the following, we show a much more efficient way to perform the computational procedure:
\begin{lemma}\label{w-inverse-lemma}
Matrix $\big(\A^{T}\A+\epsilon\I_{N}\big)^{-1}$ is block diagonal. Specifically, it can be denoted by
\begin{equation}\label{w-I-inverse-defi}
\begin{split}
\big(\A^{T}\A+\epsilon\I_{N}\big)^{-1}
={\rm diag}\Big(\big({\A}_1^{T}\A_1+\epsilon\I_{2^q-1}\big)^{-1},\ldots,\\
\big(\A_{n+\Gamma_{a}}^{T}\A_{n+\Gamma_{a}}+\epsilon\I_{2^q\!\!-\!\!1}\big)^{-1}\Big),
\end{split}
\end{equation}
where sub-matrix $\mathbf{A}_{i}$, ${\forall} i\in\{1,\dotsb, n+\Gamma_{a}\}$, is formed by  column vectors indexed from $\big((2^q-1)(i-1)+1\big)$ to $(2^q-1)i$ in matrix $\mathbf{A}$. Moreover,
\begin{equation}\label{dv-phi-I-inverse1}
\big(\A_i^{T}\A_i+\epsilon\I_{2^q-1}\big)^{-1} =\left[
  \begin{array}{cccc}
    \theta_{i} & \omega_{i} & \cdots & \omega_{i} \\
    \omega_{i} & \theta_{i} & \cdots & \omega_{i} \\
    \vdots & \vdots & \vdots & \vdots \\
    \omega_{i} & \omega_{i} & \cdots & \theta_{i} \\
  \end{array}
\right],
\end{equation}
where
\begin{equation}\label{omega_theta}
\begin{split}
  &\displaystyle\omega_{i}\!\!=\!\!\frac{-2^{q}d_{i}\!\!-\!\!1}{(2^{q}d_{i}\!+\!\epsilon)\big((2^{q+1}d_{i}\!+\!\epsilon+1)\!+(\!2^{q}d_{i}\!+\!1)(2^{q}-2)\big)}, \\
  &\theta_{i}\!\!=\!\!\omega_{i}+\frac{1}{2^{q}d_{i}+\epsilon}.
  \end{split}
\end{equation}
Here, $d_{i}$ denotes the degree of the $i$th information symbol, i.e., the number of check equations involved.
\end{lemma}

{\it Proof:} See Appendix \ref{w-inverse-lemma-proof}.

The equation \eqref{w-I-inverse-defi} indicates that $\mathbf{v}^{k+1}$ in \eqref{lp-x-update-solution} can be obtained through the following parallel implementations
\begin{equation}\label{proximal-xi-calcu}
\begin{split}
& \hspace{0.0cm} \v_{i}^{k+1}\!\!=\!\!\big(\A_i^{T}\A_i\!+\!\epsilon\I_{2^q-1}\big)^{-1}\pmb{\varphi}_i^{k}\!,\!\!\  {\forall} i\!\in\!\{1,\!2,\!\dotsb\!,\!n\!+\!\Gamma_a\},
\end{split}
\end{equation}
where $\pmb{\varphi}_i^{k}$ is the corresponding $(2^q-1)$-length sub-vector in $\pmb{\varphi}^{k}$. Moreover, based on \eqref{dv-phi-I-inverse1}, \eqref{proximal-xi-calcu} can be further derived as follows:
{\setlength\abovedisplayskip{2pt}
 \setlength\belowdisplayskip{2pt}
  \setlength\jot{2pt}
\begin{equation}\label{proximal-xi-calcu-simp}
\begin{split}
 \mathbf{v}_{i}^{k+1}
& =  \left[
  \begin{array}{cccc}
    \theta_{i}-\omega_i & 0 & \cdots & 0 \\
    0& \theta_{i}-\omega_i & \cdots & 0 \\
    \vdots & \vdots & \vdots & \vdots \\
    0& 0 & \cdots & \theta_{i}-\omega_i \\
  \end{array}
\right]\pmb{\varphi}_i^{k} \\
&\,+
 \left[
  \begin{array}{cccc}
    \omega_{i} & \omega_{i} & \cdots & \omega_{i} \\
    \omega_{i} & \omega_{i} & \cdots & \omega_{i} \\
    \vdots & \vdots & \vdots & \vdots \\
    \omega_{i} & \omega_{i} & \cdots & \omega_{i} \\
  \end{array}
\right]\pmb{\varphi}_i^{k} \\
&=(\theta_{i}-\omega_{i})\pmb{\varphi}_i^{k}+\big(\omega_{i}\sum_{\iota=1}^{2^q-1}\varphi_{i,\iota}^{k}\big)\mathbf{1}_{2^q-1},
\end{split}
\end{equation}}
where $\varphi_{i,\iota}^{k}$ denotes the $\iota th$ entry in $\pmb{\varphi}_i^{k}$.

\subsection{Solving subproblems \eqref{proximal-v1-update} and \eqref{proximal-v2-update}}

Solving \eqref{proximal-v1-update} is equivalent to solving the following $M$ subproblems in parallel
\begin{equation}\label{proximal-v1-min}
\begin{split}
& \hspace{0cm} \!\mathop {\min }\limits_{e_{1,j}} \hspace{0.1cm} y_{1, j}^{k} e_{1, j}
\!\!+\!\!\frac{\mu}{2}\!\left(\!\mathbf{a}_{j}^{T} \mathbf{v}^{k\!+\!1}\!\!+\!\!e_{1, j}\!\!-\!\!\varrho_{j}\!\right)\!^{2}\!+\!\frac{\rho}{2}\left(e_{1, j}\!\!-\!\!z_{1, j}^{k}\!\right)\!^{2}  \\
& \hspace{0.15cm} \textrm{s.t.} \hspace{0.3cm} e_{1, j} \geq 0, \ {\forall} j\in\{1,\dotsb,M\},
\end{split}
\end{equation}
where $\mathbf{a}_{j}^{T}$ denotes the $jth$ row vector of matrix $\A$.
Obviously, the optimal solution of the above problem
can be obtained by setting the gradient of its objective function to be zero and then projecting the solution
of the corresponding linear equation to region $[0,+\infty]$. Then, we can obtain
\begin{equation}\label{v1-solution-component}
e_{1,j}^{k+1} = \underset{[0,+\infty)}\Pi \frac{\mu}{\rho+\mu} \big(\varrho_{j}-\mathbf{a}_{j}^{T}\mathbf{v}^{k+1}-\frac{y_{1,j}^{k}}{\mu}+\frac{\rho}{\mu}z_{1,j}^{k}\big).
\end{equation}

Similar to \eqref{proximal-v1-update}, problem \eqref{proximal-v2-update} can be separated into the following $N$ independent subproblems
\begin{equation}\label{v2-min}
\begin{split}
& \hspace{0cm} \mathop {\min }\limits_{e_{2,\ell}} \hspace{0.1cm} -\mathrm{y}_{2, \ell}^{k} e_{2, \ell}+\frac{\mu}{2}\left(v_{\ell}^{k}-e_{2, \ell}\right)^{2}
+\frac{\rho}{2}\left(e_{2, \ell}-z_{2, \ell}^{k}\right)^{2},  \\
& \hspace{0.15cm} \textrm{s.t.} \hspace{0.3cm} 0 \leq e_{2, \ell} \leq 1, \ {\forall} \ell\in\{1,\dotsb,N\}.
\end{split}
\end{equation}
 Their optimal solutions can be expressed as
\begin{equation}\label{v2-solution-component}
e_{2,\ell}^{k+1} = \underset{[0,1]}\Pi \frac{\mu}{\rho+\mu}\big(v_{\ell}^{k+1}+\frac{y_{2,\ell}^{k}}{\mu}+\frac{\rho}{\mu}z_{2,\ell}^{k}\big).
\end{equation}

In \emph{Algorithm \ref{proxiaml-ADMM-QP}}, we summarize the proposed proximal-ADMM decoding algorithm for nonbinary LDPC codes in $\mathbb{F}_{2^{q}}$.

\begin{algorithm}[t]
\caption{Proximal-ADMM decoding algorithm}
\label{proxiaml-ADMM-QP}
\begin{algorithmic}[1]
\STATE Initializations: decompose each check equation into three-variables parity-check equations based on the parity-check matrix $\mathbf{H}$. {Construct matrix $\mathbf{T}_i$ and $\mathbf{D}$ via \eqref{Ti} and \eqref{D} respectively. Construct matrix $\mathbf{B}$ via \eqref{B} and the variable-selecting matrices $\mathbf{Q}_{\tau}$, $ \tau=1,\ldots,\Gamma_c$.  Construct matrices $\hat{\mathbf{W}}_{j}$, $j=1,\ldots, \Gamma_c$, via \eqref{Ww_hat}. Construct $\mathbf{A}$ and $\boldsymbol\varrho$ via  \eqref{Ab-construct}}. Let $\mu>0$, $\rho>\alpha$ and $0 <\beta \leq1$. For all $i \in \{1,\dotsb, n+\Gamma_{a}\}$, compute $\theta_{i}$ and $\omega_{i}$ via \eqref{omega_theta}. Initialize variables $\{\mathbf{v}, \mathbf{e}_1, \mathbf{e}_2, \mathbf{p}, \mathbf{z}_1, \mathbf{z}_2, \mathbf{y}_1, \mathbf{y}_2\}$ as all-zeros vectors. \\
\STATE  \textbf{Repeat}  \\
\STATE \hspace{0.3cm} Compute $\pmb{\varphi}^{k}$ via \eqref{epsilon_varphi}. Then, update $\v_i^{k+1}$ {via \eqref{proximal-xi-calcu-simp}} in parallel, ${\forall} i\in \{1,\ldots,n+\Gamma_{a}\}$. \\
\STATE \hspace{0.3cm} Update $e_{1,j}^{k+1}$, ${\forall} j\in\{1,\ldots,M\}$, via \eqref{v1-solution-component} in parallel.
\STATE \hspace{0.3cm} Update $e_{2,\ell}^{k+1}$, ${\forall} \ell\in\{1,\ldots,N\}$, via \eqref{v2-solution-component} in parallel.
\STATE \hspace{0.3cm} Update $\mathbf{p}^{k+1}$, $\mathbf{z}_{1}^{k+1}$, $\mathbf{z}_{2}^{k+1}$, $\y_1^{k+1}$ and $\y_2^{k+1}$ in parallel via \eqref{proximal-z1-update} -- \eqref{proximal-y2-update} respectively.
\STATE \hspace{0.2cm} $k \leftarrow k+1$. \\
\STATE \textbf{Until} some preset conditions are satisfied. \\
\end{algorithmic}
\end{algorithm}

\section{Performance Analysis}\label{Analysis-admm-qp-decoding}
{In this section, we analyze the convergence property and computational complexity of the proposed proximal-ADMM decoding algorithm.}

\subsection{Convergence}
We have the following theorem to characterize the convergence property of the proposed proximal-ADMM decoding algorithm.
\begin{theorem}\label{converge-proof-theorem}
Assume that $\rho>\alpha>0$ and
$\frac{\alpha}{\lambda_{\min}(\A^T\A)}\leq\mu\leq \frac{\rho(\rho-\alpha)^{2}}{4\delta_{\A\I}^2(\rho+L+2)^2-(\rho-\alpha)^2}$,
where $\delta_{\A\I}$ is the spectral norm of matrix
$\left[\begin{array}{ccc}{\A} & {\I_{M}} & {\mathbf{0}} \\ {~\I_{N}} & {\mathbf{0}} & {-\I_{N}}\end{array}\right]$ and $L$ denotes the Lipschitz constant for the gradient $\nabla_{\v} \mathcal{L}_{\mu}$.
Let $\{\v^{k},\e_{1}^{k},\e_{2}^{k},\mathbf{p}^{k},\z_{1}^{k},\z_{2}^{k},\y_1^{k},\y_2^{k}\}$ be the tuples generated by \emph{Algorithm \ref{proxiaml-ADMM-QP}}.
Then, we have the following convergence results
\begin{equation}\label{converge-limit}
\begin{split}
& \mathop {\lim }\limits_{k\rightarrow +\infty} \v^{k} \!\!=\!\v^{*}\!,\!~\mathop {\lim }\limits_{k\rightarrow +\infty}\e_{1}^{k} \!\!=\! \e_{1}^{*}\!,\!
  ~\mathop {\lim }\limits_{k\rightarrow +\infty} \e_{2}^{k}\!\! =\! \e_{2}^{*}, \\
& \mathop {\lim }\limits_{k\rightarrow +\infty} \p^{k} \!\!=\! \p^{*}\!,\!~\mathop {\lim }\limits_{k\rightarrow +\infty}\z_{1}^{k}\!\! =\! \z_{1}^{*}\!,\!
  ~\mathop {\lim }\limits_{k\rightarrow +\infty} \z_{2}^{k} \!\!=\! \z_{2}^{*}, \\
& \mathop {\lim }\limits_{k\rightarrow +\infty} \y_{1}^{k}\!\! =\! \y_{1}^{*}\!,\!\hspace{0.21cm} \mathop {\lim }\limits_{k\rightarrow +\infty} \y_{2}^{k} \!\!=\! \y_{2}^{*}\!,\!~~ \A\v^{*}\!+\!\e_1^{*}\!-\!\boldsymbol\varrho\!\! =\!\mathbf{0}, \\
&\ \ \v^{*}=\e_2^{*},~~ \v^{*}=\p^{*}\!, ~~ \e_1^{*}=\z_1^{*}, ~~ \e_2^{*}=\z_2^{*}.
\end{split}
\end{equation}
Moreover, $\v^{*}$ is a stationary point of the original problem \eqref{ML-decoding-all}, i.e.,
\begin{equation}\label{stationary-point}
(\v-\v^{*})^T\nabla_{\v} g(\v^*) \geq 0, ~~~\forall \v \in X,
\end{equation}
where $g(\v)=\pmb{\lambda}^{T}\v-\frac{\alpha}{2}\|\v-0.5\|_2^2$.
\end{theorem}

 {\it Proof:} {See Appendix \ref{converge-proof}.}

{{\it Remark:} The above \emph{Theorem \ref{converge-proof-theorem}} demonstrates that the proposed proximal-ADMM decoder is theoretically-guaranteed convergent to some stationary point of the non-convex decoding problem \eqref{ML-decoding-all}. Moreover, we should note that the convergence of the state-of-the-art decoders, including ADMM-based decoders and BP-like decoders, do not have this kind of convenient property. Furthermore, the value of parameter $\mu$ can be determined  efficiently since $\mathbf{A}^T\mathbf{A}$ is block diagonal. In the following subsection, we show that the proposed proximal-ADMM decoder's computational complexity is also competitive.}

\subsection{Computational Complexity}

{In the subsection, we show the complexity analysis of the proposed proximal-ADMM decoding algorithm in each iteration. Moreover, the presented result only includes multiplications since they occupy a dominant computation resource in practice. Before providing the analysis on the computational complexity of \emph{Algorithm \ref{proxiaml-ADMM-QP}}, we show matrix $\A$ has the following property.}

\begin{fact}\label{W-property}
 The elements in matrix $\A$ are 0, 1 or -1.
\end{fact}

{\it Proof:} See Appendix \ref{W-property-proof}.

Based on the above fact of matrix $\mathbf{A}$,
we can see that all multiplications with regard to $\mathbf{A}$ {can be  replaced by additions.}
Moreover, $(\pmb{\lambda}+0.5\alpha)/\mu$, $\theta_{i}$, $\omega_{i}$ and $\frac{\mu}{\rho+\mu}$ can be calculated in advance before we start the proximal-ADMM iterations.

\begin{table}[t]
\caption{{Summary of Computation Complexity in each proximal-ADMM iteration.}}
\label{notation-table}
\renewcommand{\arraystretch}{1.3}
\begin{center}
{\begin{tabular}{|c|c|c|}
\hline
\hline
Variables  & Equations &  {Multiplication Number}                                    \\\hline
$\mathbf{v}^{k+1}$  & \eqref{lp-x-update-solution} & $(2^{q+1}-1)(n+\Gamma_{a})$ \\ \hline
$\mathbf{e}_1^{k+1}$  & \eqref{v1-solution-component} & $2(4(2^q-1)\Gamma_{c}+n+\Gamma_{a})$  \\ \hline
$\mathbf{e}_2^{k+1}$  & \eqref{v2-solution-component} & $2(2^q-1)(n+\Gamma_{a})$  \\ \hline
$\mathbf{p}^{k+1}$    & \eqref{proximal-p-update}& $(2^q-1)(n+\Gamma_{a})$  \\ \hline
$\mathbf{z}_1^{k+1}$  & \eqref{proximal-z1-update}& $4(2^q-1)\Gamma_{c}+n+\Gamma_{a}$  \\ \hline
$\mathbf{z}_2^{k+1}$  & \eqref{proximal-z2-update} & $(2^q-1)(n+\Gamma_{a})$  \\ \hline
$\mathbf{y}_1^{k+1}/\mu_1$  & \eqref{proximal-y1-update} & free  \\ \hline
$\mathbf{y}_2^{k+1}/\mu_2$  & \eqref{proximal-y2-update} & free  \\ \hline
\multicolumn{2}{|c|}{Total}    & $(6\cdot2^q - 2)(n + \Gamma_{a}) + 12(2^q - 1)\Gamma_{c}$ \\
\hline\hline
\end{tabular}}
\end{center}
\end{table}

Consider $\mathbf{v}^{k+1}$ first. Calculating $\pmb{\varphi}_i^k$ in \eqref{epsilon_varphi} only requires $2^q-1$ multiplications, which comes from computing $\frac{\rho}{\mu}\mathbf{p}_i^k$. Here, notice $\mathbf{p}_i^k$ is the $(2^q-1)$-length sub-vector of $\mathbf{p}^k$. Moreover, from \eqref{proximal-xi-calcu-simp}, we can observe that computing $(\theta_i-\omega_i)\pmb{\varphi}_i^k$ and $\omega_i\displaystyle\sum_{\iota=1}^{q-1}\varphi_{i,\iota}^{k}$ requires $2^q-1$ multiplications and one multiplication respectively.
Therefore, updating $\mathbf{v}^{k+1}_i$ requires $2^{q+1}-1$ multiplications. This implies that the $\v^{k+1}$-update needs $(n+\Gamma_{a})(2^{q+1}-1)$ multiplications;
next, consider updating $\e_{1}^{k+1}$ and $\e_{2}^{k+1}$.
From \eqref{v1-solution-component}, we observe that each $e_{1,j}^{k+1}$ can be updated only via two multiplication operations. Thus, the multiplication number of the $\e_{1}^{k+1}$-update is 2M.
Similarly, observing \eqref{v2-solution-component}, we can find that computing $e_{2,\ell}^{k+1}$ also requires only two multiplications.
As a result, it takes $2N$ multiplications to obtain $\e_{2}^{k+1}$;
third, from {\eqref{proximal-p-update}--\eqref{proximal-z2-update}}, we easily observe that updating $\mathbf{p}^{k+1}$, $\z_1^{k+1}$, and $\z_2^{k+1}$ requires $N$, $M$, and $N$ multiplications respectively.
In addition, observing variables $\y_1$ and $\y_2$ in \eqref{proximal-xi-calcu-simp}, \eqref{v1-solution-component}, and \eqref{v2-solution-component},
one can find that if their scaled forms $\frac{\y_1}{\mu}$ and $\frac{\y_2}{\mu}$ are updated {in the iteration procedure}, they are free of multiplications,
i.e., calculating $\frac{\y_1^{k+1}}{\mu}$ and $\frac{\y_2^{k+1}}{\mu}$ only requires some addition operations.
From the above analysis, one can see that the overall multiplications of \emph{Algorithm \ref{proxiaml-ADMM-QP}} in each iteration are $(n + \Gamma_{a})(2^{q+1}-1) + 3N + 2M$.
Since $M=4(2^q-1)\Gamma_{c}+n+\Gamma_{a}$ and $N=(2^q-1)(n+\Gamma_{a})$ (see \eqref{pADMM-frame-problem}), it can be rewritten as
\begin{equation}
\begin{split}
 {(6\cdot2^{q} - 2)(n + \Gamma_{a}) + 12(2^q - 1)\Gamma_{c}}.
\end{split}
\end{equation}
Moreover, {\eqref{gamma_a_c} indicates $\Gamma_{a}\leq m(d-3)=n(1-R)(d-3)$ and $\Gamma_{c}\leq m(d-2)=n(1-R)(d-2)$, where $R$ denotes the code rate and $d$ is the largest check node degree. Then, one can see that either $\Gamma_{a}$ or $\Gamma_{c}$ is proportional to the code length $n$ since $d \ll n$ in the case of nonbinary LDPC codes. Therefore, we can conclude that the computational complexity of the proposed proximal-ADMM decoding algorithm in every iteration scales linearly with nonbinary LDPC code length and the size of the Galois field, or roughly $\mathcal{O}(2^qn)$. It is comparable to the BP-like algorithm \cite{burst-error} and cheaper than ADMM decoders proposed in \cite{Liu-nonbinary-journal} especially when $q$ is large. Moreover, we should note that sub-vectors $\mathbf{v}_i$ and entries in $\mathbf{e}_1$, $\mathbf{e}_2$, $\mathbf{p}$, $\mathbf{z}_1$, and $\mathbf{z}_2$ can be updated in parallel. In Table II, we summarize the above analysis on computational complexity in the decoding procedure.}

\section{Simulation results}\label{simulation-result}
In this section, several numerical results are presented for the proposed proximal-ADMM decoder. First, we show their error-correction performance (frame error rate (FER) and symbol error rate (SER)) and decoding efficiency, which are compared with several state-of-the-art nonbinary LDPC decoders. Second, we present how to select the proper value of the parameters in the proximal-ADMM decoders, which can improve their error-correction performance and convergence rate.

\subsection{Comparison of Decoding Performance}
{We consider four LDPC codes, which are Tanner (1055,424)-$\mathcal{C}_{1}$ from \cite{Tanner-code}, irregular PEG (504,252)-$\mathcal{C}_{2}$ from \cite{Mackay-code}, rate-0.5 MacKay (204, 102)-$\mathcal{C}_{3}$ from \cite{Mackay-code}, and Tanner (155,64)-$\mathcal{C}_{4}$ from \cite{Tanner-code} respectively.
For $\mathcal{C}_{1}$ and $\mathcal{C}_{2}$, we use the same parity-check matrix as the binary case, but its entries are cast as elements in $\mathbb{F}_{4}$. The two codes are modulated by quadrature phase shift keying (QPSK). Similar to the approach in \cite{LCLP-nonbinary}, we set the first/second/third/fourth/fifth/sixth nonzero entries in each row of the parity check matrix of $\mathcal{C}_{3}$  to 1/4/6/5/2/1$\in\mathbb{F}_8$ respectively. The code symbols of $\mathcal{C}_{3}$ are modulated by 8 phase-shift keying (8PSK).
For $\mathcal{C}_{4}$, its parity-check matrix is the same to the binary case, but its entries are cast as elements in $\mathbb{F}_{16}$.}
The corresponding modulation is sixteen quadrature amplitude modulation (16QAM). The modulated symbols are transmitted over the AWGN channel. The considered decoders include the proposed two proximal-ADMM algorithms, logarithm-domain fast-fourier-transforms-based Q-ary sum-product algorithm (Log-FFT-QSPA) \cite{burst-error}, and {non-penalized/penalized ADMM decoders in \cite{Liu-nonbinary-journal}. The transmitted information symbols are generated randomly.}
The parameters of the two proximal-ADMM algorithms are set the same, where penalty parameter $\mu$ is chosen as 0.8, 0.6, 0.7, and 0.6 for codes $\mathcal{C}_{1}$-$\mathcal{C}_{4}$ respectively; parameters $\alpha$, $\rho$, and $\beta$ are set to be 0.5, 0.52, and 0.9 respectively for both of the two codes; all the input codewords are generated randomly; we stop the iteration when either $\|\mathbf{A}\mathbf{v}^{k}+\mathbf{e}_{1}^{k}-\pmb{\varrho}\|_{2}^{2}\leq 10^{-5}$ or $\|\mathbf{v}^{k}-\mathbf{e}_{2}^{k}\|_{2}^{2}\leq 10^{-5}$ is satisfied, or the maximum iteration number $t_{\rm max}=500$ is reached. {All of the simulations are performed in MATLAB
2019b/Windows 7 environment on a computer with 3GHz
Intel i5-9500 CPU and 8GB RAM.}

 \begin{figure*}[tp]
\subfigure[Tanner (1055,424)-$\mathcal{C}_{1}$ in $\mathbb{F}_4$ with QPSK modulation.]{
    \begin{minipage}{8.5cm}
    \centering
        \includegraphics[width=3.5in,height=2.7in]{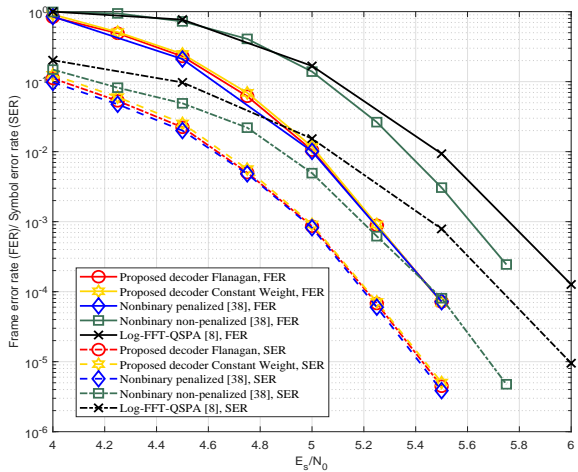}
            \label{fer1055}
    \end{minipage}%
    }
    \subfigure[{PEG (504,252)-$\mathcal{C}_{2}$ in $\mathbb{F}_4$ with QPSK modulation.}]{
    \begin{minipage}{8.5cm}
    \centering
        \includegraphics[width=3.5in,height=2.7in]{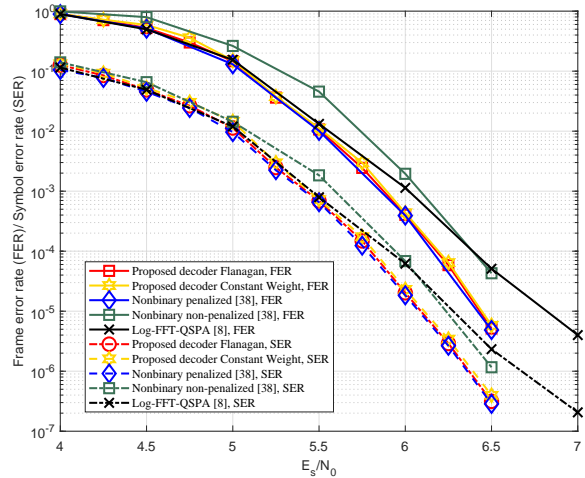}
            \label{fer504}
    \end{minipage}%
    }\
 \subfigure[{MacKay (204,102)-$\mathcal{C}_{3}$ in $\mathbb{F}_8$ with 8PSK modulation.}]{
    \begin{minipage}{8.5cm}
    \centering
        \includegraphics[width=3.5in,height=2.7in]{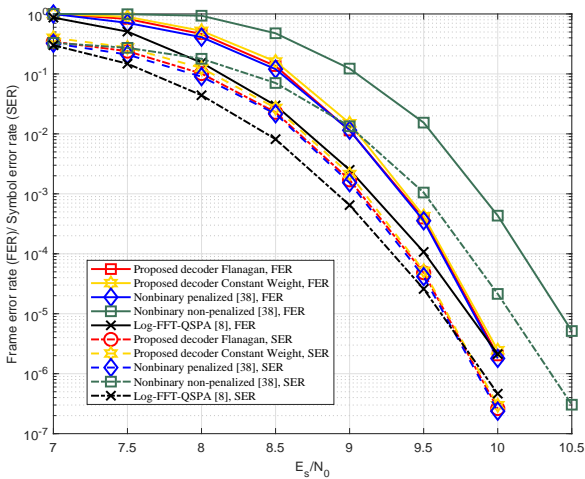}
            \label{fer204}
    \end{minipage}%
    }
    \subfigure[Tanner (155,64)-$\mathcal{C}_{4}$ in $\mathbb{F}_{16}$ with 16QAM.]{
    \begin{minipage}{8.5cm}
    \centering
        \includegraphics[width=3.5in,height=2.7in]{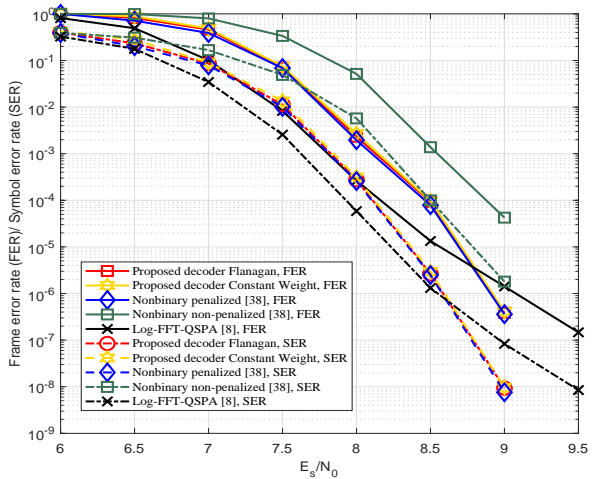}
            \label{fer155}
    \end{minipage}%
    }
     \centering
    \caption{{Comparisons of FER/SER performance for four LDPC codes from \cite{Mackay-code} and \cite{Tanner-code} in different Galois fields  with different modulations.}}
    \label{fer_ser}
 \end{figure*}

 \begin{figure*}[tp]
\subfigure[Tanner (1055,424) code $\mathcal{C}_{1}$ in $\mathbb{F}_4$ with QPSK modulation.]{
    \begin{minipage}{8.5cm}
    \centering
        \includegraphics[width=3.5in,height=2.6in]{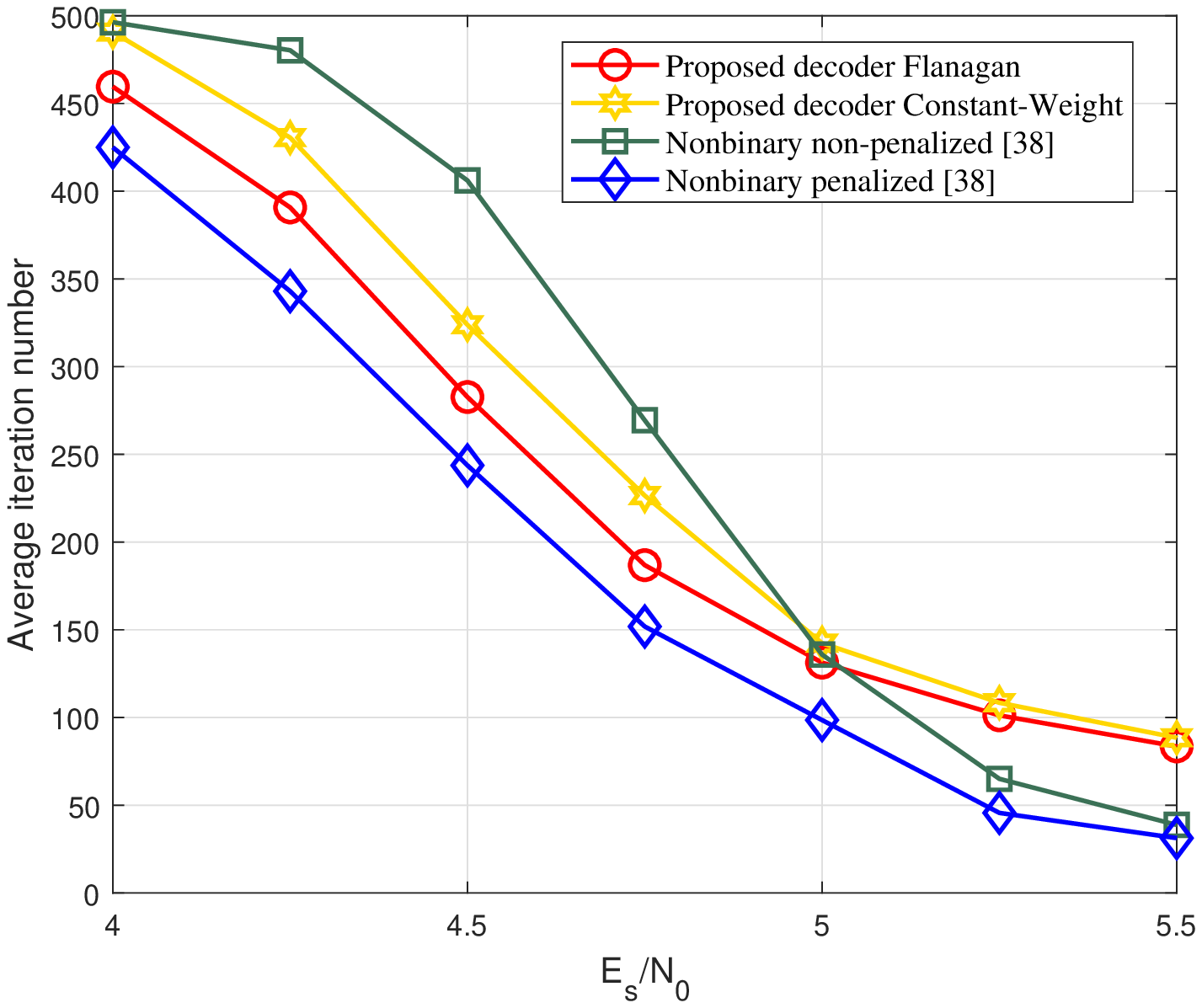}
            \label{iter1055}
    \end{minipage}%
    }
    \subfigure[{PEG (504,252) code $\mathcal{C}_{2}$ in $\mathbb{F}_4$ with QPSK modulation.}]{
    \begin{minipage}{8.5cm}
    \centering
        \includegraphics[width=3.5in,height=2.6in]{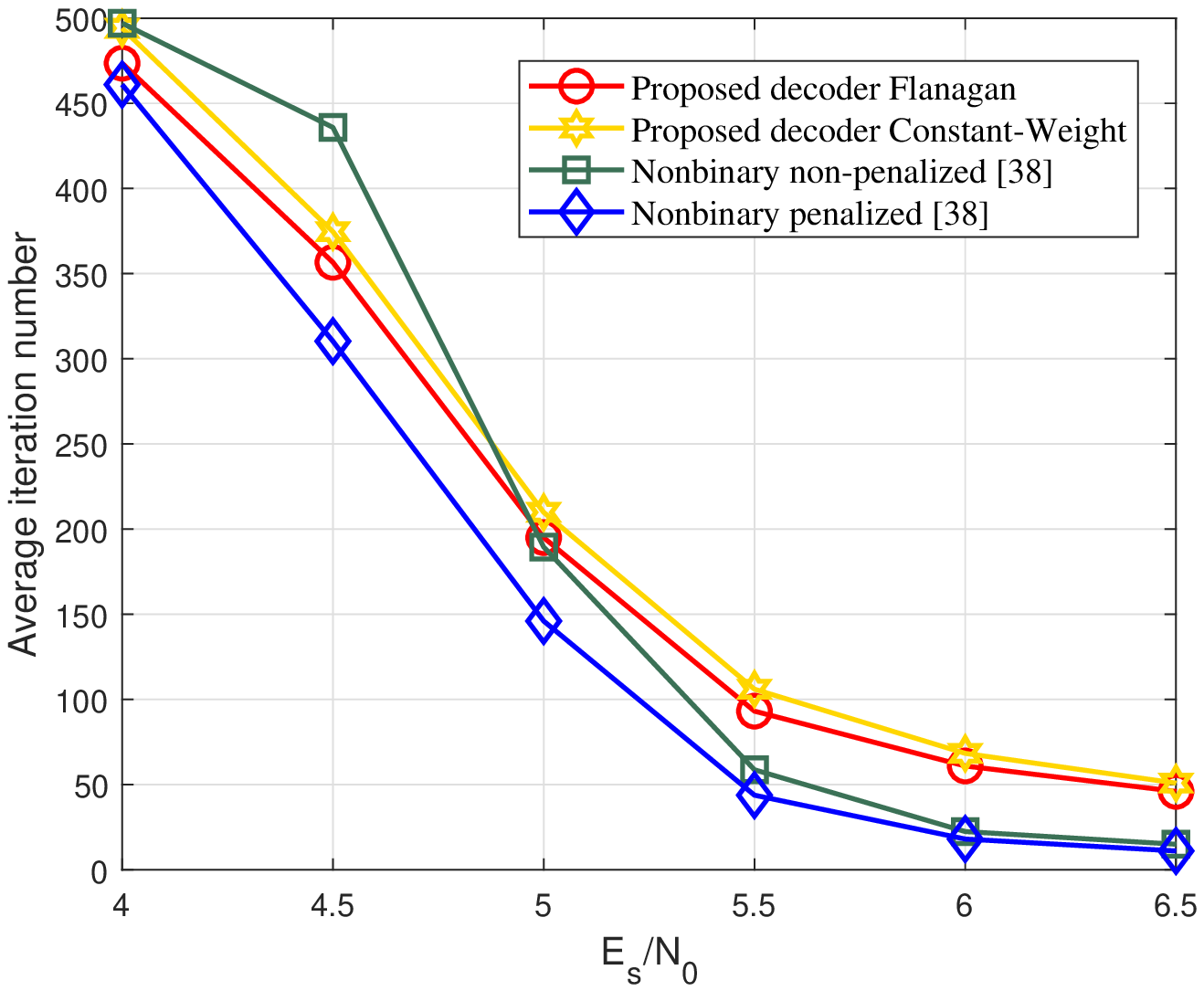}
            \label{iter504}
    \end{minipage}%
    }\
\subfigure[{MacKay (204,102) code $\mathcal{C}_{3}$ in $\mathbb{F}_8$ with 8PSK modulation.}]{
    \begin{minipage}{8.5cm}
    \centering
        \includegraphics[width=3.5in,height=2.6in]{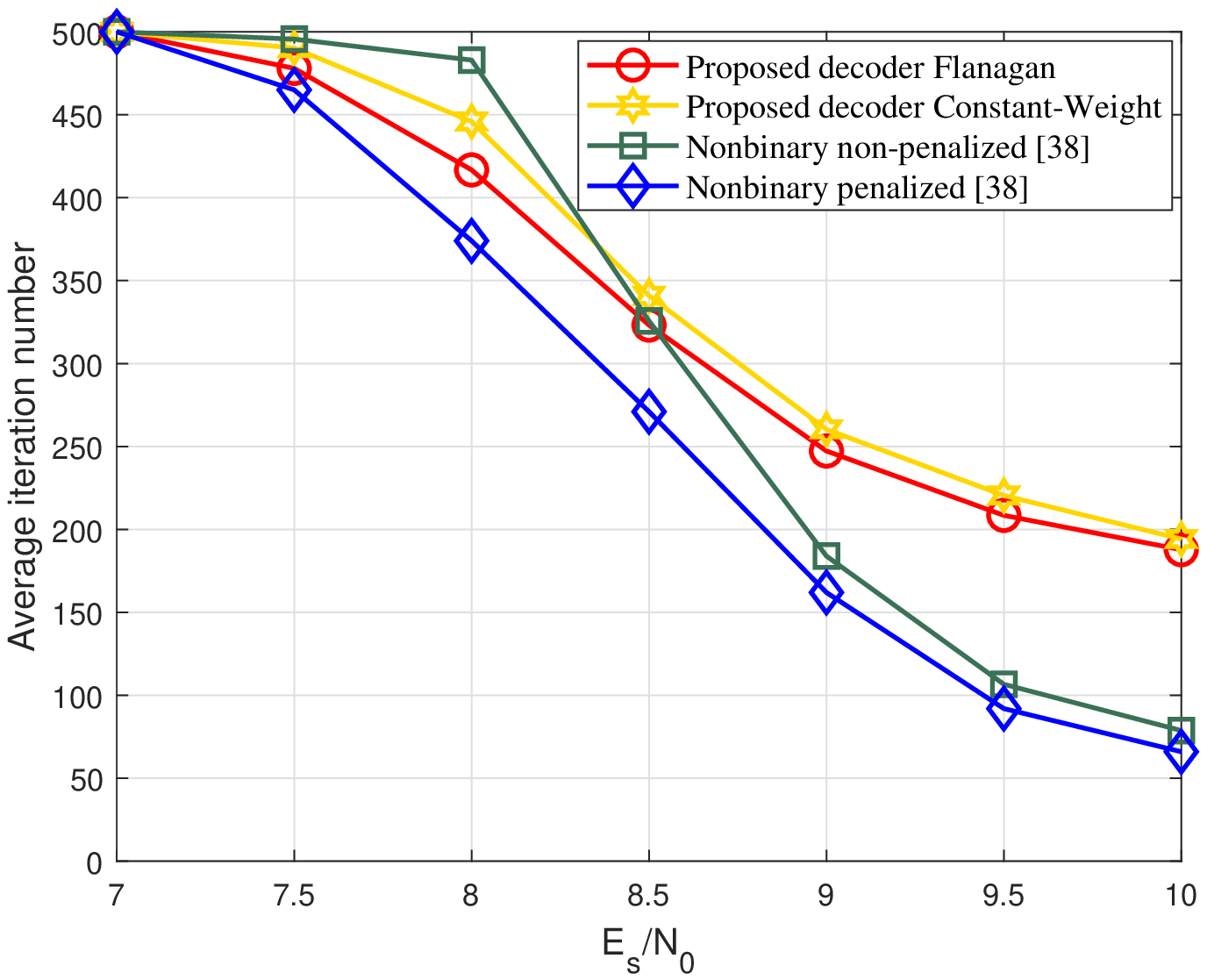}
            \label{iter204}
    \end{minipage}%
    }
    \subfigure[Tanner (155,64) code $\mathcal{C}_{4}$ in $\mathbb{F}_{16}$ with 16QAM modulation.]{
    \begin{minipage}{8.5cm}
    \centering
        \includegraphics[width=3.5in,height=2.6in]{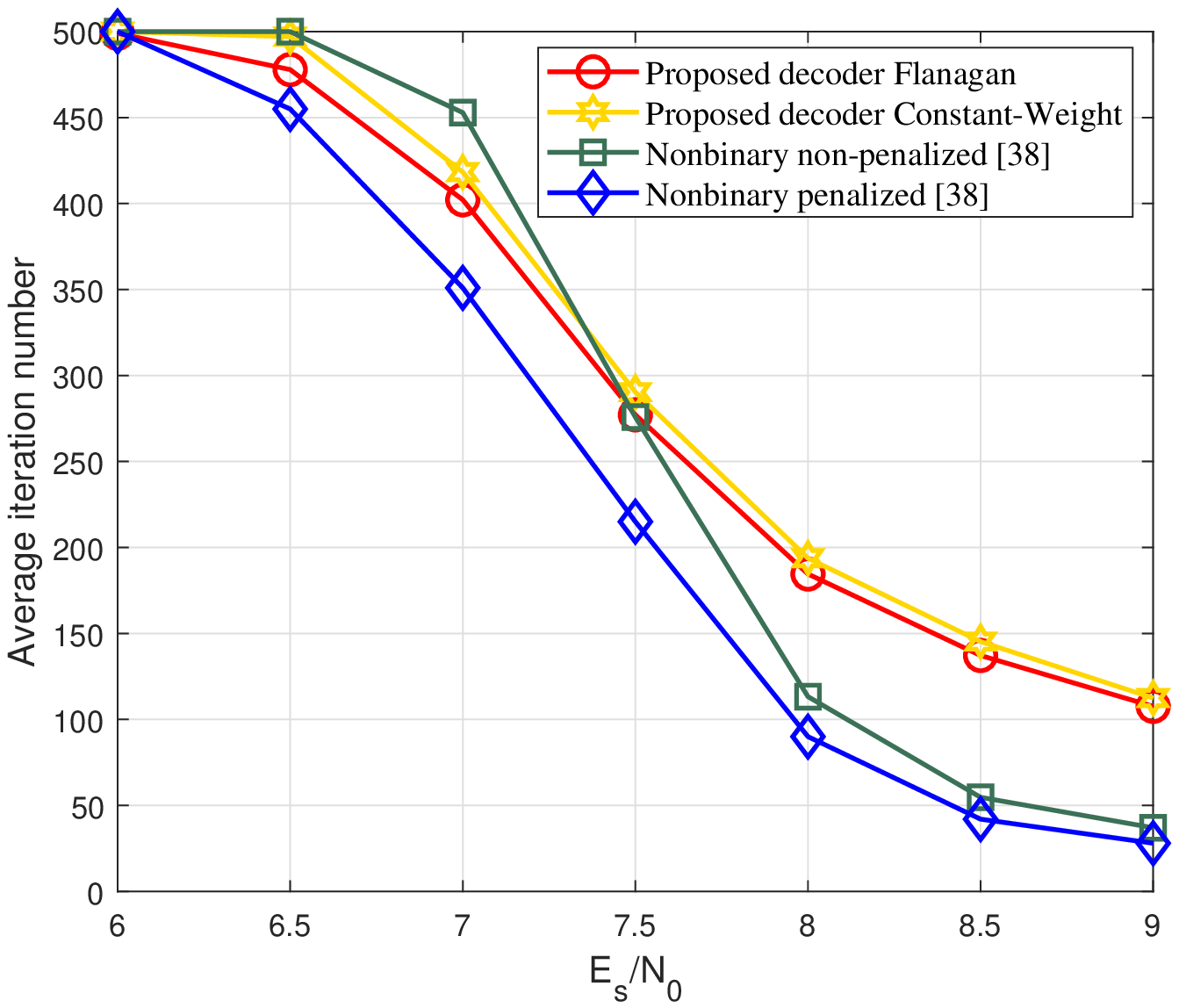}
            \label{iter155}
    \end{minipage}%
    }
     \centering
    \caption{Comparisons of the average number of iterations for four LDPC codes from \cite{Mackay-code} and \cite{Tanner-code} using different decoders with different modulations, where $\mathcal{C}_{1}$ denotes the Tanner (1055,424) code, $\mathcal{C}_{2}$ denotes the PEG (504,252) code, $\mathcal{C}_{3}$ denotes the MacKay (204,102) code, and $\mathcal{C}_{4}$ denotes the Tanner (155,64) code.}
    \label{iter4}
 \end{figure*}

Figure \ref{fer_ser} shows FER/SER curves of code {$\mathcal{C}_{1}-\mathcal{C}_{4}$ when different decoders are applied, where all data points are based on generating at least 200 error frames, except for the last two points where 50 error frames are observed due to limited computational resources.} {From figures \ref{fer1055}-\ref{fer155}, one can observe that the proposed decoders have similar FER/SER performance to the penalized ADMM decoder in \cite{Liu-nonbinary-journal}, but are much better than the non-penalized one. These indicate that the penalty term plays an important role in the decoding procedure. Moreover, from figures \ref{fer1055}, \ref{fer504}, and \ref{fer155}, one can see that the proposed decoders' FER/SER performance is better than Log-FFT-QSPA \cite{FHT-nonBP} in the high SNR region, where FER/SER curves of the proposed two decoders continue to drop in a waterfall manner while Log-FFT-QSPA decoder's decrease slowly. Specifically, in figure \ref{fer204},  the proposed decoders' FER/SER performance is inferior to the Log-FFT-QSPA decoder at considered SNRs. However, we note that the slopes of the FER/SER curves of the proposed decoders are steeper than those of the Log-FFT-QSPA decoder at high SNRs. This implies that they could surpass the Log-FFT-QSPA decoder in much higher regions, for which we did not simulate due to limited computational resources. In summary, we see that the proposed proximal-ADMM decoders have better error-correction performance than the Log-FFT-QSPA decoder at a high SNR region, especially for long nonbinary LDPC codes. The reason could be that both of them are always trying to determine the global optimal solution of the decoding problem, but the Log-FFT-QSPA decoder is designed to search the solution locally.}

\begin{figure*}[tp]
\subfigure[{Tanner (1055,424)-$\mathcal{C}_{1}$ in $\mathbb{F}_4$ with QPSK modulation.}]{
    \begin{minipage}{8.5cm}
    \centering
        \includegraphics[width=3.5in,height=2.6in]{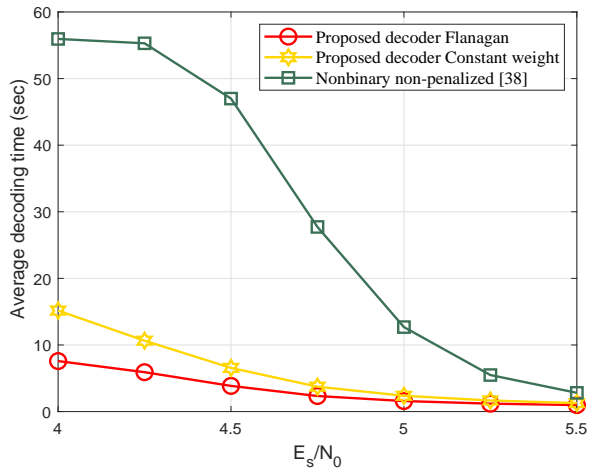}
            \label{time1055}
    \end{minipage}%
    }
    \subfigure[{PEG (504,252)-$\mathcal{C}_{2}$ in $\mathbb{F}_4$ with QPSK modulation.}]{
    \begin{minipage}{8.5cm}
    \centering
        \includegraphics[width=3.5in,height=2.6in]{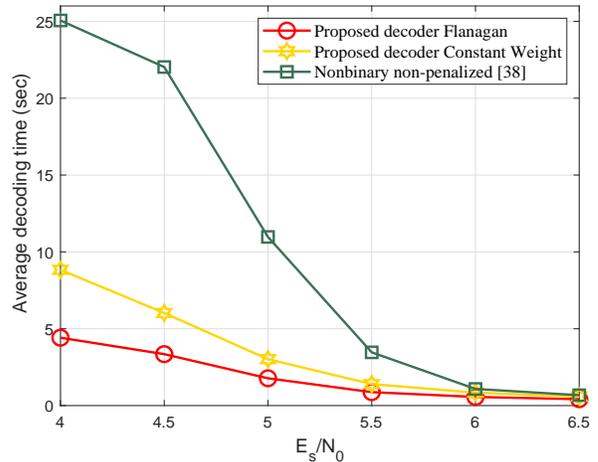}
            \label{time504}
    \end{minipage}%
    }\
\subfigure[{MacKay (204,102)-$\mathcal{C}_{3}$ in $\mathbb{F}_8$ with 8PSK modulation.}]{
    \begin{minipage}{8.5cm}
    \centering
        \includegraphics[width=3.5in,height=2.6in]{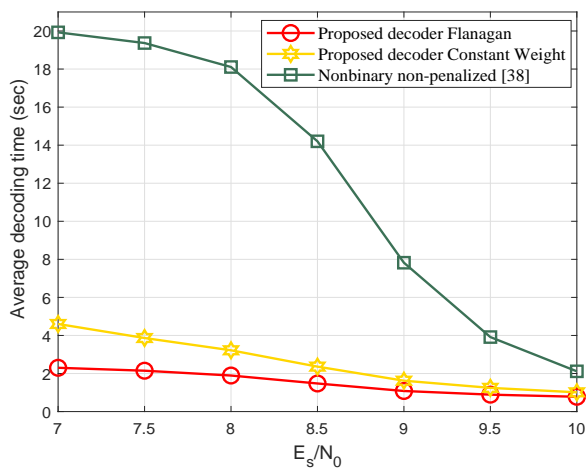}
            \label{time204}
    \end{minipage}%
    }
    \subfigure[{Tanner (155,64)-$\mathcal{C}_{4}$ in $\mathbb{F}_{16}$ with 16QAM modulation.}]{
    \begin{minipage}{8.5cm}
    \centering
        \includegraphics[width=3.5in,height=2.6in]{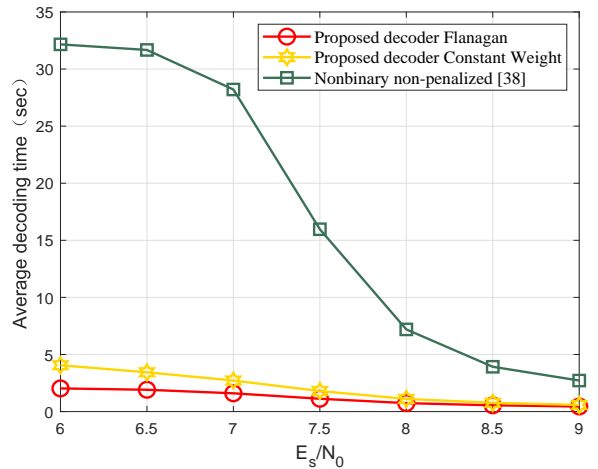}
            \label{iter155}
    \end{minipage}%
    }
     \centering
    \caption{{Comparisons of the average decoding time of the proposed decoders and non-penalized decoder in \cite{Liu-nonbinary-journal}, where LDPC codes are selected from \cite{Mackay-code} and \cite{Tanner-code}. Here, we should note that, at the very low SNR region, both of the decoders reach their  maximum iteration number of 500. Notice that execution time of the penalized decoder in \cite{Liu-nonbinary-journal} is not presented here since it will show the difference unclearly among the three decoders. In \cite{Liu-nonbinary-journal}, authors observed empirically that the penalized decoder is around 20 times slower than the non-penalized one, which was also observed in our simulations.}}
    \label{time4}
 \end{figure*}

{Figure\ \ref{iter4} and Figure\ \ref{time4} show the averaged iteration number and decoding time of the proposed two proximal-ADMM decoders and the non-penalized/penalized ADMM decoders in \cite{Liu-nonbinary-journal} respectively. In figure \ref{iter4}, all data points are averaged over one million LDPC frames. From them, it can be observed that the averaged number of iterations required by the ADMM decoders in \cite{Liu-nonbinary-journal} are less than the proposed proximal-ADMM decoders in high SNR regions for the considered codes $\mathcal{C}_{1}$- $\mathcal{C}_{4}$. The reason could be that more auxiliary variables are involved in the decoding model\footnotemark.}
 {In figure \ref{time4},  all data points of the curves are also averaged over one million LDPC frames. From the figures, one can see that the proposed two decoders cost less decoding time than the competing ADMM decoders \cite{Liu-nonbinary-journal} for the considered codes $\mathcal{C}_{1}$- $\mathcal{C}_{4}$. Moreover, the proximal-ADMM decoder based on the Constant-Weight embedding technique takes a little bit longer for decoding than the one based on the Flanagan embedding technique, which verifies the computational analysis in Appendix E.}

\footnotetext{{In comparison with the proposed proximal-ADMM decoders, non-penalized/penalized decoders in \cite{Liu-nonbinary-journal} need less variables/checks. Usually, this merit leads the decoder to costing less dynamic power\cite{hardward_ADMM} when implementing it using an FPGA chip. Their decoding procedure involves high-dimensional Euclidean projections, which have to be implemented in series. This means that the corresponding working-frequency is higher when they desire the same decoding throughput to the proximal-ADMM decoders. From a practical viewpoint, dynamic power is an important parameter, which is related to many factors, such as variables (wires/logical resource), working frequency, voltage, etc. involved in the decoding procedure. Therefore, to evaluate dynamic power of a decoder is a complex (but important) task, which should be considered carefully in practice.}}

\subsection{Parameter choices of the proposed proximal-ADMM decoder}
{Here, we just focus on the proximal-ADMM decoder based on the Flanagan embedding technique.} There are several parameters in the proposed proximal-ADMM decoding \emph{Algorithm \ref{proxiaml-ADMM-QP}}, including the penalty parameter $\mu$, the 2-norm penalty parameter $\alpha$, the ending tolerance $\xi$, the maximum iteration number $t_{\rm max}$, and parameters $\rho$ and $\beta$.
Proper parameters can make \emph{Algorithm \ref{proxiaml-ADMM-QP}} achieve favorable error-correction performance and reduce the iteration number.
It is easy to see that a sufficiently large $t_{\rm max}$ and sufficiently small $\xi$ can lead to good error-correction performance for \emph{Algorithm \ref{proxiaml-ADMM-QP}}.
Thus, we fix the ending tolerance $\xi=10^{-5}$ and the maximum number of iterations $t_{max}=500$ in the simulations.
Moreover, a proper large $\beta$ is favorable because it can update variables $\mathbf{v}$, $\mathbf{e}_1$, and $\mathbf{e}_2$ in every proximal-ADMM iteration and not deviate too much from the stabilized $\mathbf{p}$, $\mathbf{z}_1$, and $\mathbf{z}_2$ respectively (c.f.\cite{proximal-admm}). Therefore, we set $\beta$ to be 0.9 in the simulations.
Moreover, \emph{Algorithm \ref{proxiaml-ADMM-QP}} is also sensitive to the values of parameters $\mu$, $\alpha$, and $\rho$.
In order to guarantee that subproblem \eqref{proximal-x-update} is strongly convex with respect to variable $\x$, we let $\rho>\alpha$.

\begin{figure}[tp]
  \centering
  \centerline{\psfig{figure=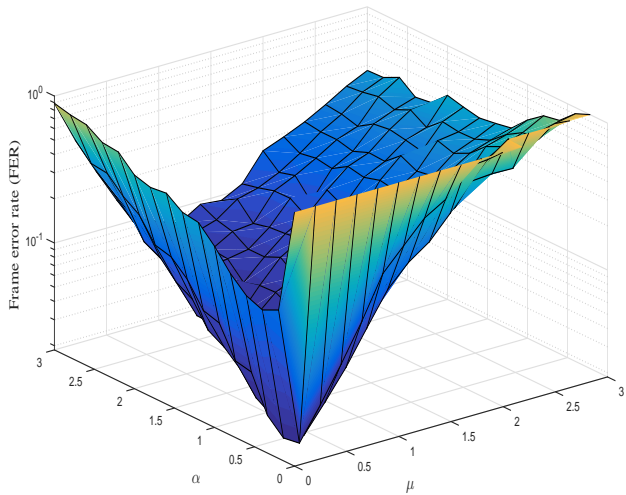,width=9.5cm,height=7.5cm}}
  \caption{FER performance plotted as a function of $\mu$ and $\alpha$ for the Tanner [1055,424] code $\mathcal{C}_{1}$ in $\mathbb{F}_{4}$.}
  \label{miu-alpha-fer}
\end{figure}
\begin{figure}[tp]
  \centering
  \centerline{\psfig{figure=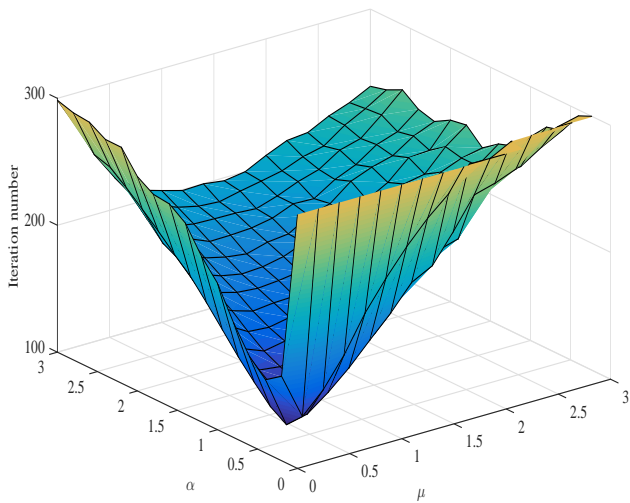,width=9.5cm,height=7.5cm}}
  \caption{Number of iterations plotted as a function of $\mu$ and $\alpha$ for the Tanner [1055,424] code $\mathcal{C}_{1}$ in $\mathbb{F}_{4}$.}
  \label{miu-alpha-iter}
\end{figure}

Next, we focus on how to choose parameters $\mu$ and $\alpha$. Figure \ref{miu-alpha-fer} and figure \ref{miu-alpha-iter} plot FER performance and iteration numbers for code $\mathcal{C}_{1}$ as a function of parameters $\mu$ and $\alpha$ at $E_s/N_0=5$\,dB respectively.
In the figures, we set $\rho=\alpha+0.02$ to ensure that $\rho>\alpha$ always holds.
Observing figure \ref{miu-alpha-fer}, one can find that \emph{Algorithm \ref{proxiaml-ADMM-QP}} achieves better FER performance when $\mu \in (0.3,1)$ and $\alpha \in (0.2,0.5)$.
Moreover, from Figure\,\ref{miu-alpha-iter}, one can see that the decoder takes fewer iterations when $\mu \in (0.3,1)$ and $\alpha \in (0.2,0.5)$.
This means that $\mu\in(0.3,1)$ and $\alpha\in(0.2,0.5)$ are good choices in terms of error-correction performance and decoding efficiency of \emph{Algorithm \ref{proxiaml-ADMM-QP}}.

\section{Conclusion} \label{Conclusion}
{In this paper, two efficient decoders are developed for nonbinary LDPC codes in $\mathbb{F}_{2^q}$ via proximal-ADMM techniques based on the Flanagan/Constant-Weight embedding technique respectively. We show that both of their decoding complexities scale linearly with block length and Galois field's size of the nonbinary LDPC codes and the corresponding iteration algorithms converge to some stationary point of the formulated decoding model. Besides, the latter one has a codeword symmetry property. Simulation results demonstrate the effectiveness of the proposed proximal-ADMM decoders in comparison with several state-of-the-art decoders. }

\appendices

\section{Proof of Fact 1}\label{check-equ-proof}
{\it Proof:} By adding both sides of equations \eqref{decom-1}-\eqref{decom-3}, we obtain
\begin{equation}
\begin{split}
 h_{\sigma_1}c_{\sigma_1}\!+\!h_{\sigma_2}c_{\sigma_2}\!+\!g_1\!+\!g_1\!+\!h_{\sigma_3}c_{\sigma_3}\!+\!g_2\!+\!g_2\!+\!\ldots\!+\!g_{d_i\!-\!3}
 \!\\
 +g_{d_i\!-3}\!+\!h_{\sigma_{d_j-1}}c_{\sigma_{d_j-1}}\!+\!h_{\sigma_{d_j}}c_{\sigma_{d_j}}\!=\!0,
 \end{split}
\end{equation}
where the sum is in $\mathbb{F}_{2^q}$.
{Since $g_t+g_t=0$ in $\mathbb{F}_{2^q}$, the above equation can be reduced to}
\begin{equation}\label{a}
 h_{\sigma_1}\!c_{\sigma_1}\!+\!h_{\sigma_2}\!c_{\sigma_2}\!+\!h_{\sigma_3}\!c_{\sigma_3}\!+\!\ldots\!+\!h_{\sigma_{d_j-1}}\!c_{\sigma_{d_j-1}}\!+\!h_{\sigma_{d_j}}\!c_{\sigma_{d_j}} \!=\!0,
\end{equation}
which is just the $j$th check equation in \eqref{C}. It can be equivalent to
\begin{subequations}\label{a}
 \begin{align}
  &h_{\sigma_1}c_{\sigma_1}+h_{\sigma_2}c_{\sigma_2}+g_1=0,\\
  &g_1+h_{\sigma_3}c_{\sigma_3}+\ldots+h_{\sigma_{d_j-1}}c_{\sigma_{d_j-1}}+h_{\sigma_{d_j}}c_{\sigma_{d_j}} =0. \label{a_2}
  \end{align}
\end{subequations}
Moreover, \eqref{a_2} can be further divided into
\begin{subequations}\label{b}
  \begin{align}
  &g_1+h_{\sigma_3}c_{\sigma_3}+g_2=0, \label{b_1}\\
  &g_2+\ldots+h_{\sigma_{d_j-1}}c_{\sigma_{d_j-1}}+h_{\sigma_{d_j}}c_{\sigma_{d_j}} =0. \label{b_2}
  \end{align}
\end{subequations}
Through similar derivations, we obtain
\begin{equation}
\begin{split}
& h_{\sigma_1}c_{\sigma_1}+h_{\sigma_2}c_{\sigma_2}+g_1=0, \\
& g_{t-1}+h_{\sigma_{t+1}}c_{\sigma_{t+1}}+g_{t}=0,~~ t=2, \dotsb, d_j-3,\\
& g_{d_i-3}+h_{\sigma_{d_j-1}}c_{\sigma_{d_j-1}}+h_{\sigma_{d_j}}c_{\sigma_{d_j}}=0.
\end{split}
\end{equation}
Thus, one can conclude that any general check equation in \eqref{C} can be equivalent to the three-variables parity-check equations \eqref{decom-1}-\eqref{decom-3}.

\section{Proof of Lemma \ref{hu-dD-u-define}}\label{hu-dD-u-define-proof}
{\it Proof:} Let $\mathbb{I}(\cdot)$ be the indicator function defined in $\mathbb{F}_{2^q}$, i.e., $\mathbb{I}(c, \hat{c})=1$ when $c=\hat{c}$ and $\mathbb{I}(c, \hat{c})=0$ otherwise. Then, from the mapping rule defined in \eqref{Mq2b}, binary vector codeword $\x_k$, corresponding to $c_k \in \mathbb{F}_{2^q}$, can be expressed by
\begin{equation}\label{x_k-express}
\x_k =[\mathbb{I}(1, c_k);\ldots;\mathbb{I}(j, c_k);\ldots;\mathbb{I}(2^q-1, c_k)].
\end{equation}

Assuming $j=c_k$, {it follows} that $\mathbb{I}(j, c_k)=1$ and other elements in $\x_k$ are zeros.
Moreover, based on the definition of $D(2^q,h_k)_{ij}$ in \eqref{permu-matrix}, matrix $\mathbf{D}(2^q,h_k)$ can be rewritten as \eqref{permu-matrix-express}.
\begin{figure*}
\begin{equation}\label{permu-matrix-express}
\begin{split}
&\mathbf{D}(2^q,h_k)
=
\left[
  \begin{array}{ccccc}
    \mathbb{I}(1, h_k)    &    \cdots   &   \mathbb{I}(1,jh_k )   &   \cdots &   \mathbb{I}(1, (2^q - 1)h_k) \\
    \vdots   &   \vdots & \vdots & \vdots & \vdots \\
    \mathbb{I}(i, h_k)    &    \cdots   &   \mathbb{I}(i, jh_k)   &   \cdots &   \mathbb{I}(i, (2^q - 1)h_k) \\
    \vdots    &    \vdots   &   \vdots   &   \vdots   &   \vdots \\
    \mathbb{I}(2^q  -  1 ,  h_k )    &     \cdots   &     \mathbb{I}(2^q - 1 ,  jh_k)    &   \cdots    &    \mathbb{I}(2^q  -  1 , (2^q -  1)h_k  ) \\
  \end{array}
    \right].
\end{split}
\end{equation}
\end{figure*}
Then, we have
\begin{equation}\label{Dx-express}
\mathbf{D}(2^q,h_k)\x_k =\left[
                       \begin{array}{c}
                         \mathbb{I}(1,jh_k)\mathbb{I}(j,c_k) \\
                         \vdots \\
                         \mathbb{I}(i,jh_k)\mathbb{I}(j,c_k) \\
                         \vdots \\
                         \mathbb{I}(2^q  -  1,jh_k)\mathbb{I}(j,c_k) \\
                       \end{array}
                       \right].
\end{equation}

{Letting} $j=c_k$ and $i=h_kc_k$, we can derive $\mathbf{D}(2^q,h_k)\x_k$ as
\begin{equation}\label{Dx-express}
\mathbf{D}(2^q,h_k)\x_k =  \left[
                       \begin{array}{c}
                         \mathbb{I}(1,h_kc_k) \\
                         \vdots \\
                         \mathbb{I}(i,h_kc_k) \\
                         \vdots \\
                         \mathbb{I}(2^q-1,h_kc_k) \\
                       \end{array}
                         \right]
  =  \left[     \begin{array}{c}0\\ \vdots \\ 1\\ \vdots \\ 0\end{array}   \right],
\end{equation}
which completes the proof.

\section{Proof of Lemma \ref{w-inverse-lemma}}\label{w-inverse-lemma-proof}
{\it Proof:} According to the definition of matrix $\A$ in \eqref{Ab-construct_A}, we have
\vspace{-0.0cm}
\begin{equation}\label{ATA-derive1}
\begin{split}
\A^T\!\A
\!=\!&\!\bigg(\!\sum_{\tau=1}^{\Gamma_{c}}\!\left(\!\hat{\mathbf{W}}_{\tau}\!(\!\mathbf{Q}_{\tau}\!\otimes\! \mathbf{I}_{2^q\!-\!1}\!)\!\right)\!^{T}\!\left(\!\hat{\mathbf{W}}_{\tau}\!(\!\mathbf{Q}_{\tau}\!\otimes \! \mathbf{I}_{2^q\!-\!1})\!\right)\!\!\!\bigg)\!\!\!+\!\!\mathbf{S}^T\!\mathbf{S} \\
=&\!\bigg(\!\sum_{\tau=1}^{\Gamma_{c}}\!(\! \mathbf{Q}_{\tau}\!\otimes \! \mathbf{I}_{2^q-1}\!)\!^T\hat{\mathbf{W}}_{\tau}^T\hat{\mathbf{W}}_{\tau}\!(\!\mathbf{Q}_{\tau}\!\otimes\! \mathbf{I}_{2^q-1}\!)\!\!\bigg)\!+\!\mathbf{S}^T\mathbf{S},
\end{split}
\end{equation}
where
$\hat{\mathbf{W}}_{\tau}=\bigg[ \mathbf{P}\hat{\mathbf{T}}_1\mathbf{D}_\tau;\dotsb; \mathbf{P}\displaystyle \hat{\mathbf{T}}_{2^q-1}\mathbf{D}_{\tau}\bigg]$, $\hat{\mathbf{T}}_{\ell}=\displaystyle\sum_{i\in\mathcal{K}_{\ell}}\mathbf{T}_{i}$, $\ell=1,\dotsb, 2^q-1$,
and $\mathcal{K}_1=\{1\}$, $\mathcal{K}_2=\{2\}$, $\mathcal{K}_3=\{1, 2\}$, $\dotsb, \mathcal{K}_{2^q-1}=\{1,\dotsb,q\}$.\footnotemark

\footnotetext{{\textbf{Example:} In $\mathbb{F}_{8}$,
$\mathcal{K}_1=\{1\}$, $\mathcal{K}_2=\{2\}$, $\mathcal{K}_3=\{1,2\}$, $\mathcal{K}_{4}=\{3\}$, $\mathcal{K}_5=\{1,3\}$, $\mathcal{K}_6=\{2,3\}$, $\mathcal{K}_{7}=\{1,2,3\}$.}}

Moreover, we have the following derivations for $\hat{\mathbf{W}}_{\tau}^T\hat{\mathbf{W}}_{\tau}$
{
\begin{equation}\label{wTw-derive1}
\begin{split}
\hat{\mathbf{W}}_{\tau}^T\hat{\mathbf{W}}_{\tau}
&= \sum_{\ell=1}^{2^q-1}\mathbf{D}_{\tau}^T\hat{\mathbf{T}}_{\ell}^T \mathbf{P}^T \mathbf{P}\hat{\mathbf{T}}_{\ell}\mathbf{D}_{\tau}  \\
&\overset{a}{=}4 \mathbf{D}_{\tau}^T\bigg(\sum_{\ell=1}^{2^q-1}\hat{\mathbf{T}}^T_{\ell}\hat{\mathbf{T}}_{\ell}\bigg) \mathbf{D}_{\tau},
\end{split}
\end{equation}
where} the equality ``$\overset{a}{=}$'' holds since $\mathbf{P}^T\mathbf{P}=4\mathbf{I}_4$.
Since $
   \hat{\mathbf{T}}_{\ell} = {\rm diag}(\displaystyle\sum_{i\in\mathcal{K}_{\ell}}\hat{\mathbf{b}}_i^T, \displaystyle\sum_{i\in\mathcal{K}_{\ell}}\hat{\mathbf{b}}_i^T, \displaystyle\sum_{i\in\mathcal{K}_{\ell}}\hat{\mathbf{b}}_i^T)
$, $\displaystyle\sum_{\ell=1}^{2^q-1}\hat{\mathbf{T}}^T_{\ell}\hat{\mathbf{T}}_{\ell}$ is also diagonal and expressed as
\begin{equation}
  \displaystyle\sum_{\ell=1}^{2^q-1}\hat{\mathbf{T}}^T_{\ell}\hat{\mathbf{T}}_{\ell}=\textrm{diag}(\pmb{\Phi},\pmb{\Phi},\pmb{\Phi}),
\end{equation}
where
\begin{equation}\label{phi-value}
\pmb{\Phi}\!\! =\!\!\! \displaystyle\sum_{\ell=1}^{2^q-1}\!\!\!\bigg(\!\!\displaystyle\sum_{i\in\mathcal{K}_{\ell}}\hat{\mathbf{b}}_i\!\bigg)\!\!\bigg(\!\!\displaystyle\sum_{i\in\mathcal{K}_{\ell}}\hat{\mathbf{b}}_i^T\!\bigg)\! \!\!=\!\! \!\left[\!\!
               \begin{array}{cccc}
                 2^{q\!-\!1} & 2^{q\!-\!2} & \cdots & 2^{q\!-\!2} \\
                 2^{q\!-\!2} & 2^{q\!-\!1} & \cdots & 2^{q\!-\!2} \\
                 \vdots & \vdots & \ddots & \vdots \\
                 2^{q\!-\!2} & 2^{q\!-\!2} & \cdots & 2^{q\!-\!1} \\
               \end{array}
             \!\!\right]\!\!,
\end{equation}
and $\pmb\Phi\in \mathbb{R}^{(2^q-1)\times(2^q-1)}$.
Then, we have
\begin{equation}\label{wTw-derive2}
\hat{\mathbf{W}}_{\tau}^T\hat{\mathbf{W}}_{\tau}
=4\mathbf{D}_{\tau}^T\textrm{diag}(\pmb{\Phi},\pmb{\Phi},\pmb{\Phi})\mathbf{D}_{\tau}.
\end{equation}
Since matrices $\mathbf{D}(2^q,h_{\tau_k})$, $k=1,2,3$, are elementary, we can get
\begin{equation}\label{DphiD-phi}
\mathbf{D}(2^q,h_{\tau_k})^{T}\pmb{\Phi}\mathbf{D}(2^q,h_{\tau_k})=\pmb{\Phi}, \ k=1,2,3,
\end{equation}
which implies that \eqref{wTw-derive2} can be further simplified to
\begin{equation}\label{wTw-derive3}
\hat{\mathbf{W}}_{\tau}^T\hat{\mathbf{W}}_{\tau}
=4\textrm{diag}(\pmb{\Phi},\pmb{\Phi},\pmb{\Phi}).
\end{equation}
\vspace{-0.0cm}
Plugging \eqref{wTw-derive3} into \eqref{ATA-derive1} and {noticing} $\mathbf{S}={\rm diag}(\underbrace{\mathbf{1}_{2^q-1}^T,\dotsb,\mathbf{1}_{2^q-1}^T}_{n+\Gamma_a})$, we {can conclude}\footnotemark
\footnotetext{
{\textbf{Example 5:} In $\mathbb{F}_{4}$, assume there are three variable-selecting matrices $\mathbf{Q}_{\tau}\in\{0,1\}^{3\times 4}$. They are $\mathbf{Q}_{1}=\begin{bmatrix}
                    1 & 0 & 0 & 0 \\
                    0 & 1& 0 & 0 \\
                    0 & 0 & 0 & 1 \\
                \end{bmatrix}$,
$\mathbf{Q}_{2}=\begin{bmatrix}
                    0 & 1 & 0 & 0 \\
                    0 & 0& 1 & 0 \\
                    0 & 0 & 0 & 1 \\
                \end{bmatrix}$,
and
$\mathbf{Q}_{3}=\begin{bmatrix}
                    1 & 0 & 0 & 0 \\
                    0 & 0& 1 & 0 \\
                    0 & 0 & 0 & 1 \\
                \end{bmatrix}$, respectively. Notice there are four variables involved in the check equations. The first three variables are involved in two check equations and the fourth variable is involved in three check equations, i.e., $d_1=d_2=d_3=2, \  d_4=3.$
Moreover, since ${\mathbf{S}=\rm diag}\underbrace{(\mathbf{1}_{2^q-1}^T,\dotsb,\mathbf{1}_{2^q-1}^T)}_{n+\Gamma_a}$, we have
\[
\begin{split}
 \mathbf{A}^T\mathbf{A}
=&4\bigg(\sum_{\tau=1}^{3}(\mathbf{Q}_{\tau}\otimes \mathbf{I}_{3})^T\textrm{diag}(\pmb{\Phi},\pmb{\Phi},\pmb{\Phi})(\mathbf{Q}_{\tau}\otimes \mathbf{I}_{3})\bigg)\!+\!\mathbf{S}^T\mathbf{S} \\
=&4\left(\begin{bmatrix}
              \mathbf{I}_{3}\! &\!\mathbf{0}   \! &\! \mathbf{0}\! &\! \mathbf{0} \\
              \mathbf{0}    \! &\!\mathbf{I}_{3}\!&\! \mathbf{0} \!& \!\mathbf{0} \\
              \mathbf{0}     \!&\!\mathbf{0}    \!&\! \mathbf{0} \!&\! \mathbf{I}_{3} \\
         \end{bmatrix}^T
         \textrm{diag}(\pmb{\Phi},\pmb{\Phi},\pmb{\Phi})\!\!
         \begin{bmatrix}
              \mathbf{I}_{3} \!&\!\mathbf{0}   \! & \!\mathbf{0} \!&\! \mathbf{0} \\
              \mathbf{0}     \!&\!\mathbf{I}_{3}\!& \!\mathbf{0} \!& \!\mathbf{0} \\
              \mathbf{0}     \!&\!\mathbf{0}    \!&\! \mathbf{0} \!&\! \mathbf{I}_{3} \\
         \end{bmatrix}\right. \\
        &+\begin{bmatrix}
             \mathbf{0} \!&\! \mathbf{I}_{3} \!&\! \mathbf{0} \!& \!\mathbf{0} \\
             \mathbf{0} \!&\! \mathbf{0}\!&\! \mathbf{I}_{3} \!&\! \mathbf{0} \\
             \mathbf{0} \!&\! \mathbf{0} \!&\! \mathbf{0} \!&\! \mathbf{I}_{3} \\
          \end{bmatrix}^T
                \textrm{diag}(\pmb{\Phi},\pmb{\Phi},\pmb{\Phi})\!\!
          \begin{bmatrix}
             \mathbf{0} & \mathbf{I}_{3} & \mathbf{0}     &\mathbf{0} \\
             \mathbf{0} & \mathbf{0}     & \mathbf{I}_{3} &\mathbf{0} \\
             \mathbf{0} & \mathbf{0}     & \mathbf{0}     &\mathbf{I}_{3} \\
          \end{bmatrix}\\
        &+\begin{bmatrix}
             \mathbf{I}_{3}\! &\! \mathbf{0} \!&\!\mathbf{0}     \!&\! \mathbf{0} \\
             \mathbf{0}    \! &\! \mathbf{0} \!&\!\mathbf{I}_{3} \!&\! \mathbf{0} \\
             \mathbf{0}    \! &\! \mathbf{0} \!&\!\mathbf{0}    \! &\! \mathbf{I}_{3} \\
          \end{bmatrix}^T
             \textrm{diag}(\!\pmb{\Phi},\pmb{\Phi},\pmb{\Phi}\!)\!\!\!
          \begin{bmatrix}
             \mathbf{I}_{3} & \mathbf{0} &\mathbf{0}     & \mathbf{0} \\
             \mathbf{0}     & \mathbf{0} &\mathbf{I}_{3} & \mathbf{0} \\
             \mathbf{0}     & \mathbf{0} &\mathbf{0}     & \mathbf{I}_{3} \\
          \end{bmatrix}\!\!\bigg)\!\!+\!\! \mathbf{S}^T\!\mathbf{S} \\
=&4\left(\textrm{diag}(\pmb{\Phi},\pmb{\Phi},\mathbf{0},\pmb{\Phi})+
         \textrm{diag}(\mathbf{0},\pmb{\Phi},\pmb{\Phi},\pmb{\Phi})+
         \textrm{diag}(\pmb{\Phi},\mathbf{0},\pmb{\Phi},\pmb{\Phi})\right)\\
         &+
  \begin{bmatrix}
  \mathbf{1}_{3\times3} & \mathbf{0}  & \mathbf{0} & \mathbf{0}\\
  \mathbf{0}  & \mathbf{1}_{3\times3} &\mathbf{0}  & \mathbf{0} \\
  \mathbf{0}  & \mathbf{0} & \mathbf{1}_{3\times3} & \mathbf{0} \\
  \mathbf{0}  & \mathbf{0} & \mathbf{0} & \mathbf{1}_{3\times3}
  \end{bmatrix} \\
=&\textrm{diag}(8\pmb{\Phi}+\mathbf{1}_{3\times3},
8\pmb{\Phi}+\mathbf{1}_{3\times3},
8\pmb{\Phi}+\mathbf{1}_{3\times3},
12\pmb{\Phi}+\mathbf{1}_{3\times3})
  \end{split}\]}
i.e.,
$\mathbf{A}^T\mathbf{A}
=\textrm{diag}(4d_1\pmb{\Phi}+\mathbf{1}_{3\times3},
4d_2\pmb{\Phi}+\mathbf{1}_{3\times3},
4d_3\pmb{\Phi}+\mathbf{1}_{3\times3},
4d_4\pmb{\Phi}+\mathbf{1}_{3\times3}).
$ }

\vspace{-0.4cm}
\begin{equation}\label{ATA-derive2}
\begin{split}
 \A^T\!\!\!\A
\!\!=&4\!\bigg(\!\sum_{\tau=1}^{\Gamma_{c}}\!(\!\mathbf{Q}_{\tau}\!\!\otimes\!\! \mathbf{I}_{2^q\!-\!1}\!)\!^T\textrm{diag}\!(\!\pmb{\Phi},\pmb{\Phi},\pmb{\Phi}\!)\!(\!\mathbf{Q}_{\tau}\!\otimes\! \mathbf{I}_{2^q\!-\!1})\!\!\bigg)\!\!\!+\!\!\mathbf{S}^T\!\mathbf{S} \\
\!\!=&\textrm{diag}\!(\!4d_{1}\pmb{\Phi}\!\!+\!\!\mathbf{1}_{2^q\!-\!1}\mathbf{1}_{2^q\!-\!1}^T\!,\!\!\dotsb\!,\!4d_{n\!+\!\Gamma_{a}}\!\pmb{\Phi}\!\!+\!\!\mathbf{1}_{2^q\!-\!1}\mathbf{1}_{2^q\!-\!1}^T\!),
\end{split}
\end{equation}
where $d_{i}$, $i=1,\dotsb,n+\Gamma_a$, {denotes the number of three-variables parity-check equations that the $i$th information symbol is involved in.}

From \eqref{ATA-derive2}, it is easy to see that matrix $\big(\A^{T}\A+\epsilon\I_{(n+\Gamma_a)(2^q-1)}\big)^{-1}$ is block diagonal, whose $i$th diagonal block sub-matrix can be expressed as
\begin{equation}\label{AAe}
  \mathbf{A}_{i}^T\!\!\mathbf{A}_i\!\!+\!\!\epsilon\mathbf{I}_{2^q\!-\!1}\!\!=\!\!4d_{i}\pmb{\Phi}\!\!+\!\!\mathbf{1}_{2^q\!-\!1} \mathbf{1}_{2^q\!-\!1}^T\!\!+\!\epsilon\mathbf{I}_{2^q\!-\!1}\!,\!\!\ i\!=\!\!1,\!\dotsb\!,n\!+\!\Gamma_a.
\end{equation}
Plugging LHS of \eqref{AAe} into \eqref{ATA-derive2}, we obtain the first part of Lemma \ref{w-inverse-lemma}, i.e.,
\[
\begin{split}
 \!\big(\!\A^{T}\!\!\A+\epsilon\I_{(n\!+\Gamma_a)(2^q\!-\!1\!)}\!\big)\!^{-1}
\!\!=\!\! &\ {\rm diag}\!\Big(\big(\!{\A}_1^{T}\A_1\!\!+\!\! \epsilon\I_{2^q-1}\!\big)\!^{-1}\!\!,\ldots, \\ &\;\big(\A_{n+\Gamma_{a}}^{T}\A_{n+\Gamma_{a}}+\epsilon\I_{2^q-1}\big)^{-1}\Big).
\end{split}
\]
For the second part of Lemma 2, we first plug \eqref{phi-value} into \eqref{AAe}, which leads to
\begin{equation}\label{AAematrix}
  \begin{split}
  \mathbf{A}_{i}^T\!\!\mathbf{A}_i&+\epsilon\mathbf{I}_{2^q\!-\!1}\!\!
=  \\
   &\left[\!\!\!\!
               \begin{array}{cccc}
                 4d_{i}2^{q\!-\!1} \!\! +\!\! 1\!\! +\!\! \epsilon & 4d_{i}2^{q\!-\!2} \!\! +\!\! 1 &  \!\! \cdots  \!\! & 4d_{i}2^{q\!-\!2}  \!\!+ \!\!1 \\
                 4d_{i}2^{q\!-\!2} \!\! + \!\!1  &\!\!  4d_{i}2^{q\!-\!1}\!  \!+\!\! 1\!\! +\!\! \epsilon & \!\!\cdots  \!\! & \!4d_{i}2^{q\!-\!2} \!\! +\!\! 1 \\
                 \vdots& \vdots & \!\!\ddots \!\!& \vdots \\
                 4d_{i}2^{q\!-\!2}\!\!  +\!\! 1   &   4d_{i}2^{q\!-\!2}\!\!  +\!\! 1   & \!\!\!   \!\!\cdots\!\! & 4d_{i}2^{q\!-\!1} \!\! +\!\! 1 \!\!+ \epsilon \\
               \end{array}
                \!\! \!\!\right]\!\!
   \end{split}.
\end{equation}
Observing \eqref{AAematrix}, it is easy to see that the elements in the diagonal line of matrix $\mathbf{A}_{i}^T\mathbf{A}_i+\epsilon\mathbf{I}_{2^q-1}$ are the same and other elements in the off-lines are the same. So its inverse matrix can be written as \cite{Matrix-analysis}
\begin{equation}\label{AAematrix inv}
(\mathbf{A}_{i}^T\mathbf{A}_i+\epsilon\mathbf{I}_{2^q-1})^{-1} =\left[
  \begin{array}{cccc}
     \theta_{i} & \omega_{i} & \cdots & \omega_{i} \\
    \omega_{i} & \theta_{i} & \cdots & \omega_{i} \\
    \vdots & \vdots & \vdots & \vdots \\
    \omega_{i} & \omega_{i} & \cdots & \theta_{i} \\
  \end{array}
\right].
\end{equation}
Multiplying both sides of \eqref{AAematrix} and \eqref{AAematrix inv}, we have the following equalities
\begin{equation}\label{cal-a-b-equations}
\begin{split}
&\!\!(4d_{i}2^{q\!-\!1}\!\!+\!\!1\!\!+\!\!\epsilon)\theta_{i}\!\! +\!\! (4d_{i}2^{q\!-\!2}\!\!+\!\!1)(2^{q}\!-\!2)\omega_{i}\!\!=\!\!1, \\
&\!\!(4d_{i}2^{q\!-\!1}\!\!\!\!+\!\!1\!\!+\!\!\epsilon\!)\omega_{i} \!\!+\!\!(4d_{i}2^{q\!-\!2}\!\!+\!\!1)\theta_{i}\!\! +\!\! (4d_{i}2^{q\!-\!2}\!\!+\!\!1\!)(2^{q}\!\!\!-\!\!3)\omega_{i}\!\!=\!\!0,
\end{split}
\end{equation}
\footnotetext{{Here, we should note that equation \eqref{cal-a-b-equations} is only suitable for nonbinary LDPC codes, i.e., $q\geq2$. When $q=1$, $\pmb{\Phi}=[1]$ and $(\mathbf{A}_{i}^T\mathbf{A}_i+\epsilon\mathbf{I})=4d_i+1+\epsilon$ (notice no $\omega_i$ in the equation), which indicates $\theta_i=\frac{1}{4d_i+1+\epsilon}$.}}
which lead to\footnotemark
\[
\begin{split}
& \theta_{i} = \omega_{i}+\frac{1}{2^{q}d_{i}+\epsilon}, \\
& \omega_{i} = \frac{ - 2^{q}d_{i}-1}{(2^qd_{i} + \epsilon)\big((2^{q+1}d_{i} + \epsilon + 1) + (2^{q}d_{i} + 1)(2^{q} - 2)\big)}.
\end{split}
\]
This completes the proof.

\section{Proof of Fact \ref{W-property}}\label{W-property-proof}
{\it Proof:} {To be clear, we rewrite $\mathbf{A}$ in \eqref{Ab-construct_A} as follows:
\[
\begin{split}
{\mathbf{A}} = &[\hat{\mathbf{W}}_1(\mathbf{Q}_1 \otimes \mathbf{I}_{2^q-1}); \cdots;\hat{\mathbf{W}}_{\tau} (\mathbf{Q}_\tau \otimes \mathbf{I}_{2^q-1});\cdots;\\
&\quad\hat{\mathbf{W}}_{\Gamma_c}(\mathbf{Q}_{\Gamma_c} \otimes \mathbf{I}_{2^q-1});\mathbf{S}],
 \end{split}
\]
where $\hat{\mathbf{W}}_{\tau}=\bigg[ \mathbf{P}\hat{\mathbf{T}}_1\mathbf{D}_\tau;\dotsb; \mathbf{P}\displaystyle \hat{\mathbf{T}}_{2^q-1}\mathbf{D}_{\tau}\bigg]$ and $   \hat{\mathbf{T}}_{\ell} = {\rm diag} (\displaystyle\sum_{i\in\mathcal{K}_{\ell}}\hat{\mathbf{b}}_i^T, \displaystyle\sum_{i\in\mathcal{K}_{\ell}}\hat{\mathbf{b}}_i^T, \displaystyle\sum_{i\in\mathcal{K}_{\ell}}\hat{\mathbf{b}}_i^T)$, $\ell=1,\dotsb, 2^q-1$. Here, we should note that $\hat{\mathbf{T}}_i$ is binary since the sum in $\displaystyle\sum_{i\in\mathcal{K}_i}\hat{\mathbf{b}}_i^T$ is in $\mathbb{F}_2$.}

{Since $\mathbf{D}(2^q,h_{\tau_j})$ is elementary, there is at most one nonzero element 1 in each column of $\mathbf{T}_{\ell}\mathbf{D}(2^q,h_{\tau_j})$.
Moreover, since elements in matrix $\mathbf{P}$ are $1$ or $-1$, elements in $\mathbf{P} \displaystyle\bigg(\sum_{i\in\mathcal{K}_\ell}\hat{\mathbf{T}}_{i}\bigg)\mathbf{D}_\tau$ are 0, $1$, or $-1$.
Therefore, we can conclude that elements in $\hat{\mathbf{W}}_{\tau}$ are also either 0, $1$, or $-1$. Furthermore, since variable-selecting matrix $\mathbf{Q}_{\tau}$ has one nonzero element ``1'' at most in its each row/column, $\mathbf{Q}_{\tau}\otimes \mathbf{I}_{2^q-1}$ should have the same property. Therefore, it is obvious that elements in $\hat{\mathbf{W}}_{\tau}(\mathbf{Q}_{\tau}\otimes \mathbf{I}_{2^q-1})$ are also 1, -1, or 0. Besides, since matrix $\mathbf{S}$ only includes one nonzero element ``1'', we can conclude that matrix $\A$ consists of elements 1, -1, and 0.}

{\section{Brief presentation on the design of the LDPC decoder via the Constant-Weight embedding technique}\label{CR}}
{In the beginning, we should note that the formulation procedure of the decoding model via the Constant-Weight embedding technique is almost the same as the Flanagan one. A few differences are shown in Table III. To facilitate understanding them, we use the same notations for the referred parameters.}

\begin{table*}[tbp]
\renewcommand \arraystretch{1.2}
\newcommand{\tabincell}[2]{\begin{tabular}{@{}#1@{}}#2\end{tabular}}
\caption{Parameter comparison of two different embedding techniques}
\label{table comparisons}
\centering
\renewcommand{\arraystretch}{1.3}
\begin{center}
{\begin{tabular}{|c|c|c|}
\hline
\hline
       & Flanagan embedding  & Constant-Weight embedding                                     \\\hline
$\gamma_{i,\sigma}$ & \tabincell{c}{$\log\frac{p(r_i|x_{i,\sigma}=0)}{p(r_i|x_{i,\sigma}=1)}$ \\   $i=1,\dotsb,n$,\ $\sigma=1,\dotsb,2^q-1$}  & \tabincell{c}{$\log\frac{1}{p(r_i|x_{i,\sigma}=1)}$ \\   $i=1,\dotsb,n$,\ $\sigma=0,1,\dotsb,2^q-1$} \\ \hline
\tabincell{c}{Rotation \\matrix} & $\mathbf{D}(2^q,h_k)$ & $\begin{bmatrix} 1\ &\mathbf{0}_{2^q-1}^T \\ \mathbf{0}_{2^q-1}\ & \mathbf{D}(2^q,h_k) \end{bmatrix}$  \\ \hline
$\mathbf{B}$     & $\begin{bmatrix} \tilde{\mathbf{b}}(1), \ldots,\tilde{\mathbf{b}}(2^q-1)\end{bmatrix}$ & \tabincell{l}{$\begin{bmatrix}\tilde{\mathbf{b}}(0), \tilde{\mathbf{b}}(1), \ldots,\tilde{\mathbf{b}}(2^q-1)\end{bmatrix}$ \\ $\tilde{\mathbf{b}}(0)$: $q$-length all-zeros vector.} \\ \hline
$\hat{\mathbf{w}}$     &   $\mathbf{1}_{2^q-1} \otimes\mathbf{t}$ & $\mathbf{1}_{2^q}\otimes\mathbf{t}$ \\
\hline
$\hat{\mathbf{F}}$  & $[\hat{\mathbf{W}}_1(\mathbf{Q}_1\otimes\mathbf{I}_{{2^q-1}}); \cdots;\hat{\mathbf{W}}_{\Gamma_c}(\mathbf{Q}_{\Gamma_c}\otimes\mathbf{I}_{{2^q-1}})]$ & $[\hat{\mathbf{W}}_1(\mathbf{Q}_1\otimes\mathbf{I}_{{2^q}}); \cdots;\hat{\mathbf{W}}_{\Gamma_c}(\mathbf{Q}_{\Gamma_c}\otimes\mathbf{I}_{{2^q}})]$ \\ \hline
$\mathbf{S}$     &   ${\rm diag}(\underbrace{\mathbf{1}_{2^q-1}^T,\dotsb,\mathbf{1}_{2^q-1}^T}_{n+\Gamma_a})$ & ${\rm diag}(\underbrace{\mathbf{1}_{2^q}^T,\dotsb,\mathbf{1}_{2^q}^T}_{n+\Gamma_a})$ \\
\hline
\eqref{Sv} & $\mathbf{Sv}\preceq\mathbf{1}_{n+\Gamma_a}$ & $\mathbf{Sv}=\mathbf{1}_{n+\Gamma_a}$ \\ \hline
$\mathbf{A}$     &   $[\hat{\mathbf{W}}_1(\mathbf{Q}_1\otimes\mathbf{I}_{{2^q-1}}); \cdots;\hat{\mathbf{W}}_{\Gamma_c}(\mathbf{Q}_{\Gamma_c}\otimes\mathbf{I}_{{2^q-1}});\mathbf{S}]$ & $[\hat{\mathbf{W}}_1(\mathbf{Q}_1\otimes\mathbf{I}_{{2^q}}); \cdots;\hat{\mathbf{W}}_{\Gamma_c}(\mathbf{Q}_{\Gamma_c}\otimes\mathbf{I}_{{2^q}});\mathbf{S}]$ \\ \hline
\eqref{proximal-v1-min} & $j\in\{1,\dotsb, M\}$  & $j\in\{1,\dotsb, 4(2^q-1)\Gamma_{c}\}$\\ \hline
N     &   $(2^q-1)(n+\Gamma_a)$  & $2^q(n+\Gamma_a)$ \\
\hline
\hline
\end{tabular}}
\end{center}
\end{table*}

{
Moreover, we consider the implementation of the proximal-ADMM solving algorithm when the Constant-Weight embedding technique is also almost the same as the Flaganan one. Besides the different dimensions of the vectors and matrices ($2^q-1$ to $2^q$), the only difference is how to update variable $\mathbf{v}$. To be clear, we rewrite \eqref{lp-x-update-solution} as follows
\begin{equation}\label{lp-x-update-solution_CW}
\mathbf{v}^{k+1} = \big(\A^{T}\A+\epsilon\I_{N})^{-1}\pmb{\varphi}^{k}.
\end{equation}
Similar to \eqref{w-I-inverse-defi}, we have
\[
\begin{split}
 \big(\!\A^{T}\!\!\A\!\!+\!\epsilon\I_N\!\big)\!^{-1}
\!\!\!\!=\!\!  {\rm diag}\!\Big(\!\!\big(\!{\A}_1^{T}\A_1\!\!+\!\!\epsilon\I_{2^q}\!\big)\!^{-1}\!\!\!\!,\ldots,\!\big(\!\!\A_{n\!+\!\Gamma_{a}}^{T}\A_{n\!+\!\Gamma_{a}}\!\!+\!\!\epsilon\I_{2^q}\!\big)\!\!^{-1}\!\!\Big)\!,
\end{split}
\]
where $\big({\A}_i^{T}\A_i+\epsilon\I_{2^q}\big)=\left[
               \begin{array}{ccccc}
                 1 \!\!+\!\! \epsilon & 1 & 1& \cdots & 1 \\
                 1&4d_{i}2^{q\!-\!1} \!\! +\!\! 1\!\! +\!\! \epsilon   &   4d_{i}2^{q\!-\!2} \!\! + \!\!1 &   \cdots    &   4d_{i}2^{q\!-\!2} \!\! +\!\! 1 \\
                 1&4d_{i}2^{q\!-\!2} \!\! + \!\!1   &   4d_{i}2^{q\!-\!1} \!\! + \!\!1 \!\!+\! \epsilon &   \cdots    &   4d_{i}2^{q\!-\!2} \!\! + \!\!1 \\
                 \vdots &\vdots & \vdots & \ddots & \vdots \\
                 1&4d_{i}2^{q\!-\!2}\!\!  +\!\! 1   &   4d_{i}2^{q\!-\!2} \!\! + \!\!1   &    \cdots & 4d_{i}2^{q\!-\!1} \!\! +\!\! 1 \!\!+ \!\epsilon \\
               \end{array}
                 \right].$
}
{Since its inverse has the following special structure
\begin{equation}\label{AAematrix inv_CW}
(\mathbf{A}_{i}^T\mathbf{A}_i+\epsilon\mathbf{I}_{2^q})^{-1} =\left[
  \begin{array}{ccccc}
     \varsigma_i&\kappa_i    & \kappa_i    & \cdots & \kappa_i  \\
     \kappa_i &\theta_{i} & \omega_{i} & \cdots & \omega_{i} \\
     \vdots  &\omega_{i} & \theta_{i} & \cdots & \omega_{i} \\
     \vdots  &\vdots     & \vdots     & \vdots & \vdots \\
     \kappa_i &\omega_{i} & \omega_{i} & \cdots & \theta_{i} \\
  \end{array}
\right],
\end{equation}}
{we can obtain $\varsigma_i$, $\kappa_i$, $\omega_i$, and $\theta_i$ by solving a four-variables linear equation (the derivations are quite similar to the ones presented in Appendix C), which are
\[
  \begin{split}
    &\kappa_i = \frac{1}{2^q-1+(1+\epsilon)(\epsilon-1+2^q(2^qd_i+1))},\\
    &\varsigma_i = \frac{1-(2^q-1)\kappa_i}{1+\epsilon},\\
    &\theta_i = \omega_i + \frac{1}{2^qd_i+\epsilon}, \\
    &\omega_i = \frac{(2^qd_i+1)(1+\epsilon)-1}{(2^{q}d_i+\epsilon)\left(2^q-1+(1+\epsilon)(\epsilon-1+2^q(2^qd_i+1))\right)}.
  \end{split}
\]
According to \eqref{AAematrix inv_CW}, since $(\mathbf{A}_{i}^T\mathbf{A}_i+\epsilon\mathbf{I}_{2^q})^{-1}$ can be  expressed as
\[
\begin{split}
  \mathbf{v}_k^{k+1}\! \!=\!&\! \begin{bmatrix}
                       \varsigma_i\!-\!\kappa_i   &\dotsb            &\dotsb  &0      \\
                       0                  &\theta_i\!-\!\omega_i &\dotsb  &0      \\
                       \vdots             &\dotsb            &\ddots  &\vdots \\
                       0                  &\dotsb            &\dotsb  &\theta_i\!-\!\omega_i
                       \end{bmatrix}\!\!+\!\!
                       \begin{bmatrix}
                       \omega_i   &\dotsb            &\dotsb  &\omega_i      \\
                       \omega_i                  &\omega_i &\dotsb  &\omega_i      \\
                       \vdots             &\dotsb            &\ddots  &\vdots \\
                       \omega_i                  &\omega_i            &\dotsb  &\omega_i
                       \end{bmatrix}\!\\
                       &+\!
                       \begin{bmatrix}
                       \kappa_i\!-\!\omega_i   &\kappa_i\!-\!\omega_i  &\dotsb  &\kappa_i\!-\!\omega_i      \\
                       \kappa_i\!-\!\omega_i   &0                 &\dotsb  &0      \\
                       \vdots             &\dotsb            &\ddots  &\vdots \\
                       \kappa_i\!-\!\omega_i   &0                 &\dotsb  &0
                       \end{bmatrix},
\end{split}
\]
\eqref{lp-x-update-solution_CW} can be written as
\begin{equation}\label{vi_CW}
  \mathbf{v}_k^{k+1} \!\!=\!\! \begin{bmatrix} \varsigma_i\!-\!\kappa_i \\ \theta_i\!-\!\omega_i \\ \vdots \\ \theta_i\!-\!\omega_i \end{bmatrix}\!\!\!\circ\!\pmb{\varphi}_i^{k}\!\!+\!\!\big(\omega_i\displaystyle\sum_{\iota=1}^{2^q} {\varphi}_{i,\iota}^{k}\big)\mathbf{1}_{2^q}\!\!+\!\!(\kappa_i\!-\!\omega_i)\!\!\begin{bmatrix} \displaystyle\sum_{\iota=1}^{2^q} \varphi_{i,\iota}^{k} \\ \varphi_{i,1}^{k} \\ \vdots \\ \varphi_{i,1}^{k} \end{bmatrix},
\end{equation}
where operator ``$\circ$'' denotes the Hardmard product. Using \eqref{vi_CW} to take the place of \eqref{proximal-xi-calcu-simp} in Algorithm 1, we can obtain the complete proximal-ADMM decoding algorithm based on the Constant-Weight embedding technique.}

{
    Furthermore, we consider the performance of the presented decoding algorithm above. The decoder's convergence property can also be characterized by {\it Theorem 1}. The computational complexity in each ADMM iteration is still scaled linearly with nonbinary LDPC code length and the size of the Galois field. However, we should say that its decoding complexity is slightly larger than the one using the Flanagan embedding technique since the size of the matrices and vectors involved is scaled linearly in terms of $2^q$. To be clear, we show the number of multiplications used in the implementation in Table IV. Besides the guaranteed convergence and similar computational complexity, the proposed decoder based on Constant-Weight embedding satisfies a favorable property of the \emph{all-zeros assumption}, which is described as follows.
\begin{theorem}\label{all zero assumption}
  assume that the noisy channel is symmetrical. Then, the probability that the decoding algorithm based on the Constant-Weight embedding technique fails is independent of the transmitted codeword.
\end{theorem}
}

{\it Proof:} {See Appendix \ref{proof of all zero assumption}.}

{This property is also called codeword symmetry, which is very favorable in either practice or theory since it guarantees that all of the codewords have the same error probability when they are transmitted through the AWGN channel. Theorem 2 holds since all of the nonbinary symbols in $GF(2^q)$ are mapped in the same way. Moreover, since the Flanagan embedding technique treats symbol 0 differently, the corresponding decoder does not satisfy the property of codeword symmetry. However, we should note that the presented simulation results show that both of the two proximal-ADMM decoders have almost the same error-correction performance.
}

\begin{table}[t]
\caption{{Summary of Computation Complexity in each proximal-ADMM iteration using the Constant-Weight embedding technique.}}
\renewcommand{\arraystretch}{1.3}
\begin{center}
{\begin{tabular}{|c|c|c|}
\hline
\hline
Variables  & Equations &  {Multiplications Number}                                    \\\hline
$\mathbf{v}^{k+1}$  & \eqref{lp-x-update-solution_CW} & $3\cdot2^q(n+\Gamma_{a})$ \\ \hline
$\mathbf{e}_1^{k+1}$  & \eqref{v1-solution-component} & $2(4\cdot2^q\Gamma_{c}+n+\Gamma_{a})$  \\ \hline
$\mathbf{e}_2^{k+1}$  & \eqref{v2-solution-component} & $2^{q+1}(n+\Gamma_{a})$  \\ \hline
$\mathbf{p}^{k+1}$    & \eqref{proximal-p-update}& $2^q(n+\Gamma_{a})$  \\ \hline
$\mathbf{z}_1^{k+1}$  & \eqref{proximal-z1-update}& $4\cdot2^q\Gamma_{c}+n+\Gamma_{a}$  \\ \hline
$\mathbf{z}_2^{k+1}$  & \eqref{proximal-z2-update} & $2^q(n+\Gamma_{a})$  \\ \hline
$\mathbf{y}_1^{k+1}/\mu_1$  & \eqref{proximal-y1-update} & free  \\ \hline
$\mathbf{y}_2^{k+1}/\mu_2$  & \eqref{proximal-y2-update} & free  \\ \hline
\multicolumn{2}{|c|}{Total}    & $(7\cdot2^{q} + 3)(n + \Gamma_{a}) + 12\cdot2^q\Gamma_{c}$ \\
\hline\hline
\end{tabular}}
\end{center}
\end{table}

\section{Proof of Theorem \ref{converge-proof-theorem}}\label{converge-proof}
Before we show its proof, we give one definition and three lemmas that are used to establish \emph{Theorem \ref{converge-proof-theorem}}.

we have the following lemma to show that the gradient of the considered augmented Lagrangian $\mathcal{L}_{\mu}(\cdot)$ is Lipschitz continuous.

\begin{lemma}\label{x-Lipschitz-constant}
Suppose $\alpha>0$ and $\mu>0$ and let $X:=\{\v|\A\v \preceq {\boldsymbol\varrho}, \mathbf{0} \preceq \v \preceq \mathbf{1}\}$. Then, the gradient of the augmented Lagrangian $\mathcal{L}_{\mu}$  with respect to variable $\v$ is Lipschitz continuous,
i.e., for any $\v,\v^{\prime}\in X$,
\begin{equation}\label{x-Lipschitz-defi}
\begin{split}
& \|\nabla_{\v} \mathcal{L}_{\mu}(\v,\! \e_{1}, \! \e_{2},\! \y_{1}, \! \y_{2})\!-\!\nabla_{\v} \mathcal{L}_{\mu}(\v^{\prime},\!\e_{1},\!\e_{2},\!\y_{1},\!\y_{2})\|_{2} \\
\leq &L\|\v-\v^{\prime}\|_{2},
\end{split}
\end{equation}
where $L\geq \alpha+\mu+\mu\delta_{\A}^2$ and ``$\delta_{\A}$'' denotes the spectral norm of matrix $\A$.
\end{lemma}

{\it Proof:} {See Appendix \ref{Lipschitz-continuous}.}

\begin{definition}\label{F-D-P-define}
Define the following local functions
 {\setlength\abovedisplayskip{1pt}
 \setlength\belowdisplayskip{1pt}
  \setlength\jot{1pt}
\begin{equation}\label{D-define}
\begin{split}
& \mathcal{D}(\p,\! \z_{1},\! \z_{2}, \!\y_{1}, \!\y_{2})
\!\! =\!\!\min\limits_{\v, \e_{1} \succeq \mathbf{0}, \atop \mathbf{0} \preceq \e_{2} \preceq \mathbf{1}}\!\!\!\mathcal{F}(\v,\!\e_{1},\! \e_{2}, \!\p, \!\z_{1}, \!\z_{2},\! \y_{1},\! \y_{2}),
\end{split}
\end{equation}
\begin{equation}\label{x-D-define}
\begin{split}
& \hspace{-0.45cm} \left[\!\!\!\!\begin{array}{l}{\v(\p,\!\z_{1},\!\z_{2},\!\y_{1},\! \y_{2})} \\ {\e_{1}(\p,\! \z_{1},\! \z_{2},\! \y_{1},\! \y_{2})} \\ {\e_{2}(\p,\! \z_{1},\! \z_{2},\! \y_{1},\! \y_{2})}\end{array}\!\!\!\!\right]
\!\!\!=\!\underset{\v, \e_{1} \succeq \mathbf{0}, \atop \mathbf{0} \preceq \e_{2} \preceq \mathbf{1}}{\arg \min }~\!\mathcal{F}\!(\v,\! \e_{1},\! \e_{2},\! \p,\! \z_{1},\! \z_{2},\! \y_{1},\! \y_{2}\!),
\end{split}
\end{equation}
\begin{equation}\label{P-define}
\begin{split}
 & \mathcal{P}(\p,\! \z_{1},\! \z_{2})
\! =\!\!\!\min\limits_{\A\v+\e_1={\boldsymbol\varrho}, \v=\e_2, \atop \e_{1} \succeq \mathbf{0},  \mathbf{0} \preceq \e_{2} \preceq \mathbf{1}}\!\!\mathcal{F}(\v,\!\e_{1},\! \e_{2}, \!\p, \!\z_{1}, \!\z_{2},\! \y_{1},\! \y_{2}),
\end{split}
\end{equation}
\begin{equation}\label{x-P-define}
\begin{split}
& \left[\!\!\!\!\begin{array}{l}{\v(\p,\!\z_{1},\!\z_{2})\!} \\ {\e_{1}(\p,\! \z_{1},\! \z_{2})} \\ {\e_{2}(\p,\! \z_{1},\! \z_{2})}\end{array}\!\!\!\!\right]
\!\!\! =\!\!\!\!\underset{\A\v+\e_1={\boldsymbol\varrho}, \v=\e_2, \atop \e_{1} \succeq \mathbf{0},  \mathbf{0} \preceq \e_{2} \preceq \mathbf{1}}{\arg \min }\!\!\!\!\mathcal{F}(\v,\!\e_{1},\! \e_{2}, \!\p,\z_{1}, \!\z_{2},\! \y_{1},\! \y_{2}),
\end{split}
\end{equation}
where} function $\mathcal{F}(\cdot)$ is expressed by
 {\setlength\abovedisplayskip{1pt}
 \setlength\belowdisplayskip{1pt}
  \setlength\jot{1pt}
\begin{equation}\label{F-define}
\hspace{-0.2cm}\begin{split}
\mathcal{F}(\v,\e_{1},\e_{2},\p, \z_{1}, \z_{2}, \y_{1}, \y_{2})=&\mathcal{L}_{\mu}(\v,\e_{1}, \e_{2}, \p, \z_{1}, \z_{2}, \y_{1},\y_{2})   \\
&+\frac{\rho}{2}\|\v\!-\!\p\|_{2}^{2}\!+\!\frac{\rho}{2}\left\|\e_{1}-\z_{1}\right\|_{2}^{2}+\frac{\rho}{2}\left\|\e_{2}-\z_{2}\right\|_{2}^{2}.
\end{split}
\end{equation}}
\end{definition}

Based on \emph{Definition 1}, the following inequalities hold.

\begin{lemma}\label{error-bound-lemma}
Suppose $\rho>\alpha>0$. then we have
\begin{equation}\label{error-bound1}
\begin{split}
& \|\!\!\!\left[\!\!\!\begin{array}{l}{\v^{k}-\v^{k+1}} \\ {\e_{1}^{k}-\e_{1}^{k+1}} \\ {\e_{2}^{k}-\e_{2}^{k+1}}\end{array}\!\!\!\right]\!\!\!\|_{2}^{2}
\geq \varepsilon_{1}\|\!\!\left[\!\!\begin{array}{c}{\v^{k}\! -\!\v\left(\p^{k}, \z_{1}^{k}, \z_{2}^{k}, \y_{1}^{k}, \y_{2}^{k}\right)}
 \\ {\e_{1}^{k}\!-\!\e_{1}\left(\p^{k}, \z_{1}^{k}, \z_{2}^{k}, \y_{1}^{k}, \y_{2}^{k}\right)}
 \\ {\e_{2}^{k}\!-\!\e_{2}\left(\p^{k}, \z_{1}^{k}, \z_{2}^{k}, \y_{1}^{k}, \y_{2}^{k}\right)}\end{array}\!\!\!\right]\!\!\|_{2}^{2},
\end{split}
\end{equation}
\begin{equation}\label{error-bound2}
\begin{split}
& \|\!\!\!\left[\!\!\begin{array}{l}{\v^{k}\!-\!\v^{k+1}} \\ {\e_{1}^{k}\!-\!\e_{1}^{k+1}} \\ {\e_{2}^{k}\!-\!\e_{2}^{k+1}}\end{array}\!\!\!\right]\!\!\!\|_{2}^{2}
\geq \!\varepsilon_{2}\|\!\!\left[\!\!\begin{array}{c}{\v^{k+1}\!\! -\!\v\left(\p^{k}, \z_{1}^{k}, \z_{2}^{k}, \y_{1}^{k}, \y_{2}^{k}\right)}
 \\ {\e_{1}^{k+1}\!\!-\!\e_{1}\left(\p^{k}, \z_{1}^{k}, \z_{2}^{k}, \y_{1}^{k}, \y_{2}^{k}\right)}
 \\ {\e_{2}^{k+1}\!\!-\!\e_{2}\left(\p^{k}, \z_{1}^{k}, \z_{2}^{k}, \y_{1}^{k}, \y_{2}^{k}\right)}\end{array}\!\!\right]\!\!\|_{2}^{2},
\end{split}
\end{equation}
\begin{equation}\label{error-bound3}
\begin{split}
&\hspace{-0.45cm}\|\!\!\left[\!\!\!\begin{array}{l}{\y_{1}\!\!-\!\y_{1}^{\prime}} \\ {\y_{2}\!\!-\!\y_{2}^{\prime}}\end{array} \!\!\right]\!\!\! \|_{2}^{2}
\!\geq\! \varepsilon_{3}\|\!\!\left[\!\!\begin{array}{l}{\v\left(\p, \z_{1}, \z_{2}, \y_{1}, \y_{2}\!\right)\!-\!\v\!\left(\p, \z_{1}, \z_{2}, \y_{1}^{\prime}, \y_{2}^{\prime}\right)}
\\ {\e_{1}\left(\p, \z_{1}, \z_{2}, \y_{1}, \y_{2}\right)\!-\!\e_{1}\left(\p, \z_{1}, \z_{2}, \y_{1}^{\prime}, \y_{2}^{\prime}\right)}
\\ {\e_{2}\left(\p, \z_{1}, \z_{2}, \y_{1}, \y_{2}\right)\!-\!\e_{2}\left(\p, \z_{1}, \z_{2}, \y_{1}^{\prime}, \y_{2}^{\prime}\right)}\end{array}\!\!\right]\!\!\|_2^2,
\end{split}
\end{equation}
\begin{equation}\label{error-bound4}
\begin{split}
&\hspace{-0.45cm}\|\!\!\left[\!\!\begin{array}{l}{\p^{k}\!-\!\p^{k+1}} \\ {\z_{1}^{k}\!-\!\z_{1}^{k+1}} \\ {\z_{2}^{k}\!-\!\z_{2}^{k+1}}\end{array}\!\!\right]\!\!\|_{2}^{2}
\!\geq \varepsilon_{4}\|\!\!\left[\!\!\begin{array}{l}{\v\left(\p^{k+1}, \z_{1}^{k+1}, \z_{2}^{k+1}\right)\!-\!\v\!\left(\p^{k}, \z_{1}^{k}, \z_{2}^{k}\right)}
\\ {\e_{1}\left(\p^{k+1}, \z_{1}^{k+1}, \z_{2}^{k+1}\right)\!-\!\e_{1}\left(\p^{k}, \z_{1}^{k}, \z_{2}^{k}\right)}
\\ {\e_{2}\left(\p^{k+1}, \z_{1}^{k+1}, \z_{2}^{k+1}\right)\!-\!\e_{2}\left(\p^{k}, \z_{1}^{k}, \z_{2}^{k}\right)}\end{array}\!\!\right],
\end{split}
\end{equation}
\begin{equation}\label{error-bound5}
\begin{split}
 \|\!\!\left[\!\!\begin{array}{l}{\p^{k}\!-\!\p^{k+1}} \\ {\z_{1}^{k}\!-\!\z_{1}^{k+1}} \\ {\z_{2}^{k}\!-\!\z_{2}^{k+1}}\end{array}\!\!\!\right]\!\!\!\|_{2}^{2}
& \!\geq \! \varepsilon_{5} \|\!\!\left[\!\!\begin{array}{l}{\v\!\left(\p^{k+1}\!, \z_{1}^{k+1}\!, \z_{2}^{k+1}\!, \y_{1}^{k+1}\!, \y_{2}^{k+1}\right)\!-\v\!\left(\u^{k}, \z_{1}^{k}, \z_{2}^{k}, \y_{1}^{k+1}, \y_{2}^{k+1}\right)}
\\ {\e_{1}\!\!\left(\p^{k+1}\!, \z_{1}^{k+1}\!, \z_{2}^{k+1}\!, \y_{1}^{k+1}\!, \y_{2}^{k+1}\!\right)-\e_{1}\!\!\left(\p^{k}, \z_{1}^{k}, \z_{2}^{k}, \y_{1}^{k+1}, \y_{2}^{k+1}\right)}
\\ {\e_{2}\!\!\left(\p^{k+1}\!, \z_{1}^{k+1}\!, \z_{2}^{k+1}\!, \y_{1}^{k+1}\!, \y_{2}^{k+1}\!\right)-\e_{2}\!\!\left(\p^{k}, \z_{1}^{k}, \z_{2}^{k}, \y_{1}^{k+1}, \y_{2}^{k+1}\right)}\end{array}\!\!\right]\!\!\|_{2}^{2},
\end{split}
\end{equation}
where
 {\setlength\abovedisplayskip{1pt}
 \setlength\belowdisplayskip{1pt}
  \setlength\jot{1pt}
\begin{equation}\label{varepsilon}
\begin{split}
&\varepsilon_1=\frac{(\rho-\alpha)^2}{(\rho+L+2)^2}, \  \varepsilon_2=\frac{(\rho-\alpha)^2}{(2\rho+L+2-\alpha)^2}, \\ &\varepsilon_3=\frac{(\rho-\alpha)^2}{\delta_{\A\I}^{2}}, \
\varepsilon_4=\frac{(\rho-\alpha)^2}{\rho^2}, \ \varepsilon_5=\frac{(\rho-\alpha)^2}{\rho^2}.
\end{split}
\end{equation}
Moreover,} if
$$\|\!\left[\!\!\begin{array}{c}{\A \v\left(\p,\z_{1},\z_{2},\y_{1}, \y_{2}\right)+\e_{1}\left(\p,\z_{1},\z_{2},\y_{1}, \y_{2}\right)\!-\!{\boldsymbol\varrho}}
\\ {\v\left(\p,\z_{1},\z_{2}, \y_{1}, \y_{2}\right)\!-\!\e_{2}\!\left(\p,\z_{1}, \z_{2}, \y_{1}, \y_{2}\right)}\end{array}\!\!\right]\!\!\|_2 \!\leq\! \Delta,$$
and
$$\|\!\!\left[\!\!\begin{array}{l}{\v-\p} \\ {\e_{1}-\z_1}
\\ {\e_{2}-\z_2}\end{array}\!\!\right]\!\!\|_2 \leq \Delta,$$
where $\Delta>0$ is some constant, then there exists $\varepsilon_{6}>0$ such that
 {\setlength\abovedisplayskip{1pt}
 \setlength\belowdisplayskip{1pt}
  \setlength\jot{1pt}
\begin{equation}\label{error-bound6}
\begin{split}
& \|\!\!\left[\!\!\begin{array}{l}{\y_{1}}-\y_1^{*}\left(\p, \z_{1}, \z_{2}\right) \\ {\y_{2}- \y_2^{*}\left(\u, \z_{1}, \z_{2}\right)}\end{array}\!\!\right]\!\!\|_2^2
\leq   \varepsilon_{6}\|\!\!\left[\!\!\!\!\begin{array}{c}{\A \v\left(\p, \z_{1}, \z_{2}, \y_{1}, \y_{2}\right)\!+\!\e_{1}\left(\p, \z_{1}, \z_{2}, \y_{1}, \y_{2}\right)\!-\!{\boldsymbol\varrho}}
\\ {\v\left(\p, \z_{1}, \z_{2}, \y_{1}, \y_{2}\right)-\e_{2}\left(\p, \z_{1}, \z_{2}, \y_{1}, \y_{2}\right)}\end{array}\!\!\right]\!\!\|_{2}^{2},
\end{split}
\end{equation}
where} ${\y_1^{*}\left(\p, \z_{1}, \z_{2}\right)}$ and ${\y_2^{*}\left(\p, \z_{1}, \z_{2}\right)}$ are the solution sets of dual multipliers for problem \eqref{P-define}.
\end{lemma}

{\it Proof:} {See Appendix \ref{error-bound-proof}.}

To save space, throughout the whole proof we denote functions $\mathcal{F}$, $\mathcal{D}$ and $\mathcal{P}$ at the $kth$ iteration by
\[
\begin{split}
&\F^{k} := \F\left(\v^{k}, \e_{1}^{k}, \e_{2}^{k}, \p^{k}, \z_{1}^{k}, \z_{2}^{k}, \y_{1}^{k}, \y_{2}^{k}\right), \\
&\D^{k} := \D\left(\p^{k}, \z_{1}^{k}, \z_{2}^{k}, \y_{1}^{k}, \y_{2}^{k}\right), \\
&\P^{k} := \P\left(\p^{k}, \z_{1}^{k}, \z_{2}^{k}\right),
\end{split}
\]
respectively. Using the above abbreviations, we further introduce the following lemma.
\begin{lemma}\label{F-D-P-change}
Let $\alpha\leq\mu\lambda_{\min}(\A^T\A)$. Then, the following inequalities hold
 {\setlength\abovedisplayskip{1pt}
 \setlength\belowdisplayskip{1pt}
  \setlength\jot{1pt}
\begin{equation}\label{F-change}
\begin{split}
 \F^{k}\!-\!\F^{k+1}\!\! \geq \!& \frac{\rho\!+\!\mu}{2}\|\!\!\left[\!\!\begin{array}{l}{\v^{k}\!-\!\v^{k+1}} \\ {\e_{1}^{k}\!-\!\e_{1}^{k+1}} \\ {\e_{2}^{k}\!-\!\e_{2}^{k+1}}\end{array}\!\!\right]\!\!\|_{2}^{2}
\!+\!\!\frac{\rho}{2 \beta}\!\|\!\!\left[\!\!\!\begin{array}{l}{\p^{k}\!-\!\p^{k+1}} \\ {\z_{1}^{k}\!-\!\z_{1}^{k+1}} \\ {\z_{2}^{k}\!-\!\z_{2}^{k+1}}\end{array}\!\!\!\right]\!\!\|_{2}^{2}
\!-\!\mu\|\!\!\left[\!\!\begin{array}{c}{\A \v^{k}+\e_{1}^{k}-{\boldsymbol\varrho}} \\ {\v^{k}-\e_{2}^{k}}\end{array}\!\!\right]\!\!\|_{2}^{2},
\end{split}
\end{equation}
\begin{equation}\label{D-change}
\begin{split}
 & \D^{k+1}\!\!-\!\!\D^{k}
\!\geq\!  \mu\!\!\left[\!\!\!\!\begin{array}{c}{\A\v^{k}\!+\!\e_{1}^{k}\!-\!{\boldsymbol\varrho}} \\ {\v^{k}\!-\!\e_{2}^{k}}\end{array}\!\!\!\right]^{T}\!\!\!\!\! \pmb{\phi}^{k}
\!\!+\!\!\frac{\rho}{2}\left[\!\!\!\!\begin{array}{c}{\p^{k+1}\!-\!\p^{k}} \\ {\z_{1}^{k+1}\!-\!\z_{1}^{k}} \\ {\z_{2}^{k+1}\!-\!\z_{2}^{k}}\end{array}\!\!\!\right]^{T}\!\!\!\!\! \pmb{\psi}^{k},
\end{split}
\end{equation}
\begin{equation}\label{P-change}
\begin{split}
 \P^{k}\!-\! \P^{k+1} \geq  & \rho\!\left[\!\!\!\begin{array}{c}{\p^{k+1}\!-\!\p^{k}} \\ {\z_{1}^{k+1}\!-\!\z_{1}^{k}} \\ {\z_{2}^{k+1}\!-\!\z_{2}^{k}}\end{array}\!\!\!\right]^{T}\!\!
\left[\!\!\begin{array}{l}{\v\left(\p^{k}, \z_{1}^{k}, \z_{2}^{k}\right)-\p^{k}} \\ {\e_{1}\left(\p^{k}, \z_{1}^{k}, \z_{2}^{k}\right)-\z_{1}^{k}}
\\ {\e_{2}\left(\p^{k}, \z_{1}^{k}, \z_{2}^{k}\right)-\z_{2}^{k}}\end{array}\!\!\right]
-\frac{\rho\eta}{2}\|\!\!\left[\!\!\begin{array}{c}{\p^{k+1}-\p^{k}} \\ {\z_{1}^{k+1}-\z_{1}^{k}} \\ {\z_{2}^{k+1}-\z_{2}^{k}}\end{array}\!\!\right]\!\!\|_2^2,
\end{split}
\end{equation}
where} $\eta$ and ``$\pmb{\phi}^{k}$'' and ``$\pmb{\psi}^{k}$'' are defined as follows
\begin{subequations}\label{eta D-part1-change D-part2-change}
 \begin{align}
  &\eta=1+\frac{1}{\sqrt{\epsilon_4}}, \\
  &\pmb{\phi}^{k}\!\!=\!\!\left[\!\!\begin{array}{c}{\A\v\left(\p^{k}, \z_{1}^{k},\z_{2}^{k},\y_{1}^{k+1}, \y_{2}^{k+1}\right)\!+\!\e_{1}\left(\p^{k},\z_{1}^{k},\z_{2}^{k}, \y_{1}^{k+1},\y_{2}^{k+1}\right)\!-\!{\boldsymbol\varrho}}
\\ {\v\left(\p^{k},\z_{1}^{k},\z_{2}^{k},\y_{1}^{k+1}, \y_{2}^{k+1}\right)\!-\!\e_{2}\left(\p^{k},\z_{1}^{k},\z_{2}^{k}, \y_{1}^{k+1},\y_{2}^{k+1}\right)}\end{array}\!\!\!\right], \label{D-part1-change}\\
  &\pmb{\psi}^{k}\!\!=\!\!\left[\!\!\!\begin{array}{c}{\p^{k+1}\!+\!\p^{k}\!-\!2 \v\left(\p^{k+1},\!\z_{1}^{k+1},\!\z_{2}^{k+1},\!\y_{1}^{k+1},\! \y_{2}^{k+1}\!\right)}
\\ {\z_{1}^{k+1}\!+\!\z_{1}^{k}\!-\!2\e_{1}\!\left(\p^{k+1},\!\z_{1}^{k+1},\! \z_{2}^{k+1},\!\y_{1}^{k+1},\! \y_{2}^{k+1}\!\right)}
\\ {\z_{2}^{k+1}\!+\!\z_{2}^{k}\!-\!2\e_{2}\!\left(\p^{k+1},\!\z_{1}^{k+1},\! \z_{2}^{k+1},\!\y_{1}^{k+1},\!\y_{2}^{k+1}\!\right)}\end{array}\!\!\!\right].
\end{align}
\end{subequations}
\end{lemma}

{\it Proof:} {See Appendix \ref{descent3-proof}.}

Now we are ready to prove \emph{Theorem \ref{converge-proof-theorem}}.

{\it Proof:} {
First, we define a potential function as follows
\begin{equation}\label{Psi-function}
\begin{split}
& \Psi=\F-2 \D+2 \P.
\end{split}
\end{equation}
The key to proving convergence of the proposed proximal-ADMM algorithm is to verify that the function $\Psi$ not only {\it decreases  sufficiently} in each iteration but also is lower-bounded.

Based on \eqref{F-change}-\eqref{P-change} in \emph{Lemma \ref{F-D-P-change}}, we obtain
\begin{equation}\label{Psi-change1}
\begin{split}
 & \Psi^{k}-\Psi^{k+1} \\
= & \left(\F^{k}-\F^{k+1}\right)+2\left(\D^{k+1}-\D^{k}\right)+2\left(\P^{k}-\P^{k+1}\right) \\
\geq & \frac{\rho\!+\!\mu}{2} \|\!\!\left[\!\!\begin{array}{l}{\v^{k}-\v^{k+1}} \\ {\e_{1}^{k}-\e_{1}^{k+1}} \\ {\e_{2}^{k}-\e_{2}^{k+1}}\end{array}\!\!\right]\!\!\|_2^2
+\!\left(\!\frac{\rho}{2 \beta}-\rho \eta\!\right)\|\!\!\left[\!\!\begin{array}{c}{\p^{k+1}-\p^{k}} \\ {\z_{1}^{k+1}-\z_{1}^{k}} \\ {\z_{2}^{k+1}-\z_{2}^{k}}\end{array}\!\!\!\right]\!\!\|_2^2\\
&\!+\!\rho\! \left[\!\!\!\begin{array}{c}{\p^{k+1}-\p^{k}} \\ {\z_{1}^{k+1}-\z_{1}^{k}} \\ {\z_{2}^{k+1}-\z_{2}^{k}}\end{array}\!\!\!\right]^{T}\!\!
\left(\!\!\pmb{\psi}^k+2\left[\!\!\!\!\begin{array}{c}{\v\left(\p^{k}, \z_{1}^{k}, \z_{2}^{k}\right)-\p^{k}} \\ {\e_{1}\left(\p^{k}, \z_{1}^{k}, \z_{2}^{k}\right)-\z_{1}^{k}}
\\ {\e_{2}\left(\p^{k}, \z_{1}^{k}, \z_{2}^{k}\right)-\z_{2}^{k}}\end{array}\!\!\!\!\right]\!\right) \\
&-\mu\left(\!\!\|\!\!\left[\!\!\!\begin{array}{c}{\A \v^{k}+\e_{1}^{k}-{\boldsymbol\varrho}} \\ {\v^{k}-\e_{2}^{k}}\end{array}\!\!\!\right]\!\!\|_{2}^{2}
\!-\!2\left[\!\!\!\begin{array}{c}{\A\v^{k}\!+\!\e_{1}^{k}\!-\!{\boldsymbol\varrho}} \\ {\v^{k}\!-\!\e_{2}^{k}}\end{array}\!\!\!\right]^{T}\!\!\!\! \pmb{\phi}^k\!\right).
\end{split}
\end{equation}
For the last term of \eqref{Psi-change1}, we have the following derivations
\[
\begin{split}
\hspace{-0.35cm}
& \|\!\!\left[\!\!\begin{array}{c}{\A \v^{k}+\e_{1}^{k}-{\boldsymbol\varrho}} \\ {\v^{k}-\e_{2}^{k}}\end{array}\!\!\!\right]\!\!\|_{2}^{2}
\!-\!2\left[\!\!\begin{array}{c}{\A\v^{k}\!+\!\e_{1}^{k}\!-\!{\boldsymbol\varrho}} \\ {\v^{k}\!-\!\e_{2}^{k}}\end{array}\!\!\right]^{T}\!\! \pmb{\phi}^k \\
=&\|\!\!\left[\!\!\begin{array}{ccc}{\A} & {\I_{M}} & {\mathbf{0}} \\ {\I_{N}} & {\mathbf{0}} & {-\I_{N}}\end{array}\!\!\right]\!\!
\left[\!\!\begin{array}{c}{\v^{k}\!-\!\v\left(\p^{k}, \z_{1}^{k}, \z_{2}^{k}, \y_{1}^{k+1}, \y_{2}^{k+1}\right)}
\\ {\e_{1}^{k}-\e_{1}\left(\p^{k}, \z_{1}^{k}, \z_{2}^{k}, \y_{1}^{k+1}, \y_{2}^{k+1}\right)}
\\ {\e_{2}^{k}-\e_{2}\left(\p^{k}, \z_{1}^{k}, \z_{2}^{k}, \y_{1}^{k+1}, \y_{2}^{k+1}\!\right)}\end{array}\!\!\!\!\right]
\!\!\|_2^2 \!-\!\|\pmb{\phi}^k\|_2^{2} \\
 \leq &  \delta_{\A\I}^{2}\|\!\!\left[\!\!\begin{array}{c}{\v^{k}\!-\!\v\left(\p^{k}, \z_{1}^{k}, \z_{2}^{k}, \y_{1}^{k+1}, \y_{2}^{k+1}\right)}
\\ {\e_{1}^{k}\!-\!\v_{1}\!\left(\p^{k}, \z_{1}^{k}, \z_{2}^{k}, \y_{1}^{k+1}, \y_{2}^{k+1}\right)}
\\ {\e_{2}^{k}\!-\!\v_{2}\!\left(\p^{k}, \z_{1}^{k}, \z_{2}^{k}, \y_{1}^{k+1}, \y_{2}^{k+1}\right)}\end{array}\!\!\right]\!\!\|_2^2\!-\! \|\pmb{\phi}^k\|_2^{2}.
\end{split}
\]
From \eqref{error-bound1} in Lemma \ref{error-bound-lemma}, we can further get
\begin{equation}\label{Psi-change11}
\begin{split}
 \|\!\!\left[\!\!\!\begin{array}{c}{\A \v^{k}+\e_{1}^{k}-{\boldsymbol\varrho}} \\ {\v^{k}-\e_{2}^{k}}\end{array}\!\!\!\right]\!\!\|_{2}^{2}
\!-\!2\left[\!\!\!\begin{array}{c}{\A\v^{k}\!+\!\e_{1}^{k}\!-\!{\boldsymbol\varrho}} \\ {\v^{k}\!-\!\e_{2}^{k}}\end{array}\!\!\!\right]^{T}\!\!\!\! \pmb{\phi}^k
\leq  \frac{\delta_{\A\I}^{2}}{\varepsilon_{1}}
\|\!\!\left[\!\!\begin{array}{l}{\v^{k}-\v^{k+1}} \\ {\e_{1}^{k}-\e_{1}^{k+1}} \\ {\e_{2}^{k}-\e_{2}^{k+1}}\end{array}\!\!\!\right]\!\!\!\|_2^2\!-\! \|\pmb{\phi}^k\|_2^{2}.
\end{split}
\end{equation}
Plugging \eqref{Psi-change11} into \eqref{Psi-change1}, the inequality can be revised as
\begin{equation}\label{Psi-change2}
\begin{split}
  & \Psi^{k}-\Psi^{k+1} \\
 \geq & \left(\!\!\frac{\rho\!+\!\mu}{2}\!-\!\mu\frac{\delta_{\A\I}^{2}}{\varepsilon_{1}}\!\!\right)\!\!
\|\!\!\left[\!\!\begin{array}{l}{\v^{k}\!-\!\v^{k+1}\!} \\ {\e_{1}^{k}\!-\!\e_{1}^{k+1}\!} \\ {\e_{2}^{k}\!-\!\e_{2}^{k+1}\!}\end{array}\!\!\right]\!\!\|_2^2
\!+\!\left(\!\!\frac{\rho}{2 \beta}\!-\!\rho \eta\!\right)\!\!\|\!\!\left[\!\!\!\!\begin{array}{c}{\p^{k+1}\!-\!\p^{k}} \\ {\z_{1}^{k+1}\!-\!\z_{1}^{k}} \\ {\z_{2}^{k+1}\!-\!\z_{2}^{k}}\end{array}\!\!\!\right]\!\!\!\|_2^2\\
 &\!+\! \mu\|\pmb{\phi}^k\|_2^{2}
\!+\!\rho\! {\left[\!\!\begin{array}{c}{\p^{k+1}\!-\!\mathbf{p}^{k}} \\ {\z_{1}^{k+1}\!-\!\z_{1}^{k}} \\ {\z_{2}^{k+1}\!-\!\z_{2}^{k}}\end{array}\!\!\right]^{T}\!\!
\left(\!\!\pmb{\psi}^k\!+\!2\!\left[\!\!\begin{array}{c}{\v\left(\p^{k}, \z_{1}^{k}, \z_{2}^{k}\right)-\p^{k}} \\ {\v_{1}\!\left(\p^{k}, \z_{1}^{k}, \z_{2}^{k}\right)-\z_{1}^{k}}
\\ {\v_{2}\!\left(\p^{k}, \z_{1}^{k}, \z_{2}^{k}\right)-\z_{2}^{k}}\end{array}\!\!\right]\!\right)}.
\end{split}
\end{equation}

To facilitate derivations later, we define
\begin{equation*}
\begin{split}
& \mathcal{X}^{k}\! :=\! \left[\!\!\begin{array}{l}{\v\left(\p^{k+1}, \z_{1}^{k+1}, \z_{2}^{k+1}, \y_{1}^{k+1}, \y_{2}^{k+1}\right)\!-\!\v\left(\p^{k}, \z_{1}^{k}, \z_{2}^{k}\right)}
\\ {\e_{1}\left(\p^{k+1}, \z_{1}^{k+1}, \z_{2}^{k+1}, \y_{1}^{k+1}, \y_{2}^{k+1}\right)\!-\!\e_{1}\left(\p^{k}, \z_{1}^{k}, \z_{2}^{k}\right)}
\\ {\e_{2}\left(\p^{k+1}, \z_{1}^{k+1}, \z_{2}^{k+1}, \y_{1}^{k+1}, \y_{2}^{k+1}\right)\!-\!\e_{2}\left(\p^{k}, \z_{1}^{k}, \z_{2}^{k}\right)}\end{array}\!\!\right].
\end{split}
\end{equation*}
Then, the last term in \eqref{Psi-change2} can be rewritten as \eqref{Psi-change22}.
Applying property $2ab\leq a^{2} / \lambda^{2}+\lambda^{2} b^{2}$, we derive ``$\eqref{Psi-change22}-(a)$'' as
\begin{figure*}
\begin{equation}\label{Psi-change22}
\begin{split}
\left[\!\!\!\!\begin{array}{c}{\p^{k+1}\!-\p^{k}} \\ {\z_{1}^{k+1}\!-\z_{1}^{k}} \\ {\z_{2}^{k+1}\!-\z_{2}^{k}}\end{array}\!\!\!\!\right]^{T}\!\!\!\!
& \left(\!\!\pmb{\psi}^k+2\!\left[\!\!\!\!\begin{array}{c}{\v\left(\p^{k}, \z_{1}^{k}, \z_{2}^{k}\right)-\p^{k}} \\ {\e_{1}\left(\p^{k}, \z_{1}^{k}, \z_{2}^{k}\right)-\z_{1}^{k}}
\\ {\e_{2}\left(\p^{k}, \z_{1}^{k}, \z_{2}^{k}\right)-\z_{2}^{k}}\end{array}\!\!\!\!\right]\!\right)
=\|\!\!\left[\!\!\!\!\begin{array}{c}{\p^{k+1}\!-\!\p^{k}} \\ {\z_{1}^{k+1}\!-\!\z_{1}^{k}} \\ {\z_{2}^{k+1}\!-\!\z_{2}^{k}}\end{array}\!\!\!\right]\!\!\!\|_2^2
-\underbrace{2\left[\!\!\!\!\begin{array}{c}{\p^{k+1}\!-\mathbf{p}^{k}} \\ {\z_{1}^{k+1}\!-\z_{1}^{k}} \\ {\z_{2}^{k+1}\!-\z_{2}^{k}}\end{array}\!\!\!\right]^{T}\!\!\!\!\!\mathcal{X}^{k}\!\!}_{\eqref{Psi-change22}-(a)}\\
&~~~~~~~~~~ -2\underbrace{\left[\!\!\!\!\begin{array}{c}{\p^{k+1}\!-\mathbf{p}^{k}} \\ {\z_{1}^{k+1}\!-\z_{1}^{k}} \\ {\z_{2}^{k+1}\!-\z_{2}^{k}}\end{array}\!\!\!\right]^{T}\!\!\!
\left[\!\!\!\begin{array}{c}{\v\left(\p^{k}, \z_{1}^{k}, \z_{2}^{k}, \y_{1}^{k+1}, \y_{2}^{k+1}\right)-\v\left(\p^{k}, \z_{1}^{k}, \z_{2}^{k}\right)}
\\ {\e_{1}\left(\p^{k}, \z_{1}^{k}, \z_{2}^{k}, \y_{1}^{k+1}, \y_{2}^{k+1}\right)-\e_{1}\left(\p^{k}, \z_{1}^{k}, \z_{2}^{k}\right)}
\\ {\e_{2}\left(\p^{k}, \z_{1}^{k}, \z_{2}^{k}, \y_{1}^{k+1}, \y_{2}^{k+1}\right)-\e_{2}\left(\p^{k}, \z_{1}^{k}, \z_{2}^{k}\right)}\end{array}\!\!\!\right]}_{\eqref{Psi-change22}-(b)}.
\end{split}
\end{equation}
\end{figure*}
\begin{equation}\label{Psi-change222}
\begin{split}
2\left[\!\!\begin{array}{c}{\p^{k+1}\!-\mathbf{p}^{k}} \\ {\z_{1}^{k+1}-\z_{1}^{k}} \\ {\z_{2}^{k+1}-\z_{2}^{k}}\end{array}\!\!\right]^{T}\!\!\mathcal{X}^{k}
&\leq \|\!\!\left[\!\!\begin{array}{c}{\p^{k+1}\!-\p^{k}} \\ {\z_{1}^{k+1}-\z_{1}^{k}} \\ {\z_{2}^{k+1}-\z_{2}^{k}}\end{array}\!\!\right]\!\!\|_2^2/\lambda^2
+ \lambda^2 \|\mathcal{X}^{k}\|_2^2.
\end{split}
\end{equation}
Moreover, according to Cauchy-Schwarz inequality and \eqref{error-bound5} in \emph{Lemma \ref{error-bound-lemma}}, inequality \eqref{Psi-change221} holds.
\begin{figure*}
\begin{equation}\label{Psi-change221}
\begin{split}
& \left[\!\!\begin{array}{c}{\p^{k+1}\!-\p^{k}} \\ {\z_{1}^{k+1}\!-\z_{1}^{k}} \\ {\z_{2}^{k+1}\!-\z_{2}^{k}}\end{array}\!\!\right]^{T}\!\!
 \left[\!\!\begin{array}{c}{\v\left(\p^{k}, \z_{1}^{k}, \z_{2}^{k}, \y_{1}^{k+1}, \y_{2}^{k+1}\right)-\v\left(\p^{k}, \z_{1}^{k}, \z_{2}^{k}\right)}
\\ {\e_{1}\left(\p^{k}, \z_{1}^{k}, \z_{2}^{k}, \y_{1}^{k+1}, \y_{2}^{k+1}\right)-\e_{1}\left(\p^{k}, \z_{1}^{k}, \z_{2}^{k}\right)}
\\ {\e_{2}\left(\p^{k}, \z_{1}^{k}, \z_{2}^{k}, \y_{1}^{k+1}, \y_{2}^{k+1}\right)-\e_{2}\left(\p^{k}, \z_{1}^{k}, \z_{2}^{k}\right)}\end{array}\!\!\right]
\leq \frac{1}{\sqrt{\varepsilon_{5}}}\|\!\!\left[\!\!\begin{array}{c}{\p^{k+1}\!-\!\p^{k}} \\ {\z_{1}^{k+1}\!-\!\z_{1}^{k}} \\ {\z_{2}^{k+1}\!-\!\z_{2}^{k}}\end{array}\!\!\right]\!\!\|_2^2.
\end{split}
\end{equation}
\hrulefill
\vspace*{4pt}
\end{figure*}
Then, plugging \eqref{Psi-change222} and \eqref{Psi-change221} into \eqref{Psi-change22}, we can obtain
\begin{equation}\label{Psi-change22-2}
\begin{split}
& \left[\!\!\begin{array}{c}{\p^{k+1}\!-\!\p^{k}} \\ {\z_{1}^{k+1}\!-\!\z_{1}^{k}} \\ {\z_{2}^{k+1}\!-\!\z_{2}^{k}}\end{array}\!\!\right]^{T}\!\!\!\!
\left(\!\!\pmb{\psi}^k\!+\!2\!\left[\!\!\begin{array}{c}{\v\left(\p^{k}, \z_{1}^{k}, \z_{2}^{k}\right)-\p^{k}} \\ {\e_{1}\left(\p^{k}, \z_{1}^{k}, \z_{2}^{k}\right)-\z_{1}^{k}}
\\ {\e_{2}\left(\p^{k}, \z_{1}^{k}, \z_{2}^{k}\right)-\z_{2}^{k}}\end{array}\!\!\!\right]\!\right)
\geq  \left(1-\frac{1}{\lambda^{2}}-\frac{2}{\sqrt{\varepsilon_{5}}}\right)
\|\!\!\left[\!\!\!\!\begin{array}{c}{\p^{k+1}\!-\p^{k}} \\ {\z_{1}^{k+1}\!-\z_{1}^{k}} \\ {\z_{2}^{k+1}\!-\z_{2}^{k}}\end{array}\!\!\!\right]\!\!\|_2
-\lambda^{2}\|\mathcal{X}^{k}\|_2^2.
\end{split}
\end{equation}
Furthermore, plugging \eqref{Psi-change22-2} into \eqref{Psi-change2}, we have
\begin{equation}\label{Psi-change3}
\begin{split}
\Psi^{k}-\Psi^{k+1}
 \geq&
\rho\left(\!\!\frac{1}{2 \beta}\!-\! \eta\!+\!1\!-\!\frac{1}{\lambda^{2}}\!-\!\frac{2}{\sqrt{\varepsilon_{5}}}\right)\!\!\|\!\!\left[\!\!\begin{array}{c}{\p^{k+1}\!-\!\p^{k}} \\ {\z_{1}^{k+1}\!-\!\z_{1}^{k}} \\ {\z_{2}^{k+1}\!-\!\z_{2}^{k}}\end{array}\!\!\!\right]\!\!\!\|_2^2\\
&+\left(\!\!\frac{\rho\!+\!\mu}{2}\!-\!\mu\frac{\delta_{\A\I}^{2}}{\varepsilon_{1}}\!\!\right)\!\!
\|\!\!\left[\!\!\begin{array}{l}{\v^{k}\!-\!\v^{k+1}\!} \\ {\e_{1}^{k}\!-\!\e_{1}^{k+1}\!} \\ {\e_{2}^{k}\!-\!\e_{2}^{k+1}\!}\end{array}\!\!\right]\!\!\|_2^2
\!+\! \mu\|\pmb{\phi}^k\|_2^{2}\!-\!\rho\lambda^{2}\|\mathcal{X}^{k}\|_2^2.
\end{split}
\end{equation}
Letting $\lambda^{2}\!=\!\omega\beta$ and noticing  $\eta\!=\!1\!+\!\frac{1}{\sqrt{\varepsilon_{4}}}$(see \eqref{eta D-part1-change D-part2-change}), one can verify $\eta-1+\frac{1}{\lambda^{2}}+\frac{2}{\sqrt{\varepsilon_{5}}} \leq\frac{1}{3\beta}$  when
$\omega \geq \frac{6 \sqrt{\varepsilon_{4} \varepsilon_{5}}}{\sqrt{\varepsilon_{4} \varepsilon_{5}}-6 \beta\left(2\sqrt{\varepsilon_{4}}+\sqrt{\varepsilon_{5}}\right)}$.
Moreover, since $\varepsilon_{1}\! =\!\frac{(\rho-\alpha)^{2}}{(\rho+L+2)^{2}}$ (see \eqref{epsilon_varphi}) and the assumption $\mu \leq \frac{\rho(\rho-\alpha)^{2}}{4 \delta_{\A\I}^{2}(\rho+L+2)^{2}-(\rho-\alpha)^{2}}$ in \emph{Theorem \ref{converge-proof-theorem}}, one can verify that $\frac{\mu \delta_{\A\I}^{2}}{\varepsilon_{1}} \leq \frac{\rho+\mu}{4}$ holds.
Then, \eqref{Psi-change3} can be deduced as follows
\begin{equation}\label{Psi-change4}
\begin{split}
\Psi^{k}-\Psi^{k+1}
\geq & \frac{\rho}{3 \beta}\|\!\!\left[\!\!\begin{array}{c}{\p^{k+1}\!-\!\p^{k}} \\ {\z_{1}^{k+1}\!-\!\z_{1}^{k}} \\ {\z_{2}^{k+1}\!-\!\z_{2}^{k}}\end{array}\!\!\right]\!\!\|_2^2\!+\!\frac{\rho\!+\!\mu}{4}
\|\!\!\left[\!\!\begin{array}{l}{\v^{k}\!-\!\v^{k+1}\!} \\ {\e_{1}^{k}\!-\!\e_{1}^{k+1}\!} \\ {\e_{2}^{k}\!-\!\e_{2}^{k+1}\!}\end{array}\!\!\right]\!\!\|_2^2
+ \mu\|\pmb{\phi}^k\|_2^{2}-\rho \omega \beta\|\mathcal{X}^{k}\|_2^2.
\end{split}
\end{equation}

In the following, we show that term $\rho\omega\beta\|\mathcal{X}^{k}\|_2^2$ can be bounded by the previous three terms.

First, since $\mathbf{0}\preceq\v\preceq\mathbf{1}$, $\A\v$ is bounded.
Moreover, since $\e_1 \succeq \mathbf{0}$ and $\A\v+\e_1-{\boldsymbol\varrho}=\mathbf{0}$, there exists some positive vector $\pmb{\theta}$ such that $\e_1\preceq \pmb{\theta}$. Then, we can define
\begin{equation}\label{V-define}
\begin{split}
& V: =\max\limits_{\mathbf{0} \preceq \v,\v^{\prime},\e_{2},\e_2^{\prime} \preceq \mathbf{1}, \atop \mathbf{0}\preceq \e_{1},\e_1^{\prime} \preceq \pmb{\theta}}
\|\!\!\left[\!\!\begin{array}{l}{\v} \\ {\e_{1}} \\ {\e_{2}}\end{array}\!\!\right]\!\!-\!\!\left[\!\!\begin{array}{c}{\v^{\prime}} \\ {\e_{1}^{\prime}} \\ {\e_{2}^{\prime}}\end{array}\!\!\right]\!\!\|_{2}.
\end{split}
\end{equation}
Moreover, we define
 {\setlength\abovedisplayskip{0pt}
 \setlength\belowdisplayskip{2pt}
  \setlength\jot{1pt}
\begin{equation}\label{deita-function-define}
\begin{split}
&\zeta:=\min \{\Delta, \sigma(\Delta / \sqrt{6 \omega})\},
\end{split}
\end{equation}
where} $\sigma(\cdot)$ is some function satisfying $\underset{\epsilon\rightarrow0}\lim \sigma(\epsilon)=0$. Since $0<\beta\leq1$ (see \eqref{proximal-ADMM update_LP} below), we can denote $\beta$'s upper-bound as
 {\setlength\abovedisplayskip{1pt}
 \setlength\belowdisplayskip{1pt}
  \setlength\jot{1pt}
\begin{equation}\label{beita-upper-define}
\begin{split}
& \beta < \min \left\{1, \frac{(\rho+\mu) \zeta^{2}}{8 \rho \omega V^{2}}, \frac{\zeta^{2} \mu}{2 \rho \omega V^{2}}, \frac{\mu \varepsilon_{3}}{2 \rho \omega \varepsilon_{6}}\right\},
\end{split}
\end{equation}
Moreover,} we define the following inequalities
\begin{subequations}
\begin{align}
 &\|\!\!\left[\begin{array}{c}{\v^{k}-\v^{k+1}} \\ {\e_{1}^{k}-\e_{1}^{k+1}} \\ {\e_{2}^{k}-\e_{2}^{k+1}}\end{array}\right]\!\rVert_{2}^{2} \leq \frac{8 \rho \omega V^{2} \beta}{\rho+\mu},   \label{supposed-inequality-1} \\
 & \|\pmb{\phi}^k\|_2^2 \leq \frac{2 \rho \omega V^{2}}{\mu} \beta, \label{supposed-inequality-2} \\
 & \|\!\!\left[\!\!\!\begin{array}{c}{\p^{k}}-{\p^{k+1}} \\ {\z_{1}^{k}}-{\z_{1}^{k+1}} \\ {\z_{2}^{k}}-{\z_{2}^{k+1}}\end{array}\!\!\right]\!\!\|_{2}^{2}
 \leq 6 \omega \beta^{2} \|\mathcal{X}^{k}\|_2^2. \label{supposed-inequality-3}
\end{align}
\end{subequations}

Now, we are ready to check the boundness of $\mathcal{X}^{k}$.
First, we assume all of the inequalities \eqref{supposed-inequality-1}--\eqref{supposed-inequality-3} hold. Then,
plugging \eqref{V-define}-\eqref{beita-upper-define} into \eqref{supposed-inequality-2}, we can obtain \eqref{case1-inequality2} and \eqref{case1-inequality2_2} simultaneously.
\begin{equation}\label{case1-inequality2}
\begin{split}
& \|\pmb{\phi}^k\|_2 \leq \Delta,
\end{split}
\end{equation}
\begin{equation}\label{case1-inequality2_2}
\begin{split}
& \|\pmb{\phi}^k\|_2 \leq \sigma(\Delta / \sqrt{6 \omega}).
\end{split}
\end{equation}
Then, we can obtain\footnotemark \footnotetext{See proofs in Appendix \ref{case1-inequality3-proof} .}
\begin{equation}\label{case1-inequality3}
\begin{split}
& \|\mathcal{X}^{k}\|_2 \leq \Delta / \sqrt{6 \omega}.
\end{split}
\end{equation}
Then, combining \eqref{proximal-z1-update}, \eqref{supposed-inequality-3}, and \eqref{case1-inequality3}, we have
\begin{equation}\label{error-bound6-hold}
\begin{split}
 & \|\!\!\left[\!\!\begin{array}{c}{\v^{k+1}\!-\!\p^k} \\ {\e_{1}^{k+1}\!-\!\z_1^k} \\ {\e_{2}^{k+1}\!-\!\z_2^{k}}\end{array}\!\!\right]\!\!\|_{2}^{2}
=\!\|\!\!\left[\!\!\begin{array}{c}{\p^{k+1}\!-\p^{k}} \\ {\z_{1}^{k+1}\!-\z_{1}^{k}} \\ {\z_{2}^{k+1}\!-\z_{2}^{k}}\end{array}\!\!\!\right]\!\!\|_{2}^{2}/\beta^{2}
\!\leq \! 6 \omega \|\mathcal{X}^{k}\|_2^2 \! \leq\! \Delta.
\end{split}
\end{equation}
Combining it with \eqref{case1-inequality2}, we can see \eqref{error-bound6} holds.

Moreover, noticing $\mathcal{X}^k$ and $\pmb\phi^k$ are on the right side of the inequalities \eqref{error-bound6} and \eqref{error-bound3} respectively, we have the following inequality chain
\begin{equation}\label{case1-inequality4}
 \|\mathcal{X}^{k}\|_2^2
\leq \frac{1}{\varepsilon_{3}} \|\!\!\left[\!\!\begin{array}{c}{\y_{1}^{k+1}}\!-\!\y_1^{*}\left(\p^{k}, \z_{1}^{k}, \z_{2}^{k}\right) \\ {\y_{2}^{k+1}}\!\!-\!\!{\y_2^{*}\left(\p^{k}, \z_{1}^{k}, \z_{2}^{k}\right)}\end{array}\!\!\right]
\|_{2}^{2}
\!\!\leq\!\!  \frac{\varepsilon_6}{\varepsilon_{3}}\|\pmb{\phi}^k\|_2^2,
\end{equation}
where the first inequality comes from \eqref{error-bound3}, and the second inequality comes from \eqref{error-bound6}. Moreover, according to  \eqref{beita-upper-define}, \eqref{case1-inequality4} can be further derived to
\begin{equation}\label{case1-inequality5}
 \rho\omega\beta\|\mathcal{X}^{k}\|_2^2
\leq  \frac{\mu}{2}\|\pmb{\phi}^k\|_2^2.
\end{equation}

Next, we consider the case that at least one of the inequalities \eqref{supposed-inequality-1}-\eqref{supposed-inequality-3} does not hold.
There are three scenarios:
\begin{enumerate}
  \item \eqref{supposed-inequality-1} does not hold i.e.,
$
 \|\!\!\left[\begin{array}{c}{\v^{k}-\v^{k+1}} \\ {\e_{1}^{k}-\e_{1}^{k+1}} \\ {\e_{2}^{k}-\e_{2}^{k+1}}\end{array}\right]\!\!\|_{2}^{2} > \frac{8 \rho \omega V^{2} \beta}{\rho+\mu}.
$
Then, we have the following derivations
\begin{equation}\label{Psi-change5-case}
\begin{split}
&\frac{\rho+\mu}{4}
\|\!\!\left[\!\!\begin{array}{l}{\v^{k}\!-\!\v^{k+1}\!} \\ {\e_{1}^{k}\!-\!\e_{1}^{k+1}\!} \\ {\e_{2}^{k}\!-\!\e_{2}^{k+1}\!}\end{array}\!\!\right]\!\!\|_2^2
-\rho\omega\beta\|\mathcal{X}^{k}\|_2^2  \\
\geq &  \!\frac{\rho+\mu}{8}
\|\!\!\left[\!\!\begin{array}{l}{\v^{k}\!-\!\v^{k+1}} \\ {\e_{1}^{k}\!-\!\e_{1}^{k+1}} \\ {\e_{2}^{k}\!-\!\e_{2}^{k+1}}\end{array}\!\!\right]\!\!\|_2^2
\!+\! \frac{\rho\!+\!\mu}{8} \cdot \frac{8 \rho \omega V^{2} \beta}{\rho+\mu}
\!-\!\rho\omega\beta\|\mathcal{X}^{k}\|_2^2 \\
= & \frac{\rho+\mu}{8}
\|\!\!\left[\!\!\begin{array}{l}{\v^{k}\!-\!\v^{k+1}\!} \\ {\e_{1}^{k}\!-\!\e_{1}^{k+1}\!} \\ {\e_{2}^{k}\!-\!\e_{2}^{k+1}\!}\end{array}\!\!\right]\!\!\|_2^2
+ \rho\omega \beta V^{2}-\rho\omega\beta\|\mathcal{X}^{k}\|_2^2.
\end{split}
\end{equation}
Since $V\geq\|\mathcal{X}^{k}\|_2$ (see \eqref{V-define}), we can further get
\begin{equation}\label{Psi-change5-case21}
\hspace{-20pt}\frac{\rho\!+\!\mu}{4}
\|\!\!\left[\!\!\begin{array}{l}{\v^{k}\!-\!\v^{k+1}\!} \\ {\e_{1}^{k}\!-\!\e_{1}^{k+1}\!} \\ {\e_{2}^{k}\!-\!\e_{2}^{k+1}\!}\end{array}\!\!\right]\!\!\|_2^2
\!-\!\rho\omega\beta\|\mathcal{X}^{k}\|_2^2
\!\geq\!  \frac{\rho\!+\!\mu}{8}
\|\!\!\left[\!\!\begin{array}{l}{\v^{k}\!-\!\v^{k+1}\!} \\ {\e_{1}^{k}\!-\!\e_{1}^{k+1}\!} \\ {\e_{2}^{k}\!-\!\e_{2}^{k+1}\!}\end{array}\!\!\right]\!\!\|_2^2.
\end{equation}

  \item \eqref{supposed-inequality-2} does not hold, i.e.,
$
 \|\pmb{\phi}^k\|_2^2 > \frac{2 \rho \omega V^{2}}{\mu} \beta.
$
By exploiting the above inequality and $V\geq\|\mathcal{X}^{k}\|_2^2$,  we can get
\begin{equation}\label{Psi-change5-case22}
\begin{split}
&\mu\|\pmb{\phi}^k\|_2^2- \rho\omega\beta\|\mathcal{X}^{k}\|_2^2 \geq  \frac{\mu}{2}\|\pmb{\phi}^k\|_2^2.
\end{split}
\end{equation}
  \item \eqref{supposed-inequality-3} does not hold, i.e.,
$
\|\!\!\left[\!\!\begin{array}{c}{\p^{k}}\!-\!{\p^{k+1}} \\ {\z_{1}^{k}}\!-\!{\z_{1}^{k+1}} \\ {\z_{2}^{k}}\!-\!{\z_{2}^{k+1}}\end{array}\!\!\right]\!\!\|_{2}^{2}
 \!>\! 6\omega\beta^{2}\|\mathcal{X}^{k}\|_2^2.
$
Through similar derivations to \eqref{Psi-change5-case21} and \eqref{Psi-change5-case22}, we can obtain
\begin{equation}\label{Psi-change5-case23}
\hspace{-9pt}\frac{\rho}{3\beta}\|\!\!\left[\!\!\!\begin{array}{c}{\p^{k}}\!-\!{\p^{k+1}} \\ {\z_{1}^{k}}\!-\!{\z_{1}^{k+1}} \\ {\z_{2}^{k}}\!-\!{\z_{2}^{k+1}}\end{array}\!\!\!\right]\!\!\!\|_{2}^{2}
\!-\!\rho\omega\beta\|\mathcal{X}^{k}\|_2^2
\geq\frac{\rho}{6\beta}\|\!\!\left[\!\!\begin{array}{c}{\p^{k}}\!-\!{\p^{k+1}} \\ {\z_{1}^{k}}\!-\!{\z_{1}^{k+1}} \\ {\z_{2}^{k}}\!-\!{\z_{2}^{k+1}}\end{array}\!\!\right]\!\!\|_{2}^{2}.
\end{equation}
\end{enumerate}

Then, from \eqref{case1-inequality5}, \eqref{Psi-change5-case21}, \eqref{Psi-change5-case22}, and \eqref{Psi-change5-case23}, we can derive \eqref{Psi-change4} as
\begin{equation}\label{Psi-change5-case2}
\begin{split}
\Psi^{k}-\Psi^{k+1}
\geq  \frac{\rho\!+\!\mu}{8}
\|\!\!\left[\!\!\begin{array}{l}{\v^{k}\!-\!\v^{k+1}\!} \\ {\e_{1}^{k}\!-\!\e_{1}^{k+1}\!} \\ {\e_{2}^{k}\!-\!\e_{2}^{k+1}\!}\end{array}\!\!\!\right]\!\!\|_2^2
\!+\!\frac{\rho}{6 \beta}\!\left[\!\!\begin{array}{c}{\p^{k+1}\!-\!\p^{k}} \\ {\z_{1}^{k+1}\!-\!\z_{1}^{k}} \\ {\z_{2}^{k+1}\!-\!\z_{2}^{k}}\end{array}\!\!\right]\!\!\|_2^2
 \!+\! \frac{\mu}{2}\|\pmb{\phi}^k\|_2^{2}.
\end{split}
\end{equation}

Adding both sides of the above inequality from $k = 1, 2, \ldots,$ we can get
 {\setlength\abovedisplayskip{2pt}
 \setlength\belowdisplayskip{2pt}
  \setlength\jot{1pt}
\begin{equation}\label{Psi-sum}
\begin{split}
\underset{k\rightarrow+\infty}\lim\Psi^{1}\!\!-\!\!\Psi^{k+1}&
\!\geq\!\frac{\rho\!+\!\mu}{8}
\sum_{k=1}^{+\infty}\|\!\!\left[\!\!\begin{array}{l}{\v^{k}\!-\!\v^{k+1}\!} \\ {\e_{1}^{k}\!-\!\e_{1}^{k+1}\!} \\ {\e_{2}^{k}\!-\!\e_{2}^{k+1}\!}\end{array}\!\!\right]\!\!\|_2^2
\!+\!\frac{\rho}{6 \beta}\!\sum_{k=1}^{+\infty}\|\!\!\left[\!\!\begin{array}{c}{\p^{k+1}\!-\!\p^{k}} \\ {\z_{1}^{k+1}\!-\!\z_{1}^{k}} \\ {\z_{2}^{k+1}\!-\!\z_{2}^{k}}\end{array}\!\!\right]\!\!\|_2^2
 \!+\! \frac{\mu}{2}\sum_{k=1}^{+\infty}\|\pmb{\phi}^k\|_2^{2}.
\end{split}
\end{equation}
According} to \emph{Definition \ref{F-D-P-define}},
we can see that $\F^{k} \geq \D^{k}$, $\P^{k} \geq \D^{k}$ and $\P^{k}$ are lower-bounded.
Therefore, $\Psi^k=(\F^{k}-\D^{k})+(\P^{k}-\D^{k})+\P^{k}$ means that $\Psi$ is also lower-bounded. Therefore, we can obtain
 {\setlength\abovedisplayskip{1pt}
 \setlength\belowdisplayskip{1pt}
  \setlength\jot{1pt}
\begin{equation}\label{xhat-limit-divided}
\begin{split}
& \lim_{k \rightarrow +\infty}\v^{k}-\v^{k+1}=\mathbf{0}, ~~ \lim _{k \rightarrow +\infty}\e_1^{k}-\e_1^{k+1}=\mathbf{0}, \\
& \lim _{k \rightarrow +\infty}\e_2^{k}-\e_2^{k+1}=\mathbf{0}, ~~ \lim _{k \rightarrow +\infty}\p^{k}-\p^{k+1}=\mathbf{0}, \\
& \lim _{k \rightarrow +\infty}\z_1^{k}-\z_1^{k+1}=\mathbf{0}, ~~~ \lim _{k \rightarrow +\infty}\z_2^{k}-\z_2^{k+1}=\mathbf{0},
\end{split}
\end{equation}
and}
\begin{equation}\label{Wx-limit}
\begin{split}
& \lim _{k \rightarrow +\infty} \pmb{\phi}^k =\mathbf{0}.
\end{split}
\end{equation}
Plugging \eqref{xhat-limit-divided} into \eqref{proximal-z1-update}, we can obtain
 {\setlength\abovedisplayskip{1pt}
 \setlength\belowdisplayskip{1pt}
  \setlength\jot{1pt}
\begin{equation}\label{xvhat-limit-divided}
\begin{split}
& \lim_{k \rightarrow +\infty}\v^{k+1}-\p^{k}=\mathbf{0},
~~ \lim _{k \rightarrow +\infty}\e_1^{k+1}-\z_1^{k}=\mathbf{0},
~~ \lim _{k \rightarrow +\infty}\e_2^{k+1}-\z_2^{k}=\mathbf{0}.
\end{split}
\end{equation}
Plugging} \eqref{xhat-limit-divided} into \eqref{error-bound1} and \eqref{error-bound5} respectively, we have
\begin{equation}\label{limit-error-bound2}
\begin{split}
& \lim_{k \rightarrow +\infty}\left[\!\!\!\begin{array}{c}{\v^{k}\! -\!\v\left(\p^{k}, \z_{1}^{k}, \z_{2}^{k}, \y_{1}^{k}, \y_{2}^{k}\right)}
 \\ {\e_{1}^{k}\!-\!\e_{1}\left(\p^{k}, \z_{1}^{k}, \z_{2}^{k}, \y_{1}^{k}, \y_{2}^{k}\right)}
 \\ {\e_{2}^{k}\!-\!\e_{2}\left(\p^{k}, \z_{1}^{k}, \z_{2}^{k}, \y_{1}^{k}, \y_{2}^{k}\right)}\end{array}\!\!\!\right]=\mathbf{0},
\end{split}
\end{equation}
and
\begin{equation}\label{limit-error-bound5}
\begin{split}
\lim_{k\rightarrow+\infty} &\left[\!\!\begin{array}{l}{\v\left(\p^{k+1}\!, \z_{1}^{k+1}\!, \z_{2}^{k+1}\!, \y_{1}^{k+1}\!, \y_{2}^{k+1}\right)-\v\left(\u^{k}, \z_{1}^{k}, \z_{2}^{k}, \y_{1}^{k+1}, \y_{2}^{k+1}\right)}
\\ {\e_{1}\!\left(\p^{k+1}\!, \z_{1}^{k+1}\!, \z_{2}^{k+1}\!, \y_{1}^{k+1}\!, \y_{2}^{k+1}\right)-\e_{1}\!\left(\p^{k}, \z_{1}^{k}, \z_{2}^{k}, \y_{1}^{k+1}, \y_{2}^{k+1}\right)}
\\ {\e_{2}\!\left(\p^{k+1}\!, \z_{1}^{k+1}\!, \z_{2}^{k+1}\!, \y_{1}^{k+1}\!, \y_{2}^{k+1}\right)-\e_{2}\!\left(\p^{k}, \z_{1}^{k}, \z_{2}^{k}, \y_{1}^{k+1}, \y_{2}^{k+1}\right)}\end{array}\!\!\right] =\mathbf{0},
\end{split}
\end{equation}
respectively.
From \eqref{proximal-y2-update}, we have
\begin{equation}\label{Wx-limit2}
\begin{split}
\left[\begin{array}{ccc} \mathbf{y}_1^{k+1}\!-\!\mathbf{y}_1^k\\ \mathbf{y}_2^{k+1}\!-\!\mathbf{y}_2^k \end{array}\right]\!=\!\left[\!\!\begin{array}{ccc}{\A} & {\I_{M}} & {\mathbf{0}} \\ {\I_{N}} & {\mathbf{0}} & {-\I_{N}}\end{array}\!\!\right]\!\!
\left[\!\!\begin{array}{c}{\v^{k+1}} \\ {\e_{1}^{k+1}}\\ {\e_{2}^{k+1}}\end{array}\!\!\right]\!-\!\left[\!\!\begin{array}{l}{{\boldsymbol\varrho}} \\ {\mathbf{0}}\end{array}\!\!\right].
\end{split}
\end{equation}
Plugging \eqref{D-part1-change} into \eqref{Wx-limit2}, we have
\begin{equation}\label{Wx-limit3}
\begin{split}
 \left[\!\!\begin{array}{ccc} \mathbf{y}_1^{k+1}\!-\!\mathbf{y}_1^k\\ \mathbf{y}_2^{k+1}\!-\!\mathbf{y}_2^k \end{array}\!\!\right]
&= \!\!\left[\!\!\begin{array}{ccc}{\A} \!&\! {\I_{M}} \!\!&\!\! {\mathbf{0}} \\ \!{\I_{N}}\!\!&\!\!{\mathbf{0}} \!\!&\!\! {-\!\I_{N}}\end{array}\!\!\!\!\right]\!\!
 \left[\!\!\begin{array}{c}{\v^{k+1}\!\!-\!\!\v\!\left(\p^{k+1}, \z_{1}^{k+1}, \z_{2}^{k+1}, \y_{1}^{k+1}, \y_{2}^{k+1}\right)}
 \\ {\e_{1}^{k+1}\!\!-\!\e_{1}\!\left(\p^{k+1}, \z_{1}^{k+1}, \z_{2}^{k+1}, \y_{1}^{k+1}, \y_{2}^{k+\!1}\right)}
 \\ {\e_{2}^{k+1}\!\!-\!\e_{2}\!\left(\p^{k+1}, \z_{1}^{k+1}, \z_{2}^{k+1}, \y_{1}^{k+1}, \y_{2}^{k+1}\right)}\end{array}\!\!\!\!\right] \\
&+ \!\!\left[\!\!\!\begin{array}{ccc}{\A} \!&\! {\I_{M}} \!\!&\!\! {\mathbf{0}} \\ {\I_{N}} \!\!&\!\! {\mathbf{0}} \!\!&\!\! {-\I_{N}}\end{array}\!\!\!\right]\!\!\left[\!\!\begin{array}{l}{\v\!\left(\p^{k+1}\!, \z_{1}^{k+1}\!, \z_{2}^{k+1}\!, \y_{1}^{k+1}\!, \y_{2}^{k+1}\right)\!-\v\!\left(\p^{k}, \z_{1}^{k}, \z_{2}^{k}, \y_{1}^{k+1}, \y_{2}^{k+1}\right)}
\\ {\e_{1}\!\left(\p^{k+1}\!, \z_{1}^{k+1}\!, \z_{2}^{k+1}\!, \y_{1}^{k+1}\!, \y_{2}^{k+1}\right)\!-\e_{1}\!\left(\p^{k}, \z_{1}^{k}, \z_{2}^{k}, \y_{1}^{k+1}, \y_{2}^{k+1}\right)}
\\ {\e_{2}\!\left(\p^{k+1}\!, \z_{1}^{k+1}\!, \z_{2}^{k+1}\!, \y_{1}^{k+1}\!, \y_{2}^{k+1}\right)\!-\e_{2}\!\left(\p^{k}, \z_{1}^{k}, \z_{2}^{k}, \y_{1}^{k+1}, \y_{2}^{k+1}\right)}\end{array}\right] \!\! +\!\pmb{\phi}^k.
\end{split}
\end{equation}
Plugging \eqref{Wx-limit}, \eqref{limit-error-bound2}, and \eqref{limit-error-bound5} into \eqref{Wx-limit3}, we can obtain
\begin{equation}\label{dual-limit}
\lim_{k \rightarrow +\infty} \mathbf{y}_1^{k+1}\!-\!\mathbf{y}_1^k=\mathbf{0}, ~~ \lim_{k \rightarrow +\infty} \mathbf{y}_2^{k+1}\!-\!\mathbf{y}_2^k=\mathbf{0}.
\end{equation}

From \eqref{pADMM-frame-b} and \eqref{pADMM-frame-c}, we can see clearly that $\{\e_1^k\}$ and $\{\v^k\}$ are bounded sequences since $\{\e_2^k\}$ is bounded by $[0,1]^N$. Plugging these bounded results into \eqref{xvhat-limit-divided}, we can find that $\{\p^{k}\}$, $\{\z_1^{k}\}$, $\{\z_2^{k}\}$ are also bounded sequences.
Furthermore, based on the above bounded results, \eqref{v1-solution-component} and \eqref{v2-solution-component} imply that $\{\y_1^{k}\}$ and $\{\y_2^{k}\}$ are also bounded sequences.

Combining these bounded results with \eqref{xhat-limit-divided} and \eqref{dual-limit}, we can conclude
\begin{equation}\label{variables-limit}
\begin{split}
& \lim_{k \rightarrow +\infty}\v^{k}=\v^{*}, ~ \lim _{k \rightarrow +\infty}\e_1^{k}=\e_1^{*},~\lim _{k \rightarrow +\infty}\e_2^{k}=\e_2^{*},\\
& \lim _{k \rightarrow +\infty}\p^{k}=\p^{*}, ~ \lim _{k \rightarrow +\infty}\z_1^{k}=\z_1^{*}, ~\lim _{k \rightarrow +\infty}\z_2^{k}=\z_2^{*}, \\
&\lim_{k \rightarrow +\infty}\y_1^{k}=\y_1^{*}, ~\lim _{k \rightarrow +\infty}\y_2^{k}=\y_2^{*}.
\end{split}
\end{equation}

Plugging \eqref{variables-limit} into \eqref{pADMM-frame-b} and \eqref{xvhat-limit-divided}, we can get
\begin{equation}\label{Wx-limit-star}
\begin{split}
& \A\v^{*}+\e_1^{*}-{\boldsymbol\varrho} = \mathbf{0}, ~~ \v^{*}=\e_2^{*},
~~ \v^{*}=\p^{*}, ~~ \e_1^{*}=\z_1^{*}, ~~ \e_2^{*}=\z_2^{*}.
\end{split}
\end{equation}
which completes the proof of the first part of {\it Theorem 1}.

Next, we prove that $\v^*$ is a stationary point of the original problem \eqref{ML-decoding-all}. Letting $g(\v)=\pmb{\lambda}^T\v-\frac{\alpha}{2}\|\v-0.5\|_2^2$, we can obtain, $\forall \x \in X$,
\begin{equation}\label{stationary1}
\begin{split}
(\v-\v^{*})^T\nabla_{\v} g(\v^*) &= (\v-\v^{*})^T(\pmb{\lambda}-\alpha(\v-0.5)). \\
\end{split}
\end{equation}
Moreover, since $\v^{k+1}\!=\!\mathop{\rm argmin}\limits_{\v}  \F\left(\v, \e_{1}^{k},\e_{2}^{k},\p^{k},\z_{1}^{k},\z_{2}^{k},\y_{1}^{k},\y_{2}^{k}\right)$, then we have
\begin{equation}\label{grandient-x-star}
\begin{split}
0= &\nabla_{\v}\F(\v^{*},\e_{1}^{*},\e_2^{*},\p^*,\z_1^*,\z_2^*,\y_1^*,\y_2^*) \\
=&\pmb{\lambda}-\alpha(\v^*-0.5)+\rho(\v^*-\p^*)+\A^{T}\y_1^{*}+\y_2^*
 +\mu\A^T(\A\v^*+\e_1^*-{\boldsymbol\varrho})+\mu(\v^*-\e_2^*)\\
=&\nabla_{\v} g(\v^*)+\A^{T}\y_1^{*}+\y_2^*,
\end{split}
\end{equation}
i.e., $\nabla_{\v} g(\v^*)=-\A^{T}\y_1^{*}-\y_2^*$, where the last equality follows from \eqref{Wx-limit-star}. Then, we can further obtain
\begin{equation}\label{stationary2}
\begin{split}
\hspace{-0.4cm}(\v\!-\!\v^{*})^T\nabla_{\v} g(\v^*) & \!=\! \!-\!(\v-\v^{*})^T\!(\A^{T}\y_1^{*}\!+\!\y_2^*) \\
\hspace{-0.4cm} &\! =\! \!-\!(\y_1^{*})^T\!\A(\v\!-\!\v^{*})\!\!-\!\!(\y_2^*)^{T}\!(\v\!-\!\v^{*}).
\end{split}
\end{equation}
Obviously, if $(\y_1^{*})^T\A(\v-\v^{*})\leq 0$ and $(\y_2^{*})^{T}(\v-\v^{*})\leq 0$, then $(\v\!-\!\v^{*})^T\nabla_{\v} g(\v^*) \geq 0$.
In the following, we will prove that both of them hold.

First, we have
\begin{equation}\label{first-term1}
\begin{split}
(\y_1^{*})^T\A(\v-\v^{*}) & = (\y_1^{*})^T\left(({\boldsymbol\varrho}-\e_1)-({\boldsymbol\varrho}-\e_1^*)\right) \\
& = (\y_1^{*})^T(\e_1^*-\e_1).
\end{split}
\end{equation}

Then, since $\e_1^* \succeq 0$, we have the following derivations
\begin{equation}\label{y-z-star-1}
\begin{split}
 (\y_1^{*})^T\e_1^* & \!=\! \sum_{e_{1,j}^{*}>0} y_{1,j}^*e_{1,j}^{*} \\
& \!=\! \sum_{e_{1,j}^{*}>0} \! \frac{\mu}{\rho+\mu} y_{1,j}^*({\varrho}_{j}\!-\!\mathbf{a}_{j}^{T}\v^{*}\!-\!\frac{y_{1,j}^*}{\mu}\!+\!\frac{\rho}{\mu}z_{1,j}^{*}) \\
&\! =\! \sum_{e_{1,j}^{*}>0}y_{1,j}^*\big({\varrho}_{j}-\mathbf{a}_{j}^{T}\v^{*}-\frac{y_{1,j}^*}{\rho+\mu}\big),
\end{split}
\end{equation}
where the last equality holds since $\z_1^*={\boldsymbol\varrho}-\A\v^{*}$, which follows from $\e_1^*=\z_1^*$ and $\A\v^{*}+\e_1^{*}-{\boldsymbol\varrho}=0$.
Moreover, since $\A\v^{*}+\e_1^{*}-{\boldsymbol\varrho}=0$, we also have
\begin{equation}\label{y-z-star-2}
\begin{split}
 (\y_1^{*})^T\e_1^* & = (\y_1^{*})^T({\boldsymbol\varrho}-\A\v^{*}) \\
& = \sum_{e_{1,j}^{*}>0}\!\! y_{1,j}^*({\varrho}_{j}\!-\!\mathbf{a}_{j}^{T}\x^{*})\!+\!\!\sum_{e_{1,j}^{*}=0}\! y_{1,j}^*({\varrho}_{j}\!-\!\mathbf{a}_{j}^{T}\x^{*}) \\
& = \sum_{e_{1,j}^{*}>0} y_{1,j}^*({\varrho}_{j}-\mathbf{a}_{j}^{T}\v^{*}).
\end{split}
\end{equation}

Comparing \eqref{y-z-star-1} and \eqref{y-z-star-2}, we can see that when $e_{1,j}^{*}>0$,
\begin{equation}\label{z-star-dayu0-1}
\begin{split}
& y_{1,j}^* =0.
\end{split}
\end{equation}
Therefore, we can obtain
\begin{equation}\label{y-z-star-0}
(\y_1^{*})^T\e_1^* =0,
\end{equation}
since $\e_1^*\succeq\mathbf{0}$.

On the other hand, if $e_{1,j}^{*}=0$, there exists $\frac{\mu}{\rho+\mu}({\varrho}_{j}-\mathbf{a}_{j}^{T}\v^{*}-\frac{y_{1,j}^*}{\mu}+\frac{\rho}{\mu}z_{1,j}^{*})\leq0$ (see \eqref{v1-solution-component}).
Moreover, since $\e_1^*=\z_1^*={\boldsymbol\varrho}-\A\v^{*}$, one can see that $\frac{y_{1,j}^*}{\mu} \geq (1+\frac{\rho}{\mu})e_{1,j}^{*}\geq0$.
Besides, since $\rho >0$ and $\mu>0$, thus we have
\begin{equation}\label{z-star-dayu0-2}
\begin{split}
& y_{1,j}^* \geq 0,
\end{split}
\end{equation}
when $e_{1,j}^{*}=0$.
From \eqref{z-star-dayu0-1} and \eqref{z-star-dayu0-2}, we conclude
\begin{equation}\label{y-star-dayu0}
\y_1^{*} \succeq 0.
\end{equation}
Plugging \eqref{y-z-star-0} and \eqref{y-star-dayu0} into \eqref{first-term1},
we obtain $(\y_1^{*})^T(\e_1^*-\e_1)=-(\y_1^{*})^T \e_1 \leq 0$,
which means
\begin{equation}\label{first-term-proof}
(\y_1^{*})^T\A(\v-\v^{*}) \leq 0.
\end{equation}

Similar to the above derivations for \eqref{first-term-proof}, we can also have
\begin{equation}\label{second-term-proof}
(\y_2^*)^{T}(\v-\v^{*})\leq 0.
\end{equation}

Therefore, we can conclude
\begin{equation}\label{stationary3}
(\v-\v^{*})^T\nabla_{\v} g(\v^*) \geq 0, \ \forall \v \in \mathcal{X}.
\end{equation}
This completes the proof.
}

\section{Proof of Lemma \ref{x-Lipschitz-constant}}\label{Lipschitz-continuous}
{\it Proof:} {Based on the definition of $\mathcal{L}_{\mu}$ in \eqref{aug-Lagrangian}, its gradient, with respect to variable $\v$, can be calculated by
\begin{equation*}
\begin{split}
 \nabla_{\v} \mathcal{L}_{\mu}(\v,\e_{1}, \e_{2},&\y_{1}, \y_{2})\! =\! (\pmb{\lambda}-\alpha(\v-0.5)) +\A^{T}\y_1 +\y_2\\
&+\mu\A^T(\A\v+\e_1-{\boldsymbol\varrho})+\mu(\v-\e_2).
\end{split}
\end{equation*}
Then, we have the following derivations
 {\setlength\abovedisplayskip{2pt}
 \setlength\belowdisplayskip{2pt}
  \setlength\jot{1pt}
\begin{equation*}
\begin{split}
& \|\nabla_{\v} \mathcal{L}_{\mu}(\v, \e_{1},  \e_{2}, \y_{1},  \y_{2})\!-\!\nabla_{\v} \mathcal{L}_{\mu}(\v^{\prime},\e_{1},\e_{2},\y_{1},\y_{2})\|_{2}\\
=&\|-\alpha(\v-\v^{\prime})+\mu(\v-\v^{\prime})+\mu\A^T\A(\v-\v^{\prime})\|_2\\
\leq & (\alpha+\mu+\mu\delta_{\A}^2)\|\v-\v^{\prime}\|_2, \\
\end{split}
\end{equation*}
where} ``$\delta_{\A}$'' is the spectral norm of matrix $\A$. Letting $L=\alpha+\mu+\mu\delta_{\A}^2$, we obtain \eqref{x-Lipschitz-defi}.}

\section{Proof of \eqref{case1-inequality3}}\label{case1-inequality3-proof}
We prove by contradiction. Suppose \eqref{case1-inequality3} does not hold. Since $\underset{\epsilon\rightarrow0}\lim \sigma(\epsilon)=0$, and suppose there exists a sequence of tuples
\begin{equation}\label{con_hat}
\begin{split}
\underset{k\rightarrow+\infty}\lim\{\p^k,\v_1^k,\v_2^k,\y_1^{k+1},\y_2^{k+1}\}= \{\hat{\p},\hat{\v}_1,\hat{\v}_2,\hat{\y}_1,\hat{\y}_2\},
\end{split}
\end{equation}
such that
\begin{equation}\label{contrary1}
\begin{split}
\underset{k\rightarrow+\infty}\lim\|\pmb{\phi}^k\|_2 = 0.
\end{split}
\end{equation}
\begin{equation}\label{contrary2}
\begin{split}
\underset{k\rightarrow+\infty}\lim\|\mathcal{X}^k\|_2 > 0.
\end{split}
\end{equation}
Next we prove \eqref{contrary1} and \eqref{contrary2} cannot hold simultaneously.

First, plugging the convergence result \eqref{con_hat} into \eqref{contrary1} and noticing
$\left[\!\!\begin{array}{l}{\v(\p^{k},\z_{1}^{k},\z_{2}^{k},\y_{1}^{k+1}, \y_{2}^{k+1})}
\\ {\e_{1}(\p^{k}, \z_{1}^{k}, \z_{2}^{k}, \y_{1}^{k+1}, \y_{2}^{k+1})}
\\ {\e_{2}(\p^{k}, \z_{1}^{k}, \z_{2}^{k}, \y_{1}^{k+1}, \y_{2}^{k+1}\!)}\end{array}\!\!\right] $ is continuous (see \eqref{x-D-define} and $\pmb\phi^k$ in \eqref{D-part1-change}), we can see
\begin{equation}\label{phi-0}
\begin{split}
\hspace{-10pt}\left[\!\!\begin{array}{c}{\A \v\left(\hat{\p},\hat{\z}_1,\hat{\z}_2,\hat{\y}_1,\hat{\y}_2\right)\!+\!\e_{1}\left(\hat{\p},\hat{\z}_1,\hat{\z}_2,\hat{\y}_1,\hat{\y}_2\right)\!-\!{\boldsymbol\varrho}}
\\ {\v\left(\hat{\p},\hat{\z}_1,\hat{\z}_2,\hat{\y}_1,\hat{\y}_2\right)\!-\!\e_{2}\left(\hat{\p},\hat{\z}_1,\hat{\z}_2,\hat{\y}_1,\hat{\y}_2\right)}\end{array}\!\!\right]\!=\!0.
\end{split}
\end{equation}
Plugging \eqref{phi-0} into KKT equations of problem \eqref{x-D-define} and notice $\hat{\y}_1$ and $\hat{\y}_2$ are the corresponding optimal Lagrangian multipliers (problem \eqref{x-D-define} is strongly convex), we can see
$$\left[\!\!\begin{array}{l}{\v(\hat{\p},\hat{\z}_{1},\hat{\z}_{2},\hat{\y}_{1}, \hat{\y}_{2})}
\\ {\e_{1}(\hat{\p},\hat{\z}_{1},\hat{\z}_{2},\hat{\y}_{1}, \hat{\y}_{2})}
\\ {\e_{2}(\hat{\p},\hat{\z}_{1},\hat{\z}_{2},\hat{\y}_{1}, \hat{\y}_{2})}\end{array}\!\!\!\!\right]
= \left[\!\!\begin{array}{l}{\v(\hat{\p},\hat{\z}_{1},\hat{\z}_{2})}
\\ {\e_{1}(\hat{\p},\hat{\z}_{1},\hat{\z}_{2})}
\\ {\e_{2}(\hat{\p},\hat{\z}_{1},\hat{\z}_{2})}\end{array}\!\!\right],$$
which indicates $\underset{k\rightarrow+\infty}\lim\|\mathcal{X}^k\|_2 = 0$. This is a contradiction.

\section{Proof of \emph{Lemma \ref{error-bound-lemma}}}\label{error-bound-proof}
{\it Proof:} {
First, we prove \eqref{error-bound1}. To simplify the proof, $\F(\v,\e_{1},\e_{2},\bullet^{k})$ is used
to denote function $\F(\v,\e_{1},\e_{2},\p^k,\z_1^k,\z_2^k,\y_1^k,\y_2^k)$.
Moreover, we give the following definitions relative to $\v$, $\e_{1}$ and $\e_{2}$ respectively
\begin{equation}\label{error3-defi}
\begin{split}
&\mathbf{r}_{\v}^{k}\!= \! \nabla_{\v}\F(\v^{k},\e_{1}^{k},\e_{2}^{k},\bullet^{k})
     \!-\!\nabla_{\v}\F(\v^{k+1}\!,\e_{1}^{k},\e_{2}^{k},\bullet^{k})\!+\!\v^{k+1}\!-\!\v^{k},\\
&\mathbf{r}_{\e_{1}}^{k}\!=\!\nabla_{\e_1}\F(\v^{k},\e_{1}^{k},\e_{2}^{k},\!\bullet^{k})
     \!-\!\nabla_{\e_1}\F(\v^{k+1},\e_{1}^{k+1},\e_{2}^{k},\bullet^{k})\!+\!\e_{1}^{k+1}\!-\!\e_{1}^{k},\\
&\mathbf{r}_{\mathbf{e}_{2}}^{k}\!=\!\nabla_{\e_{2}}\F(\v^{k},\e_{1}^{k},\e_{2}^{k},\bullet^{k})
     \!-\!\nabla_{\mathbf{e}_{2}}\F(\v^{k+1},\e_{1}^{k+1},\e_{2}^{k+1},\bullet^{k})\!+\!\e_{2}^{k+1}\!-\!\e_{2}^{k}.
\end{split}
\end{equation}

Applying the triangle inequality to $\mathbf{r}_{\v}^{k}$, we obtain
\begin{equation}\label{error1-derivative}
\begin{split}
 \|\mathbf{r}_{\v}^{k}\|_{2}
& \!\leq\! \|\nabla_{\v}\F(\v^{k},\e_{1}^{k},\e_{2}^{k},\bullet^{k})\!-\!\nabla_{\v}\F(\v^{k+1},\e_{1}^{k},\e_{2}^{k},\bullet^{k})\|_{2} +\|\v^{k+1}-\v^{k}\|_{2},\\
&  \!\leq\! (\rho+L+1)\|\v^{k+1}-\v^{k}\|_2,
\end{split}
\end{equation}
where $L$ is the Lipschitz constant defined in \eqref{x-Lipschitz-defi}.
Following similar derivations to \eqref{error1-derivative}, we can get
\begin{equation}\label{error2-derivative}
\begin{split}
\|\mathbf{r}_{\e_{1}}^{k}\!\|_{2} & \!\leq\! \|\nabla_{\e_{1}}\F(\v^{k}\!,\e_{1}^{k},\e_{2}^{k},\bullet^{k}\!)\!-\!\nabla_{\e_{1}}\F(\v^{k+1},\e_{1}^{k+1},\e_{2}^{k},\bullet^{k})\|_{2} \!+\!\|\e_{1}^{k+1}\!-\!\e_{1}^{k}\|_{2},\\
& \!=\! (\rho+\mu+1)\|\e_{1}^{k+1}-\e_{1}^{k}\|_{2},
\end{split}
\end{equation}
\begin{equation}\label{error3-derivative}
\begin{split}
\|\mathbf{r}_{\e_{2}}^{k}\|_{2} & \!\leq\! \|\nabla_{\e_{2}}\F(\v^{k},\e_{1}^{k},\e_{2}^{k},\bullet^{k})
     \!-\!\nabla_{\e_{2}}\F(\v^{k+1},\e_{1}^{k+1},\e_{2}^{k+1},\bullet^{k})\|_{2}
+\|\e_{2}^{k+1}-\e_{2}^{k}\|_{2},\\
& = (\rho+\mu+1)\|\e_{2}^{k+1}-\e_{2}^{k}\|_{2},
\end{split}
\end{equation}
Then, through \eqref{error1-derivative}-\eqref{error3-derivative}, we can obtain
 {\setlength\abovedisplayskip{2pt}
 \setlength\belowdisplayskip{2pt}
  \setlength\jot{1pt}
\begin{equation}\label{error4-derivative1}
\|\!\!\left[\!\!\begin{array}{c} {\mathbf{r}_{\v}^k} \\ {\mathbf{r}_{\e_{1}}^k} \\ {\mathbf{r}_{\e_{2}}^k}\end{array}\!\!\right]\!\!\|_2^2 \\
\leq  (\rho\!+\!L\!+\!1)^2\|\!\!\left[\!\!\!\begin{array}{c}{\v^{k+1}\!-\!\v^{k}} \\ {\e_{1}^{k+1}\!-\!\e_{1}^{k}} \\ {\e_{2}^{k+1}\!-\!\e_{2}^{k}}\end{array}\!\!\right]\!\!\|_{2}.
\end{equation}}

Moreover, since $\v^{k+1}$, $\e_{1}^{k+1}$, and $\e_{2}^{k+1}$ are minimizers of convex quadratic problems \eqref{proximal-x-update}, \eqref{proximal-v1-update}, and \eqref{proximal-v2-update} respectively, according to the fixed point theorem, we have
 {\setlength\abovedisplayskip{2pt}
 \setlength\belowdisplayskip{2pt}
  \setlength\jot{3pt}
\begin{equation}\label{x-v-other-expression}
\begin{split}
& \v^{k+1} = [\v^{k+1}-\nabla_{\v}\F(\v^{k+1},\e_{1}^{k},\e_{2}^{k},\bullet^{k})]_{+},\\
&\e_{1}^{k+1} = [\e_{1}^{k+1}-\nabla_{\e_{1}}\F(\v^{k+1},\e_{1}^{k+1},\e_{2}^{k},\bullet^{k})]_{+},\\
&\e_{2}^{k+1} = [\e_{2}^{k+1}-\nabla_{\e_{2}}\F(\v^{k+1},\e_{1}^{k+1},\e_{2}^{k+1},\bullet^{k})]_{+}.
\end{split}
\end{equation}
Plugging} \eqref{x-v-other-expression} into \eqref{error3-defi}, we have
\begin{equation}\label{hatx-other-expression}
\begin{split}
\left[\!\!\begin{array}{c}{\v^{k+1}} \\ {\e_{1}^{k+1}} \\ {\e_{2}^{k+1}}\end{array}\!\!\!\right]
\!\!=\!\!\left[\!\!\!\begin{array}{c}{\v^{k}-\nabla_{\v}\F(\v^{k},\e_{1}^{k},\e_{2}^{k},\bullet^{k})+\mathbf{r}_{\v}^k}
\\ {\e_{1}^{k}-\nabla_{\e_{1}}\F(\v^{k},\e_{1}^{k},\e_{2}^{k},\bullet^{k})+\mathbf{r}_{\e_{1}}^k}
\\ {\e_{2}^{k}-\nabla_{\e_{2}}\F(\v^{k},\e_{1}^{k},\e_{2}^{k},\bullet^{k})+\mathbf{r}_{\e_{2}}^k}\end{array}\!\!\!\right]_{+}.
\end{split}
\end{equation}
Then, we have the following derivations
 {\setlength\abovedisplayskip{1pt}
 \setlength\belowdisplayskip{1pt}
  \setlength\jot{1pt}
\begin{equation}\label{errorsum-derivative-2}
\begin{split}
 \|\!\!\left[\!\!\begin{array}{c}{\v^{k}}-\v^{k+1} \\ {\e_{1}^{k}}-{\e_{1}^{k+1}} \\ {\e_{2}^{k}}-{\e_{2}^{k+1}}\end{array}\!\!\right]\!\|_{2}
= & \|\!\!\left[\!\!\begin{array}{c}{\v^{k}} \\ {\e_{1}^{k}} \\ {\e_{2}^{k}}\end{array}\!\!\right]
\!\!-\!\!\left[\!\!\!\begin{array}{c}{\v^{k}\!-\!\nabla_{\v}\F(\v^{k},\e_{1}^{k},\e_{2}^{k},\bullet^{k})+\mathbf{r}_{\v}^k}
\\ {\e_{1}^{k}\!-\!\nabla_{\e_{1}}\F(\v^{k},\e_{1}^{k},\e_{2}^{k},\bullet^{k})+\mathbf{r}_{\e_{1}}^k}
\\ {\e_{2}^{k}\!-\!\nabla_{\e_{2}}\F(\v^{k},\e_{1}^{k},\e_{2}^{k},\bullet^{k})+\mathbf{r}_{\e_{2}}^k}\end{array}\!\!\!\right]_{+}\!\!\|_{2}\\
\geq & \|\!\!\left[\!\!\begin{array}{c}{\v^{k}} \\ {\e_{1}^{k}} \\ {\e_{2}^{k}}\end{array}\!\!\right]
-\!\left[\!\!\!\begin{array}{c}{\v^{k}-\nabla_{\v}\F(\v^{k},\e_{1}^{k},\e_{2}^{k},\bullet^{k})}
\\ {\e_{1}^{k}-\nabla_{\e_{1}}\F(\v^{k},\e_{1}^{k},\e_{2}^{k},\bullet^{k})}
\\ {\e_{2}^{k}-\nabla_{\e_{2}}\F(\v^{k},\e_{1}^{k},\e_{2}^{k},\bullet^{k})}\end{array}\!\!\!\right]_{+}\!\!\|_{2} \\
& - \|\!\left[\!\!\!\begin{array}{c}{\v^{k}-\nabla_{\v}\F(\v^{k},\e_{1}^{k},\e_{2}^{k},\bullet^{k})+\mathbf{r}_{\v}^k}
\\ {\e_{1}^{k}-\nabla_{\e_{1}}\F(\v^{k},\e_{1}^{k},\e_{2}^{k},\bullet^{k})+\mathbf{r}_{\e_{1}}^k}
\\ {\e_{2}^{k}-\nabla_{\e_{2}}\F(\v^{k},\e_{1}^{k},\e_{2}^{k},\bullet^{k})+\mathbf{r}_{\e_{2}}^k}\end{array}\!\!\!\right]_{+} \!\!-\!\left[\!\!\!\begin{array}{c}{\v^{k}-\nabla_{\v}\F(\v^{k},\e_{1}^{k},\e_{2}^{k},\bullet^{k})}
\\ {\e_{1}^{k}-\nabla_{\e_{1}}\F(\v^{k},\e_{1}^{k},\e_{2}^{k},\bullet^{k})}
\\ {\e_{2}^{k}-\nabla_{\e_{2}}\F(\v^{k},\e_{1}^{k},\e_{2}^{k},\bullet^{k})}\end{array}\!\!\!\right]_{+}\!\!\|_{2},
\end{split}
\end{equation}
where} the inequality comes from the triangle inequality. Following the non-expansiveness property of projection operations, \eqref{errorsum-derivative-2} can be deduced to
 {\setlength\abovedisplayskip{1pt}
 \setlength\belowdisplayskip{1pt}
  \setlength\jot{1pt}
\begin{equation}\label{errorsum-derivative-3}
\begin{split}
\|\!\!\left[\!\!\begin{array}{c}{\v^{k}}-\v^{k+1} \\ {\e_{1}^{k}}-{\e_{1}^{k+1}} \\ {\e_{2}^{k}}-{\e_{2}^{k+1}}\end{array}\!\!\right]\!\|_{2}
\geq \|\!\!\left[\!\!\!\begin{array}{c}{\v^{k}} \\ {\e_{1}^{k}} \\ {\e_{2}^{k}}\end{array}\!\!\right]
\!\!\!-\!\!\!\left[\!\!\begin{array}{c}{\v^{k}\!\!-\!\!\nabla_{\v}\F(\v^{k},\!\e_{1}^{k},\!\e_{2}^{k},\bullet^{k})}
\\ {\e_{1}^{k}\!\!-\!\!\nabla_{\e_{1}}\!\F(\v^{k},\!\e_{1}^{k},\!\e_{2}^{k},\bullet^{k})}
\\ {\e_{2}^{k}\!\!-\!\!\nabla_{\e_{2}}\!\F(\v^{k},\!\e_{1}^{k},\!\e_{2}^{k},\bullet^{k})}\end{array}\!\!\right]_{+}\!\!\|_{2}
\!-\!\|\!\!\left[\!\!\begin{array}{c} {\mathbf{r}_{\v}^k} \\ {\mathbf{r}_{\e_{1}}^k} \\ {\mathbf{r}_{\e_{2}}^k}\end{array}\!\!\right]\!\!\|_2.
\end{split}
\end{equation}
Moreover,} \eqref{errorsum-derivative-3} can be further derived as
 {\setlength\abovedisplayskip{1pt}
 \setlength\belowdisplayskip{1pt}
  \setlength\jot{1pt}
\begin{equation}\label{errorsum-derivative-4}
\begin{split}
 \|\!\!\left[\!\!\begin{array}{c}{\v^{k}}-\v^{k+1} \\ {\e_{1}^{k}}-{\e_{1}^{k+1}} \\ {\e_{2}^{k}}-{\e_{2}^{k+1}}\end{array}\!\!\right]\!\|_{2}
\overset{a} \geq & (\rho-\alpha) \|\!\!\left[\!\!\begin{array}{c}{\v^{k}}-{\v(\p^{k},\z_1^{k},\z_2^{k},\y_1^k,\y_2^k)} \\ {\e_{1}^{k}}-{\e_{1}(\p^{k},\z_1^{k},\z_2^{k},\y_1^k,\y_2^k)} \\ {\e_{2}^{k}}-{\e_{2}(\p^{k},\z_1^{k},\z_2^{k},\y_1^k,\y_2^k)}\end{array}\!\!\right]
\!\|_{2}
-\|\!\!\left[\!\!\begin{array}{c} {\mathbf{r}_{\v}^k} \\ {\mathbf{r}_{\e_{1}}^k} \\ {\mathbf{r}_{\e_{2}}^k}\end{array}\!\!\right]\!\!\|_2 \\
\overset{b} \geq & (\rho\!\!-\!\!\alpha) \|\!\!\left[\!\!\begin{array}{c}{\v^{k}}\!-\!{\v(\p^{k},\z_1^{k},\z_2^{k},\y_1^k,\y_2^k)} \\ {\e_{1}^{k}}\!-\!{\e_{1}(\p^{k},\z_1^{k},\z_2^{k},\y_1^k,\y_2^k)} \\ {\e_{2}^{k}}\!-\!{\e_{2}(\p^{k},\z_1^{k},\z_2^{k},\y_1^k,\y_2^k)}\end{array}\!\!\!\right]
 \!\!\|_{2}\!-\!(\rho\!+\!L\!+\!1)\|\!\!\left[\!\!\begin{array}{c}{\v^{k+1}\!-\!\v^{k}} \\ {\e_{1}^{k+1}\!-\!\e_{1}^{k}} \\ {\e_{2}^{k+1}\!-\!\e_{2}^{k}}\end{array}\!\!\right]\!\!\|_{2},
\end{split}
\end{equation}
where} ``$\overset{a} \geq$'' holds since $\mathcal{F}(\mathbf{v},\mathbf{e}_1,\mathbf{e}_2, \bullet^k)$ is a strongly convex function with modulus $(\rho-\alpha)>0$ \cite{global-error-bound} and ``$\overset{b} \geq $'' follows from \eqref{error4-derivative1}.
Then, we can obtain \eqref{error-bound1} as follows
\begin{equation*}\label{errorsum-derivative-5}
\begin{split}
& \|\!\!\left[\!\!\begin{array}{l}{\v^{k}\!-\!\v^{k+1}} \\ {\e_{1}^{k}\!-\!\e_{1}^{k+1}} \\ {\e_{2}^{k}\!-\!\e_{2}^{k+1}}\end{array}\!\!\right]\!\!\|_{2}^{2}
\!\geq\! \varepsilon_{1}\|\!\!\left[\!\!\!\begin{array}{c}{\v^{k}\! -\!\v\left(\p^{k}, \z_{1}^{k}, \z_{2}^{k}, \y_{1}^{k}, \y_{2}^{k}\right)}
 \\ {\e_{1}^{k}\!-\!\e_{1}\left(\p^{k}, \z_{1}^{k}, \z_{2}^{k}, \y_{1}^{k}, \y_{2}^{k}\right)}
 \\ {\e_{2}^{k}\!-\!\e_{2}\left(\p^{k}, \z_{1}^{k}, \z_{2}^{k}, \y_{1}^{k}, \y_{2}^{k}\right)}\end{array}\!\!\right]\!\!\|_{2}^{2},
\end{split}
\end{equation*}
where $\varepsilon_1=\frac{(\rho-\alpha)^2}{(\rho+L+2)^2}$.

Next, we prove that \eqref{error-bound2} holds.
Based on the triangle inequality and \eqref{error-bound1}, we have
\begin{equation}\label{error-bound2-proof}
\begin{split}
 \|\!\!\left[\!\!\begin{array}{c}{\v^{k+1}\! -\!\v\left(\p^{k}, \z_{1}^{k}, \z_{2}^{k}, \y_{1}^{k}, \y_{2}^{k}\right)}
 \\ {\e_{1}^{k+1}\!-\!\e_{1}\left(\p^{k}, \z_{1}^{k}, \z_{2}^{k}, \y_{1}^{k}, \y_{2}^{k}\right)}
 \\ {\e_{2}^{k+1}\!-\!\e_{2}\left(\p^{k}, \z_{1}^{k}, \z_{2}^{k}, \y_{1}^{k}, \y_{2}^{k}\right)}\end{array}\!\!\right]\!\!\|_{2}^{2}
\leq &  \|\!\!\left[\!\!\!\begin{array}{l}{\v^{k+1}\!-\!\v^{k}} \\ {\e_{1}^{k+1}\!-\!\e_{1}^{k}} \\ {\e_{2}^{k+1}\!-\!\e_{2}^{k}}\end{array}\!\!\right]\!\!\|_{2}^{2}
+ \|\!\!\left[\!\!\begin{array}{c}{\v_{k}\! -\!\v\left(\p^{k}, \z_{1}^{k}, \z_{2}^{k}, \y_{1}^{k}, \y_{2}^{k}\right)}
 \\ {\e_{1}^{k}\!-\!\e_{1}\left(\p^{k}, \z_{1}^{k}, \z_{2}^{k}, \y_{1}^{k}, \y_{2}^{k}\right)}
 \\ {\e_{2}^{k}\!-\!\e_{2}\left(\p^{k}, \z_{1}^{k}, \z_{2}^{k}, \y_{1}^{k}, \y_{2}^{k}\right)}\end{array}\!\!\right]\!\!\|_{2}^{2}  \\
\leq & \left(\!1+\!\frac{1}{\sqrt{\varepsilon_{1}}}\!\right)\!\!\|\!\!\left[\!\!\begin{array}{l}{\v^{k+1}\!-\!\v^{k}} \\ {\e_{1}^{k+1}\!-\!\e_{1}^{k}} \\ {\e_{2}^{k+1}\!-\!\e_{2}^{k}}\end{array}\!\!\!\right]\!\!\|_{2}^{2},
\end{split}
\end{equation}
i.e.,
\begin{equation*}\label{error-bound2-proof}
\begin{split}
& \|\!\!\left[\!\!\begin{array}{l}{\v^{k}\!-\!\v^{k+1}} \\ {\e_{1}^{k}\!-\!\e_{1}^{k+1}} \\ {\e_{2}^{k}\!-\!\e_{2}^{k+1}}\end{array}\!\!\right]\!\!\|_{2}^{2}
\!\geq\! \!\varepsilon_{2}\|\!\!\left[\!\!\begin{array}{c}{\v^{k+1}\! -\!\v\left(\p^{k}, \z_{1}^{k}, \z_{2}^{k}, \y_{1}^{k}, \y_{2}^{k}\right)}
 \\ {\e_{1}^{k+1}\!-\!\e_{1}\left(\p^{k}, \z_{1}^{k}, \z_{2}^{k}, \y_{1}^{k}, \y_{2}^{k}\right)}
 \\ {\e_{2}^{k+1}\!-\!\e_{2}\left(\p^{k}, \z_{1}^{k}, \z_{2}^{k}, \y_{1}^{k}, \y_{2}^{k}\right)}\end{array}\!\!\right]\!\!\|_{2}^{2},
\end{split}
\end{equation*}
where $\varepsilon_{2}=\frac{(\rho-\alpha)^2}{(2\rho+L+2-\alpha)^2}$.

Moreover, through similar proofs for (3.6)-(3.8) in \cite{proximal-admm}, we can verify that inequalities \eqref{error-bound3}-\eqref{error-bound5} hold.
In addition, since $g(\v, \e_{1}, \e_{2})=g(\v)=\pmb{\lambda}^T\v-\frac{\alpha}{2}\|\v-0.5\|_2^2$ is Lipschitz differentiable corresponding to variables $\v$, $\e_{1}$, and $\e_{2}$ with constant $L_{g}>\alpha$, based on \emph{Proposition~2.3} in \cite{proximal-admm}, we can see that problem \eqref{pADMM-frame-problem} satisfies the strict complementary condition. Then, we can prove that inequality \eqref{error-bound6} holds through similar derivations to (3.9) in \cite{proximal-admm}. This ends the proof.
}

\section{Proof of \emph{Lemma \ref{F-D-P-change}}}\label{descent3-proof}
{\it Proof:} {
First, we define the following quantities
\begin{subequations}\label{define-divide-F}
\begin{align}
& \mathcal{F}_\v^k=\F^k-\F(\v^{k+1},\e_{1}^{k},\e_{2}^{k},\p^k,\z_1^k,\z_2^k,\y_1^k,\y_2^k),\label{define-divide-Fv} \\
& \mathcal{F}_{\e_1}^k= \mathcal{F}(\v^{k+1},\e_{1}^{k},\e_{2}^{k},\p^k,\z_1^k,\z_2^k,\y_1^k,\y_2^k)
 -\mathcal{F}(\v^{k+1},\e_{1}^{k+1},\e_{2}^{k},\p^k,\z_1^k,\z_2^k,\y_1^k,\y_2^k),\label{define-divide-Fe1} \\
& \mathcal{F}_{\e_2}^k= \mathcal{F}(\v^{k+1},\e_{1}^{k+1},\e_{2}^{k},\p^k,\z_1^k,\z_2^k,\y_1^k,\y_2^k) -\mathcal{F}(\v^{k+1},\e_{1}^{k+1},\e_{2}^{k+1},\p^k,\z_1^k,\z_2^k,\y_1^k,\y_2^k),\label{define-divide-Fe2} \\
& \mathcal{F}_{\p\z}^k= \mathcal{F}(\v^{k+1},\e_{1}^{k+1},\e_{2}^{k+1},\p^k,\z_1^k,\z_2^k,\y_1^k,\y_2^k) - \F(\v^{k+1}\!,\e_{1}^{k+1}\!,\e_{2}^{k+1}\!,\p^{k+1}\!,\z_1^{k+1}\!,\z_2^{k+1}\!,\y_1^k,\y_2^k), \label{define-divide-Fpz} \\
& \mathcal{F}_{\y}^k\!=\! \F(\v^{k+1}\!,\e_{1}^{k+1}\!,\e_{2}^{k+1}\!,\p^{k+1}\!,\z_1^{k+1}\!,\z_2^{k+1}\!,\y_1^k,\y_2^k)\!-\!\F^{k+1}. \label{define-divide-Fy}
\end{align}
\end{subequations}
It is easy to see
\begin{equation}\label{primal-descent1}
\F^k-\F^{k+1}  =  \mathcal{F}_\v^k+\mathcal{F}_{\e_1}^k+\mathcal{F}_{\e_2}^k+\mathcal{F}_{\p\z}^k+\mathcal{F}_{\y}^k.
\end{equation}
From \eqref{proximal-x-update} and \eqref{F-define}, we can find
\[\v^{k+1}=\underset{\v}{\rm argmin}\F(\v, \e_{1}^{k}\!,\e_{2}^{k\!+\!1},\p^{k},\z_1^{k},\z_2^k,\y_1^k,\y_2^k).
\]
Since $\F(\v, \e_{1}^{k}\!,\e_{2}^{k},\p^{k},\z_1^{k},\z_2^k,\y_1^k,\y_2^k)$ is a strongly convex quadratic function, we have
\[
\begin{split}
\mathcal{F}_\v^k=& \frac{\rho+\mu-\alpha}{2}\|\v^{k}-\v^{k+1}\|_2^2 +\frac{\mu}{2}\|\A(\v^{k}-\v^{k+1})\|_2^2 \\
\geq & \frac{\rho+\mu-\alpha+\mu\lambda_{\min}(\A^T\A)}{2}\|\v^{k}-\v^{k+1}\|_2^2.
\end{split}
\]
Under the assumption of $\mu\lambda_{\min}(\A^T\A)\geq\alpha$, we can obtain
\begin{equation}\label{primal-descent1-a1}
\mathcal{F}_{\v}^k
\geq  \frac{\rho+\mu}{2}\|\v^{k}-\v^{k+1}\|_2^2.
\end{equation}

Since $\F(\v^{k+1}\!, \e_{1}\!,\e_{2}^{k},\p^{k},\z_1^{k},\z_2^k,\y_1^k,\y_2^k)$ is a strongly convex quadratic function, $\e_1\succeq\mathbf{0}$, and $\e_1^{k+1}$ is the minimizer of problem \eqref{proximal-v1-update}, we have
\begin{equation}\label{primal-descent1-e1}
  \mathcal{F}_{\e_1}^k \geq \frac{\rho+\mu}{2}\|\e_{1}^{k}-\e_1^{k+1}\|_2^2.
\end{equation}
Similarly, we can also obtain
\begin{equation}\label{primal-descent1-e2}
  \mathcal{F}_{\e_2}^k \geq \frac{\rho+\mu}{2}\|\e_{2}^{k}-\e_2^{k+1}\|_2^2.
\end{equation}

Plugging \eqref{proximal-z1-update} into \eqref{define-divide-Fpz}, we can derive it as \eqref{primal-descent1-b}, where the last inequality holds since $0<\beta \leq 1$.
\begin{figure*}
\begin{equation}\label{primal-descent1-b}
\begin{split}
\mathcal{F}_{\p\z}^k =&\frac{\rho}{2}\!\Big(\!\big(\p^{k\!+\!1}\!-\!\p^k\big)\!^T\!\big(2\v^{k\!+\!1}\!-\!\p^{k\!+\!1}\!-\!\p^k\!\big)\!\!+\!\!\big(\z_1^{k\!+\!1}\!\!-\!\z_1^k\!\big)\!^T\!\big(2\e_1^{k\!+\!1}\! \!-\!\z_1^{k+1}\!-\!\z_1^k\big)\!\!+\!\!\big(\z_2^{k+1}\!\!-\!\z_2^k\big)^T\!\big(2\e_2^{k+1}\!-\!\z_2^{k+1}\!-\!\z_2^k\big)\!\Big) \\
= & \frac{p}{2}\!\Big(\!\frac{2}{\beta}\!\!-\!\!1\!\Big)\!\Big(\!\|\p^{k\!+\!1}\!\!\!-\!\p^k\|_2^2\!+\!\!\|\z_1^{k\!+\!1}\!\!-\!\z_1^k\|_2^2\!+\!\!\|\z_2^{k\!+\!1}\!\!-\!\z_2^k\|_2^2\!\Big) \\
\geq & \frac{p}{2\beta}\Big(\!\|\p^{k+1}\!\!-\!\p^k\|_2^2\!+\!\|\z_1^{k+1}\!\!-\!\z_1^k\|_2^2\!+\!\|\z_2^{k+1}\!\!-\!\z_2^k\|_2^2\Big)
\end{split}
\end{equation}
\hrulefill
\vspace*{4pt}
\end{figure*}

Plugging \eqref{proximal-y2-update} into \eqref{define-divide-Fy}, we can obtain
\begin{equation}\label{primal-descent1-c}
\begin{split}
\mathcal{F}_{\y}^k =& \big(\y_1^k-\y_1^{k+1}\big)^T\big(\A\v^{k+1}+\e_{1}^{k+1}-{\boldsymbol\varrho}\big)
 +\big(\y_2^k-\y_2^{k+1}\big)^T\big(\v^{k+1}-\e_{2}^{k+1}\big) \\
=& \!-\mu\|\A\v^{k+1}+\e_{1}^{k+1}-{\boldsymbol\varrho}\|_2^2\!\! -\!\!\mu\|\v^{k+1}\!-\!\e_{2}^{k+1}\|_2^2.
\end{split}
\end{equation}

Then, plugging \eqref{primal-descent1-a1}--\eqref{primal-descent1-c} into \eqref{primal-descent1}, we have
\begin{equation}\label{F-change-proof}
\begin{split}
 \F^{k}\!-\!\F^{k+1} \geq & \frac{\rho+\mu}{2}\|\!\!\left[\!\!\begin{array}{l}{\v^{k}\!-\!\v^{k+1}} \\ {\e_{1}^{k}\!-\!\e_{1}^{k+1}} \\ {\e_{2}^{k}\!-\!\e_{2}^{k+1}}\end{array}\!\!\right]\!\!\|_{2}^{2}
\!+\!\!\frac{\rho}{2 \beta}\|\!\!\left[\!\!\begin{array}{l}{\p^{k}\!-\!\p^{k+1}} \\ {\z_{1}^{k}\!-\!\z_{1}^{k+1}} \\ {\z_{2}^{k}\!-\!\z_{2}^{k+1}}\end{array}\!\!\!\right]\!\!\|_{2}^{2}
\!-\!\mu\|\!\!\left[\!\!\begin{array}{c}{\A \v^{k}+\e_{1}^{k}-{\boldsymbol\varrho}} \\ {\v^{k}-\e_{2}^{k}}\end{array}\!\!\right]\!\!\|_{2}^{2}.
\end{split}
\end{equation}
That completes the proof of \eqref{F-change}.

Next, we consider to prove inequality \eqref{D-change}. To facilitate discussions later, we define
\begin{equation}\label{D-divide}
\begin{split}
& \D_{\p\z}^k=\D^{k+1}-\D(\p^k,\z_1^k,\z_2^k,\y_1^{k+1},\y_2^{k+1}), \\
& \D_{\y}^k= \D(\p^k,\z_1^k,\z_2^k,\y_1^{k+1},\y_2^{k+1})-\D^k.
\end{split}
\end{equation}
Then, we have
\begin{equation}\label{dual-descent-divide}
\begin{split}
& \D^{k+1}-\D^{k}=\D_{\p\z}^k+\D_{\y}^k.
\end{split}
\end{equation}
According to \eqref{D-define}, we can write $\D_{\p\z}^k$ as
\begin{equation}\label{dual-descent-a1-1}
\begin{split}
 \D_{\p\z}^k \!=& \F\!\Big(\!\v(\p^{k+1}\!, \z_{1}^{k+1}\!, \z_{2}^{k+1}\!, \y_{1}^{k+1}\!, \y_{2}^{k+1}),\! \e_1(\p^{k+1}\!, \z_{1}^{k+1}\!, \z_{2}^{k+1}\!,\y_{1}^{k+1}\!, \y_{2}^{k+1}), \\
&~~~  \e_2(\p^{k+1}\!, \z_{1}^{k+1}\!, \z_{2}^{k+1}\!, \y_{1}^{k+1}\!, \y_{2}^{k+1}), \p^{k+1}\!, \z_{1}^{k+1}\!, \z_{2}^{k+1}\!, \y_{1}^{k+1}\!, \y_{2}^{k+1}\Big) \\
-\!& \F\Big(\v(\p^{k}, \z_{1}^{k}, \z_{2}^{k}, \y_{1}^{k+1}\!, \y_{2}^{k+1}),\! \e_1(\p^{k}, \z_{1}^{k}, \z_{2}^{k}, \y_{1}^{k+1}\!\!, \y_{2}^{k+1}), \\
&~~~~ \e_2(\p^{k}\!, \z_{1}^{k}, \z_{2}^{k}, \y_{1}^{k+1}\!, \y_{2}^{k+1}), \p^{k}, \z_{1}^{k}, \z_{2}^{k}, \y_{1}^{k+1}\!, \y_{2}^{k+1}\!\Big).
\end{split}
\end{equation}
According to \eqref{x-D-define}, \eqref{dual-descent-a1-1} can be rewritten as
 {\setlength\abovedisplayskip{1pt}
 \setlength\belowdisplayskip{1pt}
  \setlength\jot{1pt}
\begin{equation}\label{dual-descent-a1-2}
\begin{split}
 \D_{\p\z}^k \!\!
\geq& \F\!\Big(\!\v(\p^{k+1}\!, \z_{1}^{k+1}\!, \z_{2}^{k+1}\!, \y_{1}^{k+1}\!, \y_{2}^{k+1}),\! \e_1(\p^{k+1}\!, \z_{1}^{k+1}\!, \z_{2}^{k+1}\!,\y_{1}^{k+1}\!, \y_{2}^{k+1}), \\
&~~~  \e_2(\p^{k+1}\!, \z_{1}^{k+1}\!, \z_{2}^{k+1}\!, \y_{1}^{k+1}\!, \y_{2}^{k+1}), \p^{k+1}\!, \z_{1}^{k+1}\!, \z_{2}^{k+1}\!, \y_{1}^{k+1}\!, \y_{2}^{k+1}\Big) \\
-\!& \F\Big(\v(\p^{k+1}, \z_{1}^{k+1}, \z_{2}^{k+1}, \y_{1}^{k+1}\!, \y_{2}^{k+1}),\! \e_1(\p^{k+1}, \z_{1}^{k+1}, \z_{2}^{k+1}, \y_{1}^{k+1}\!\!, \y_{2}^{k+1}), \\
&~~~~ \e_2(\p^{k+1}\!, \z_{1}^{k+1}, \z_{2}^{k+1}, \y_{1}^{k+1}\!, \y_{2}^{k+1}), \p^{k}, \z_{1}^{k}, \z_{2}^{k}, \y_{1}^{k+1}\!, \y_{2}^{k+1}\!\Big).
\end{split}
\end{equation}
Plugging} \eqref{F-define} into \eqref{dual-descent-a1-2}, we can obtain
 {\setlength\abovedisplayskip{2pt}
 \setlength\belowdisplayskip{2pt}
  \setlength\jot{1pt}
\begin{equation}\label{dual-descent-a1-3}
\begin{split}
\D_{\p\z}^k \!
\geq &\frac{\rho}{2}(\p^{k+1}\!-\p^{k})^T\left(\p^{k+1}\!+\p^{k}-2\v\!\left(\p^{k+1}, \z_{1}^{k+1}, \z_{2}^{k+1}, \y_{1}^{k+1}, \y_{2}^{k+1}\right)\right) \\
+ &  \frac{\rho}{2}(\z_{1}^{k+1}\!-\z_{1}^{k})^T\left(\z_{1}^{k+1}\!+\z_{1}^{k}-2 \e_{1}\!\left(\p^{k+1}, \z_{1}^{k+1}, \z_{2}^{k+1}, \y_{1}^{k+1}, \y_{2}^{k+1}\right)\right) \\
+ &  \frac{\rho}{2}(\z_{2}^{k+1}\!-\z_{2}^{k})^T\left(\z_{2}^{k+1}\!+\z_{2}^{k}-2 \e_{2}\!\left(\p^{k+1}, \z_{1}^{k+1}, \z_{2}^{k+1}, \y_{1}^{k+1}, \y_{2}^{k+1}\right)\right),
\end{split}
\end{equation}
which} can be further written as
\begin{equation}\label{dual-descent-a1-4}
\D_{\p\z}^k\geq\frac{\rho}{2}\!\left[\!\!\begin{array}{l}{\p^{k+1}-\p^{k}} \\ {\z_{1}^{k+1}-\z_{1}^{k}} \\ {\z_{2}^{k+1}-\z_{2}^{k}}\end{array}\!\!\right]^{T}\!\!\!\pmb{\psi}^k.
\end{equation}
Through similar derivations, we can also get
\begin{equation}\label{dual-descent-a2}
\begin{split}
\D_{\y}^k
\!\geq\!  \mu\! \left[\!\!\begin{array}{c}{\A\v^{k}+\e_{1}^{k}-{\boldsymbol\varrho}} \\ {\v^{k}-\e_{2}^{k}}\end{array}\!\!\right]^{T}\!\!\pmb{\phi}^k.
\end{split}
\end{equation}

Thus, combining \eqref{dual-descent-a1-4} and \eqref{dual-descent-a2}, we obtain \eqref{D-change}.

Finally, we consider to prove \eqref{P-change}.
According to Danskin's theorem \cite{convex-analysis}, we have
\begin{equation}\label{P-gradient-formula}
  \begin{split}
& \nabla \P\left(\p^{k}, \z_{1}^{k}, \z_{2}^{k}\right)
= \rho\left[\!\!\begin{array}{l}{\p^{k}}-\v(\p^k,\z_1^k,\z_2^k) \\ {\z_{1}^{k}}-{\e_{1}(\p^k,\z_1^k,\z_2^k)} \\ {\z_{2}^{k}}-{\e_{2}(\p^k,\z_1^k,\z_2^k)}\end{array}\!\!\!\right].
  \end{split}
\end{equation}
Applying the triangle inequality property and \eqref{error-bound4} to \eqref{P-gradient-formula}, we have the following derivations
\begin{equation}\label{P-gradient-cha}
  \begin{split}
& \left\|\nabla \P\left(\p^{k}, \z_{1}^{k}, \z_{2}^{k}\right)\!-\!\nabla \P\left(\p^{k+1}, \z_{1}^{k+1}, \z_{2}^{k+1}\right)\right\|_{2} \\
 =&\rho\|\!\!\left[\!\!\begin{array}{l}{\p^{k}\!-\!\p^{k+1}} \\ {\z_{1}^{k}\!-\!\z_{1}^{k+1}} \\ {\z_{2}^{k}\!-\!\z_{2}^{k+1}}\end{array}\!\!\right]
\!-\!\left[\!\!\begin{array}{l}{\v(\p^{k+1}\!,\z_1^{k+1}\!,\z_2^{k+1})\!-\!\v(\p^{k},\z_1^{k},\z_2^{k})}
\\ {\e_1(\p^{k+1}\!,\z_1^{k+1}\!,\z_2^{k+1})\!-\!\e_1(\p^{k},\z_1^{k},\z_2^{k})}
\\ {\e_2(\p^{k+1}\!,\z_1^{k+1},\z_2^{k+1})\!-\!\e_2\!(\p^{k},\z_1^{k},\z_2^{k})}\end{array}\!\!\right]\!\!\|_2 \\
\leq & \rho\|\!\left[\!\!\begin{array}{l}{\p^{k}\!-\!\p^{k+1}} \\ {\z_{1}^{k}\!-\!\z_{1}^{k+1}} \\ {\z_{2}^{k}\!-\!\z_{2}^{k+1}}\end{array}\!\!\right]\!\!\|_2
\!+\!\rho\|\!\!\left[\!\!\begin{array}{l}{\v(\p^{k+1},\z_1^{k+1},\z_2^{k+1})-\v(\p^{k},\z_1^{k},\z_2^{k})}
\\ {\e_1(\p^{k+1},\z_1^{k+1},\z_2^{k+1})-\e_1(\p^{k},\z_1^{k},\z_2^{k})}
\\ {\e_2(\p^{k+1},\z_1^{k+1},\z_2^{k+1})-\e_2(\p^{k},\z_1^{k},\z_2^{k})}\end{array}\!\!\right]\!\!\|_2 \\
\leq & \rho\left(1+\frac{1}{\sqrt{\varepsilon_{4}}}\right)
\|\!\!\left[\!\!\begin{array}{l}{\p^{k}\!-\!\p^{k+1}} \\ {\z_{1}^{k}\!-\!\z_{1}^{k+1}} \\ {\z_{2}^{k}\!-\!\z_{2}^{k+1}}\end{array}\!\!\right]\!\!\|_2,
  \end{split}
\end{equation}
which means the gradient of function $\P$ is Lipschitz continuous with respect to variables $\p$, $\z_1$, and $\z_2$. The corresponding Lipschitz constant is $\rho\left(1+\frac{1}{\sqrt{\varepsilon_{4}}}\right)$.
Then, according to property of the Lipschitz continuous function \cite{nonlinear-programm}, we have
\begin{equation}\label{proximal-descent-result}
  \begin{split}
 \P^{k+1}-\P^k
\leq & \left(\!\left[\!\!\begin{array}{c}{\p^{k+1}} \\ {\z_{1}^{k+1}} \\ {\z_{2}^{k+1}}\end{array}\!\!\right]
\!-\!\left[\!\!\begin{array}{c}{\p^{k}} \\ {\z_{1}^{k}} \\ {\z_{2}^{k}}\end{array}\!\!\right]\!\right)^{T}\!\!
\left(\!\left[\!\!\begin{array}{c}{\p^{k+1}} \\ {\z_{1}^{k+1}} \\ {\z_{2}^{k+1}}\end{array}\!\!\right]
\!-\!\left[\!\!\begin{array}{c}{\v(\p^{k},\z_1^k,\z_2^k)} \\ {\e_{1}(\p^{k},\z_1^k,\z_2^k)} \\ {\e_{2}(\p^{k},\z_1^k,\z_2^k)}\end{array}\!\!\right]\!\right) \\
& +\frac{\rho}{2}\left(1+\frac{1}{\sqrt{\varepsilon_{4}}}\right)\|\!\!\left[\!\!\begin{array}{c}{\p^{k+1}} \\ {\z_{1}^{k+1}} \\ {\z_{2}^{k+1}}\end{array}\!\!\right]
\!-\!\left[\!\!\begin{array}{c}{\p^{k}} \\ {\z_{1}^{k}} \\ {\z_{2}^{k}}\end{array}\!\!\right]\!\!\|_2^2.
  \end{split}
\end{equation}
Letting $\eta = \left(1+\frac{1}{\sqrt{\varepsilon_{4}}}\right)$, we can reach  \eqref{P-change}. This ends the proof.
}

\section{Proof of Theorem \ref{all zero assumption}}\label{proof of all zero assumption}
{\it Proof:} {
We need to prove $\Pr [{\rm{error}}|{{\bf{0}}^{n + {\Gamma _a}}}] = \Pr [{\rm{error}}|{\bf{v'}}]$. Vector ${\bf{v'}} = [{\bf{c}};{\bf{s'}}] \in \mathbb{F}_{2^q}^{n + \Gamma _a}$, where  ${\bf{c}}$ is the transmitted LDPC codeword in $\mathbb{F}_{2^q}^n$  and ${\bf{s}}'$  is the corresponding binary auxiliary variable.

Let ${\bf{\hat v}}$  and ${{\bf{\hat v}}^0}$  denote the output of the proximal-ADMM algorithm when  ${\bf{r}}$ and  ${{\bf{r}}^0}$ are received respectively. Here  ${\bf{r}}$ and  ${{\bf{r}}^0}$ are received vectors when codeword  ${\bf{v'}}$ and the all-zeros codeword ${{\bf{0}}^{n + {\Gamma _a}}}$  are transmitted over the channel.

let $B({\bf{v'}})$ denote the set of received vectors $\bf{r}$ that would cause decoding failure when the codeword $\bf{v'}$ is transmitted. Then, we need to verify
\begin{equation}\label{sum-of-pr-r-v}
\sum\limits_{{\bf{r}} \in B({\bf{v'}})} {\Pr [{\bf{r}}|{\bf{v}}'] }= \sum\limits_{{{\bf{r}}^0} \in B({{\bf{0}}^{n + {\Gamma _a}}})} {\Pr [{{\bf{r}}^0}|{{\bf{0}}^{n + {\Gamma _a}}}]}.
\end{equation}
It is obvious that \eqref{sum-of-pr-r-v} holds if and only if the following two
statements hold.

\begin{itemize}
   \item [(a)] $\Pr [{\bf{r}}|{\bf{v}}'] = \Pr [{{\bf{r}}^0}|{{\bf{0}}^{n + {\Gamma _a}}}]$.
   \item [(b)] ${\bf{r}} \in B({\bf{v'}}) $  if and only if ${{\bf{r}}^0} \in B({{\bf{0}}^{n + {\Gamma _{\rm{a}}}}})$.
  \end{itemize}
Statement (a) is directly implied by the definition of the symmetry condition. We change the decoding problem \eqref{ML-decoding-all-constant-weight} to an equivalent decoding problem and show in the following that statement (b) is also true.

\subsection{Symmetry Condition}
\begin{definition}\label{Symmetry-Condition}
For $\beta \in \mathbb{F}_{2^q}$, there exists a bijection
$${\tau _\beta }:\Sigma  \mapsto \Sigma,$$
such that the channel output probability conditioned on the channel input satisfies
\begin{equation}
p\left( y | \alpha  \right) = p\left( {\tau _\beta }\left( y \right) | {\alpha  - \beta } \right),
\end{equation}
for all $y \in \Sigma $, $\alpha  \in \mathbb{F}_{{2^q}}$. In addition, the mapping $\tau _\beta $  is assumed to be isometric with respect to Euclidean distance in $\Sigma $, for every $\beta \in \mathbb{F}_{2^q}$.
\end{definition}

\subsection{Equivalent decoding problem}
From Appendix \ref{CR}, we can write the decoding problem based on Constant-Weight embedding technique as
\begin{subequations}\label{ML-decoding-all-constant-weight}
\begin{align}
&\underset{\mathbf{v}}{\rm min} \hspace{0.3cm} {\boldsymbol\lambda}^{T}\mathbf{v}-\frac{\alpha}{2}\|\mathbf{v}-0.5\|_{2}^{2},\\
& \hspace{0.1cm} \rm{s.t.} \hspace{0.24cm} \hat{\mathbf{F}}\mathbf{v} \preceq \hat{\u}, \label{ML-decoding-all-b-constant}\\
&\hspace{0.9cm}\mathbf{S}\mathbf{v}= \mathbf{1}_{n+\Gamma_a}, \label{ML-decoding-all-c-constant} \\
&\hspace{0.9cm} \mathbf{v} \in [0,1]^{2^q(n+\Gamma_a)}. \label{ML-decoding-all-d-constant}
\end{align}
\end{subequations}

Next we change the decoding problem \eqref{ML-decoding-all-constant-weight} to an equivalent decoding problem. We first define the following polytopes and one code polytope.
\begin{definition}\label{polytopes}
Let ${\mathbb{P}_3}$ denote the parity polytope of dimension 3,
\begin{equation}
{\mathbb{P}_3}: = {\mathop{\rm conv}\nolimits} \left( {\left\{ {{\bf{e}} \in {{\{ 0,1\} }^3}\mid {{\left\| {\bf{e}} \right\|}_1}{\rm{is \ even}}} \right\}} \right).
\end{equation}
Let ${\mathbb{S}_d}$ denote the standard $d$-simplex,
\begin{equation}
{\mathbb{S}_d}: = \left\{ {\left( {{x_0}; \ldots ;{x_{d - 1}}} \right) \in \mathbb{R}^d|\sum\limits_{i = 0}^{d - 1} {{x_i}}  = 1,{\rm{\; and\; }}{x_i} \ge 0{\rm{\ for \ all\ }}i} \right\}.
\end{equation}

\end{definition}
\begin{definition}\label{code-polytopes}
 Let ${\cal C}$ be a non-binary SPC code defined by check vector ${\bf{h}} = [{h_1};{h_2};{h_3}] \in \mathbb{F}_{{2^q}}^3$ corresponding to a three-variables parity-check equation. In $\mathbb{F}_{2^q}$  denote by $\mathbb{U}$  the ``code polytope'' for Constant-Weight embedding where a $(3\cdot{2^q})$-length  variable ${\bf{v}} \in \mathbb{U}$  if and only if the following constraints hold:
 \begin{itemize}
   \item [(a)] ${{v}_{k,\sigma }} \in \left[ {0,1} \right]$.
   \item [(b)] $\sum\limits_{\sigma  = 0}^{{2^q} - 1} {{{v}_{k,\sigma }}}  = 1$.
   \item [(c)] Let $\tilde {\cal B}({\cal K},h): = \left\{ {\alpha  \in {\mathbb{F}_{{2^q}}}| \sum\limits_{i \in {\cal K}} {\tilde b{{(h\alpha )}_i}}  = 1} \right\}$,where $\tilde b{(h\alpha )_i}$  denotes the  $i$-th entry of the vector $\tilde{\mathbf{b}}(h\alpha )$. Let ${{\bf{g}}^{\cal K}}$  be a 3-length vector such that $g_k^{\cal K} = \sum\limits_{\sigma\in \tilde {\cal B}({\cal K},{h_k})}{{{v}_{k,\sigma}}}$, where $k \in \left\{ {1,2,3} \right\}$, then
   \begin{equation}
{{\bf{g}}^{\cal K}} \in {\mathbb{P}_3}{\rm{ \  for \ all \ }}{\cal K} \in \left\{ {{{\cal K}_\ell }|\ell  = 1, \ldots,{2^q} - 1} \right\}.
\end{equation}
  \end{itemize}
We note that in \emph{Definition \ref{code-polytopes}} conditions (a) and (b) define the simplex ${\mathbb{S}_{{2^q}}}$.
\end{definition}

Next we change the decoding problem \eqref{ML-decoding-all-constant-weight} to an equivalent decoding problem with a constraint on polytope.

\begin{lemma}\label{equivalent-decoding-problem}
the decoding problem \eqref{ML-decoding-all-constant-weight} is equivalent to the following decoding problem
\begin{subequations}\label{LP-decoding}
\begin{align}
&\underset{\mathbf{v}}{\rm min} \hspace{0.3cm} {\boldsymbol\lambda}^{T}\mathbf{v}-\frac{\alpha}{2}\|\mathbf{v}-0.5\|_{2}^{2},\\
& \hspace{0.1cm} \rm{s.t.} \hspace{0.24cm} {\left( {{{\bf{Q}}_\tau } \otimes {{\bf{I}}_{{2^q}}}} \right){\bf{v}} \in {\mathbb{U}_\tau }, \hspace{0.24cm} \forall \tau  \in \left\{ {1, \ldots, {\Gamma _c}} \right\}},
\end{align}
\end{subequations}
where $ \mathbb{U}_\tau$ is defined by \emph{Definition \ref{code-polytopes}} for the check corresponding to the $\tau$-th decomposed three-variables parity-check equation.
\end{lemma}
{\it Proof:} {
See Appendix \ref{equivalent-proof}.
}

By introducing auxiliary variables ${{\bf{z}}_\tau }$, $\tau  = 1, \ldots, {\Gamma _c}$, the decoding problem \eqref{LP-decoding} is equal to
\begin{subequations}\label{ADMM-LP-decoding}
\begin{align}
&\underset{\mathbf{v}}{\rm min} \hspace{0.3cm} {\boldsymbol\lambda}^{T}\mathbf{v}-\frac{\alpha}{2}\|\mathbf{v}-0.5\|_{2}^{2},\\
& \hspace{0.1cm} \rm{s.t.} \hspace{0.24cm} {\left( {{{\bf{Q}}_\tau } \otimes {{\bf{I}}_{{2^q}}}} \right){\bf{v}} = {{\bf{z}}_\tau }},\\
&\hspace{1.0cm}{{\bf{z}}_\tau }\in {\mathbb{U}_\tau }, \hspace{0.24cm} \forall \tau  \in \left\{ {1, \ldots,{\Gamma _c}} \right\},\\
&\hspace{1.0cm}{{\bf{v}}_i} \in {\mathbb{S}_{{2^q}}},\hspace{0.24cm} \forall i \in \left\{ {1, \ldots, n + {\Gamma _a}} \right\},
\end{align}
\end{subequations}
where vector ${\bf{v}}_i $  is the sub-vector selected from the  $i$-th  ${2^q}$-length block of $\bf{v} $. Its augmented Lagrangian function can be written as
\begin{equation}\label{aug-lag}
{{\cal L}_\mu }\left( {{\bf{v}},{\bf{z}},{\bf{y}}} \right) = {{\boldsymbol\lambda}^T}{\bf{v}} - \frac{\alpha }{2}\left\| {{\bf{v}} - 0.5} \right\|_2^2 + \sum\limits_{\tau  = 1}^{{\Gamma _c}} {{\bf{y}}_\tau ^T\left( {\left( {{{\bf{Q}}_\tau } \otimes {{\bf{I}}_{{2^q}}}} \right){\bf{v}} - {{\bf{z}}_\tau }} \right)}
+ \frac{\mu }{2}\sum\limits_{\tau  = 1}^{{\Gamma _c}} {\left\| {\left( {{{\bf{Q}}_\tau } \otimes {{\bf{I}}_{{2^q}}}} \right){\bf{v}} - {{\bf{z}}_\tau }} \right\|_2^2},
\end{equation}
where ${{\bf{y}}_\tau } \in {\mathbb{R}^{3 \cdot {2^q}}}$, $\tau  \in \left\{ {1, \ldots, {\Gamma _c}} \right\}$, is Lagrangian multipliers and $\mu > 0$ is a preset penalty parameter. Based on the augmented Lagrangian \eqref{aug-lag}, the proximal-ADMM solving algorithm for model \eqref{ADMM-LP-decoding} can be described as follows
\begin{subequations}\label{equi-proximal-ADMM update_LP}
\begin{align}
& \v^{k+1} = \mathop{\arg \min}\limits_{{\forall i \in \left\{ {1, \ldots, n + {\Gamma _a}} \right\},{{\bf{v}}_i} \in {\mathbb{S}_{{2^q}}}}}
{{\cal L}_\mu }\left( {{\bf{v}},{{\bf{z}}^k},{{\bf{y}}^k}} \right) + \frac{\rho }{2}\left\| {{\bf{v}} - {{\bf{p}}^k}} \right\|_2^2, \label{e-proximal-x-update}  \\
& {\bf{z}}_\tau ^{k + 1} = \mathop{\arg \min}\limits_{{{\bf{z}}_\tau } \in {\mathbb{U}_\tau }}  {{\cal L}_\mu }\left( {{{\bf{v}}^{k + 1}},{\bf{z}},{{\bf{y}}^k}} \right) + \frac{\rho }{2}\left\| {{{\bf{z}}_\tau } - {\bf{q}}_\tau ^k} \right\|_2^2, \hspace{0.24cm} \forall \tau  \in \left\{ {1, \ldots, {\Gamma _c}} \right\}, \label{zt-proximal-p-update}\\
& \mathbf{p}^{k+1}=\mathbf{p}^{k}+\beta(\v^{k+1}-\mathbf{p}^{k}),
\label{p-proximal-update}\\
& {\bf{q}}_\tau ^{k + 1} = {\bf{q}}_\tau ^k + \beta ({\bf{z}}_\tau ^{k + 1} - {\bf{q}}_\tau ^k), \hspace{0.24cm} \forall \tau  \in \left\{ {1, \ldots, {\Gamma _c}} \right\}, \label{qt-proximal-update}\\
& {\bf{y}}_\tau ^{k + 1} = {\bf{y}}_\tau ^k + \mu \left( {\left( {{{\bf{Q}}_\tau } \otimes {{\bf{I}}_{{2^q}}}} \right){{\bf{v}}^{k + 1}} - {\bf{z}}_\tau ^{k + 1}} \right),\hspace{0.24cm} \forall \tau  \in \left\{ {1, \ldots, {\Gamma _c}} \right\},
\label{yt-proximal-update}
\end{align}
\end{subequations}
where  $k$ is iteration number, $\left\| {{\bf{v}} - {{\bf{p}}^k}} \right\|_2^2$  and  $\left\| {{{\bf{z}}_\tau } - {\bf{q}}_\tau ^k} \right\|_2^2$ are so-called proximal terms, $\rho $  is the corresponding penalty parameter, and $\beta $  belongs to $\left( {0,1} \right]$.

Let ${{\boldsymbol\lambda}_i}$  be the  ${2^q}$-length sub-vector of $\boldsymbol\lambda$. Let ${\bf{y}}_\tau ^{\left( i \right)}$  be the  ${2^q}$-length sub-vector of ${\bf{y}}_\tau $  that corresponds to ${{\bf{v}}_i}$. Similarly, we define ${\bf{z}}_\tau ^{\left( i \right)}$  to be the sub-vector of ${{\bf{z}}_\tau }$  that corresponds to ${{\bf{v}}_i}$. Let ${\cal N}\left( i \right)$  denote the subset of the set $\left\{ {1, \ldots, {\Gamma _c}} \right\}$  that correspond to the set of the three-variables parity-check equations involving the  $i$-th information symbol. In the  ${\bf{x}}$-update we solve the following optimization problem:
\begin{equation}
\begin{split}
{{\bf{v}}^{k + 1}} =& \mathop{\arg \min}\limits_{\forall i \in \left\{ {1, \ldots, n + {\Gamma _a}} \right\},{{\bf{v}}_i} \in {\mathbb{S}_{{2^q}}}} \sum\limits_{i = 1}^{n + {\Gamma _a}} {\left( {{\boldsymbol\lambda}_i^T{{\bf{v}}_i} - \frac{\alpha }{2}\left\| {{{\bf{v}}_i} - 0.5} \right\|_2^2 + \sum\limits_{\tau  \in {\cal N}\left( i \right)} {{\left({\bf{y}}_\tau ^{k,\left( i \right)}\right)}^T\left( {{{\bf{v}}_i} - {\bf{z}}_\tau ^{k,\left( i \right)}} \right)}  } \right.}  \\
&\left.+ \frac{\mu }{2}\sum\limits_{\tau  \in {\cal N}\left( i \right)} {\left\| {{{\bf{v}}_i} - {\bf{z}}_\tau ^{k,\left( i \right)}} \right\|_2^2}
+ \frac{\rho }{2} {\left\| {{{\bf{v}}_i} - {\bf{p}}_i^k} \right\|_2^2}  \right).
\end{split}
\end{equation}
We can decouple ${{\bf{v}}_i}$  for all $i \in \left\{ {1, \ldots, n + {\Gamma _a}} \right\}$  in the sense that they can be individually solved for. Therefore
\begin{equation}\label{v_i_k+1_update}
\begin{split}
{{\bf{v}}_i^{k + 1}} =& \mathop{\arg \min}\limits_{{{\bf{v}}_i} \in {\mathbb{S}_{{2^q}}}} {{\boldsymbol\lambda}_i^T{{\bf{v}}_i} - \frac{\alpha }{2}\left\| {{{\bf{v}}_i} - 0.5} \right\|_2^2 + \sum\limits_{\tau  \in {\cal N}\left( i \right)} {{\left({\bf{y}}_\tau ^{k,\left( i \right)}\right)}^T\left( {{{\bf{v}}_i} - {\bf{z}}_\tau ^{k,\left( i \right)}} \right)}  }   \\
&+ \frac{\mu }{2}\sum\limits_{\tau  \in {\cal N}\left( i \right)} {\left\| {{{\bf{v}}_i} - {\bf{z}}_\tau ^{k,\left( i \right)}} \right\|_2^2}
+ \frac{\rho }{2} {\left\| {{{\bf{v}}_i} - {\bf{p}}_i^k} \right\|_2^2}   \\
=&\mathop \Pi \limits_{{\mathbb{S}_{{2^q}}}} \left[ {\frac{1}{{\mu {d_i} - \alpha  + \rho }}\left( {\sum\limits_{\tau  \in {\cal N}\left( i \right)} {\left( {\mu {\bf{z}}_\tau ^{k,\left( i \right)} - {\bf{y}}_\tau ^{k,\left( i \right)}} \right)}  - {{\boldsymbol\lambda}_i} - 0.5\alpha  + \rho {\bf{p}}_i^k} \right)} \right].
\end{split}
\end{equation}
Here, ${d_i}$  denotes the degree of the  $i$-th information symbol, i.e., the number of check equations involved.
In the ${\bf{z}}$-update we can solve for each  ${\bf{z}}_\tau $ separately and obtain the following update rule:
\begin{equation}\label{z_t_k+1_update}
\begin{split}
{\bf{z}}_\tau ^{k + 1} =& \mathop{\arg \min}\limits_{{{\bf{z}}_\tau } \in {\mathbb{U}_\tau }}  {{\boldsymbol\lambda}^T}{{\bf{v}}^{k + 1}} - \frac{\alpha }{2}\left\| {{{\bf{v}}^{k + 1}} - 0.5} \right\|_2^2 + \sum\limits_{\tau  = 1}^{{\Gamma _c}} {{\left({\bf{y}}_\tau ^{k}\right)}^T\left[ {\left( {{{\bf{Q}}_\tau } \otimes {{\bf{I}}_{{2^q}}}} \right){{\bf{v}}^{k + 1}} - {{\bf{z}}_\tau }} \right]} \\
&+ \frac{\mu }{2}\sum\limits_{\tau  = 1}^{{\Gamma _c}} {\left\| {\left( {{{\bf{Q}}_\tau } \otimes {{\bf{I}}_{{2^q}}}} \right){\bf{v}}^{k + 1} - {{\bf{z}}_\tau }} \right\|_2^2} + \frac{\rho }{2}\left\| {{{\bf{z}}_\tau } - {\bf{q}}_\tau ^k} \right\|_2^2  \\
 =& \mathop \Pi \limits_{{\mathbb{U}_\tau }} \left[ {\frac{\mu }{{\mu  + \rho }}\left( {{{\bf{Q}}_\tau } \otimes {{\bf{I}}_{{2^q}}}} \right){{\bf{v}}^{k + 1}} + \frac{1}{{\mu  + \rho }}{\bf{y}}_\tau ^k + \frac{\rho }{{\mu  + \rho }}{\bf{q}}_\tau ^k} \right].
\end{split}
\end{equation}
\subsection{Proof of Statement (b)}
Before showing the proof of Statement (b), we give one definition and four lemmas in
advance.
\begin{definition}\label{relative_matrices}
Let $\alpha  \in {\mathbb{F}_{{2^q}}}$  and let ${\bf{x}}$  be a ${2^q}$-length vector. We say that ${{\mathop{\rm R}\nolimits} _\alpha }({\bf{x}})$  is the ``relative vector'' of ${\bf{x}}$  based on $\alpha $  if
\begin{equation}
{({{\mathop{\rm R}\nolimits} _\alpha }({\bf{x}}))_\sigma } = {{x}_{\sigma  + \alpha }},
\end{equation}
where the sum is in $\mathbb{F}_{2^q}$.

We reuse this notation in the context of non-binary vectors. Let ${\bf{v'}} \in \mathbb{F}_{{2^q}}^{n + {\Gamma _a}}$  and let ${\bf{v}}$ be a ${2^q}(n + {\Gamma _a})$-length vector. Vector ${{\bf{v}}_i}$  is the sub-vector selected from the  $i$-th ${2^q}$-length block of  ${\bf{v}}$, where $i = 1, \ldots, n + {\Gamma _a}$. We say that  ${{\mathop{\rm R}\nolimits} _{{\bf{v'}}}}({\bf{v}})$ is the ``relative vector'' of ${\bf{v}}$  based on ${\bf{v}}'$  if
\begin{equation}
{\left( {{{\mathop{\rm R}\nolimits} _{{\bf{v'}}}}\left( {\bf{v}} \right)} \right)_{i,\sigma }} = {{v}_{i,\sigma  + v{'_i}}},
\end{equation}
where ${\left( {{{\mathop{\rm R}\nolimits} _{{\bf{v'}}}}\left( {\bf{v}} \right)} \right)_i}$  denotes the sub-vector in ${{\mathop{\rm R}\nolimits} _{{\bf{v'}}}}\left( {\bf{v}} \right)$, ${\left( {{{\mathop{\rm R}\nolimits} _{{\bf{v'}}}}\left( {\bf{v}} \right)} \right)_{i,\sigma }}$ denotes the  $\sigma $-th entry in ${({{\mathop{\rm R}\nolimits} _{{\bf{v'}}}}({\bf{v}}))_i}$, and ${{v}_{i,\sigma  + v{'_i}}}$  denotes the  $(\sigma  + v{'_i})$-th entry in ${{\bf{v}}_i}$.
\end{definition}

\begin{lemma}\label{relative_linear}
The relative operation is linear. That is,
\begin{equation}
{{\mathop{\rm R}\nolimits} _\alpha }({\phi _x}{\bf{x}} + {\phi _y}{\bf{y}}) = {\phi _x}{{\mathop{\rm R}\nolimits} _\alpha }({\bf{x}}) + {\phi _y}{{\mathop{\rm R}\nolimits} _\alpha }({\bf{y}}),
\end{equation}
where ${\phi _x} \in \mathbb{R}$  and  ${\phi _y} \in \mathbb{R}$.

Further, the relative operator is norm preserving. That is, $\left\| {{{\mathop{\rm R}\nolimits} _\alpha }({\bf{x}})} \right\| = \left\| {\bf{x}} \right\|$.
\end{lemma}
{\it Proof:} {
 Linearity is easy to verify. We have the following derivations
 \[
\begin{split}
{\left({{\mathop{\rm R}\nolimits} _\alpha }\left({\phi _x}{\bf{x}} + {\phi _y}{\bf{y}}\right)\right)_\sigma } &= {\left({\phi _x}{\bf{x}} + {\phi _y}{\bf{y}}\right)_{\sigma  + \alpha }}\\
 &= {\phi _x}{{\bf{x}}_{\sigma  + \alpha }} + {\phi _y}{{\bf{y}}_{\sigma  + \alpha }}\\
 &= {\phi _x}{\left({{\mathop{\rm R}\nolimits} _\alpha }\left({\bf{x}}\right)\right)_\sigma } + {\phi _y}{\left({{\mathop{\rm R}\nolimits} _\alpha }\left({\bf{y}}\right)\right)_\sigma }\\
 &= {\left({\phi _x}{{\mathop{\rm R}\nolimits} _\alpha }\left({\bf{x}}\right) + {\phi _y}{{\mathop{\rm R}\nolimits} _\alpha }\left({\bf{y}}\right)\right)_\sigma }.
\end{split}
\]

We note that ${{\mathop{\rm R}\nolimits} _\alpha }( \cdot )$  permutes the input vector based on $\alpha $, and therefore the norm is preserved.
}

\begin{lemma}\label{relative_in_U}
Let  ${\bf{c}}$ be a valid SPC codeword for 3-length check ${\bf{h}}$. Then ${\bf{v}} \in \mathbb{U} $  if and only if ${{\mathop{\rm R}\nolimits} _{\bf{c}}}({\bf{v}}) \in \mathbb{U}$.
\end{lemma}
{\it Proof:} {
See Appendix \ref{relative_in_U_proof}.
}

\begin{lemma}\label{projection_relative_relation}
Suppose a convex set $\mathbb{C}$  is such that  ${\bf{x}} \in \mathbb{C}$ if and only if ${{\mathop{\rm R}\nolimits} _\alpha }\left( {\bf{x}} \right) \in \mathbb{C}$  for some $\alpha $, then ${\Pi _{\mathbb{C}}}\left( {{{\mathop{\rm R}\nolimits} _\alpha }\left( {\bf{v}} \right)} \right) = {{\mathop{\rm R}\nolimits} _\alpha }\left( {{\Pi _{\mathbb{C}}}\left( {\bf{v}} \right)} \right)$.
\end{lemma}
{\it Proof:} {
Our proof is by contradiction. Suppose that the projection of  ${{\mathop{\rm R}\nolimits} _\alpha }\left( {\bf{v}} \right)$ onto $\mathbb{C}$  is ${\bf{u}} \ne {{\mathop{\rm R}\nolimits} _\alpha }\left( {{\Pi _\mathbb{C}}\left( {\bf{v}} \right)} \right)$. Then ${\mathop{\rm R}\nolimits} _\alpha ^{ - 1}\left( {\bf{u}} \right) \in \mathbb{C}$  and ${\mathop{\rm R}\nolimits} _\alpha ^{ - 1}\left( {\bf{u}} \right) \ne {\Pi _\mathbb{C}}\left( {\bf{v}} \right)$. We have
\[
\begin{split}
\left\| {{\mathop{\rm R}\nolimits} _\alpha ^{ - 1}\left( {\bf{u}} \right) - {\bf{v}}} \right\|_2^2 &= \left\| {{\bf{u}} - {{\mathop{\rm R}\nolimits} _\alpha }\left( {\bf{v}} \right)} \right\|_2^2\\
 &< \left\| {{{\mathop{\rm R}\nolimits} _\alpha }\left( {{\Pi _\mathbb{C}}\left( {\bf{v}} \right)} \right) - {{\mathop{\rm R}\nolimits} _\alpha }\left( {\bf{v}} \right)} \right\|_2^2\\
 &= \left\| {{\Pi _\mathbb{C}}\left( {\bf{v}} \right) - {\bf{v}}} \right\|_2^2.
\end{split}
\]
This contradicts the fact that  ${\Pi _\mathbb{C}}\left( {\bf{v}} \right)$ is the projection of ${\bf{v}}$  onto $\mathbb{C}$.
}

\begin{lemma}\label{updated-relative}
Let ${{\bf{v}}^k}$, ${\bf{z}}_\tau ^k$, ${{\bf{p}}^k}$, ${\bf{q}}_\tau ^k$  and ${\bf{y}}_\tau ^k$  be the updated variables in the  $k$-th iteration when decoding ${\bf{r}}$, where $\tau  \in \left\{ {1, \ldots, {\Gamma _c}} \right\}$. Let ${{\bf{v}}^{0,k}}$,  ${\bf{z}}_\tau ^{0,k}$, ${{\bf{p}}^{0,k}}$, ${\bf{q}}_\tau ^{0,k}$  and ${\bf{y}}_\tau ^{0,k}$  be the updated variables in the  $k$-th iteration when decoding ${{\bf{r}}^0}$, where $\tau  \in \left\{ {1, \ldots, {\Gamma _c}} \right\}$. Let ${\bf{v}}{'_\tau } = {{\bf{Q}}_\tau }{\bf{v}}'$  be the variable involved in the  $\tau $-th three-variables parity-check equation.
If  ${{\bf{v}}^{0,k}} = {{\mathop{\rm R}\nolimits} _{{\bf{v}}'}}\left( {{{\bf{v}}^k}} \right)$,
${\bf{z}}_\tau ^{0,k} = {{\mathop{\rm R}\nolimits} _{{\bf{v}}{'_\tau }}}\left( {{\bf{z}}_\tau ^k} \right)$,
${{\bf{p}}^{0,k}} = {{\mathop{\rm R}\nolimits} _{{\bf{v}}'}}\left( {{{\bf{p}}^k}} \right)$,
${\bf{q}}_\tau ^{0,k} = {{\mathop{\rm R}\nolimits} _{{\bf{v}}{'_\tau }}}\left( {{\bf{q}}_\tau ^k} \right)$  and
 ${\bf{y}}_\tau ^{0,k} = {{\mathop{\rm R}\nolimits} _{{\bf{v}}{'_\tau }}}\left( {{\bf{y}}_\tau ^{k}} \right)$
 then ${{\bf{v}}^{0,k + 1}} = {{\mathop{\rm R}\nolimits} _{{\bf{v}}'}}\left( {{{\bf{v}}^{k + 1}}} \right)$,
 ${\bf{z}}_\tau ^{0,k + 1} = {{\mathop{\rm R}\nolimits} _{{\bf{v}}{'_\tau }}}\left( {{\bf{z}}_\tau ^{k + 1}} \right)$,
 ${{\bf{p}}^{0,k + 1}} = {{\mathop{\rm R}\nolimits} _{{\bf{v}}'}}\left( {{{\bf{p}}^{k + 1}}} \right)$,
 ${\bf{q}}_\tau ^{0,k + 1} = {{\mathop{\rm R}\nolimits} _{{\bf{v}}{'_\tau }}}\left( {{\bf{q}}_\tau ^{k + 1}} \right)$ and
 ${\bf{y}}_\tau ^{0,k + 1} = {{\mathop{\rm R}\nolimits} _{{\bf{v}}{'_\tau }}}\left( {{\bf{y}}_\tau ^{k + 1}} \right)$.
\end{lemma}
{\it Proof:} {
Let ${{\pmb{\gamma}} ^0} \in {\mathbb{R}^{n \cdot {2^q}}}$  be the log-likelihood
 ratio for the received vector ${{\bf{r}}^0}$  and ${{\boldsymbol\lambda}^0} = \left[ {{{\pmb{\gamma}}^0};{{\bf{0}}_{{\Gamma _a} \cdot {2^q}}}} \right] \in \mathbb{R}^{(n + {\Gamma _a}){2^q}}$. If $i \in \left\{ {n + 1, \ldots, n + {\Gamma _a}} \right\}$, it is obvious that ${\boldsymbol\lambda}_i^0 = {{\mathop{\rm R}\nolimits} _{v{'_i}}}\left( {{{\boldsymbol\lambda}_i}} \right)$. If $i \in \left\{ {1, \ldots, n} \right\}$,
 \[
\begin{split}
\lambda _{i,\sigma }^0 &= \log \frac{1}{{p \left( {r_i^0 | {{x_{i,\sigma }} = 1}} \right)}}\\
 &= \log \frac{1}{{p\left( {{r_i} | {{x_{i,\sigma  + v{'_i}}} = 1} } \right)}} = {\lambda _{i,\sigma  + v{'_i}}},
\end{split}
\]
where  $\lambda _{i,\sigma }^0$  denotes the  $\sigma $-th entry in ${\boldsymbol\lambda}_i^0$. Therefore we conclude that ${{\boldsymbol\lambda}^0} = {{\mathop{\rm R}\nolimits} _{{\bf{v}}'}}\left( {\boldsymbol\lambda} \right)$. By \eqref{v_i_k+1_update}, we have
\[
\begin{split}
{\bf{v}}_i^{0,k + 1} &= \mathop \Pi \limits_{{\mathbb{S}_{{2^q}}}} \left[ {\frac{1}{{\mu {d_i} - \alpha  + \rho }}\left( {\sum\limits_{\tau  \in {\cal N}\left( i \right)} {\left( {\mu {\bf{z}}_\tau ^{0,k,\left( i \right)} - {\bf{y}}_\tau ^{0,k,\left( i \right)}} \right)}  - {\boldsymbol\lambda}_i^0 - 0.5\alpha  + \rho {\bf{p}}_i^{0,k}} \right)} \right]\\
 &= \mathop \Pi \limits_{{\mathbb{S}_{{2^q}}}} \left[ {\frac{1}{{\mu {d_i} - \alpha  + \rho }}\left( {\sum\limits_{\tau  \in {\cal N}\left( i \right)} {\left( {\mu {{\mathop{\rm R}\nolimits} _{v{'_i}}}\left( {{\bf{z}}_\tau ^{k,\left( i \right)}} \right) - {{\mathop{\rm R}\nolimits} _{v{'_i}}}\left( {{\bf{y}}_\tau ^{k,\left( i \right)}} \right)} \right)}  - {{\mathop{\rm R}\nolimits} _{v{'_i}}}\left( {{{\boldsymbol\lambda}_i}} \right) - 0.5\alpha  + \rho {{\mathop{\rm R}\nolimits} _{v{'_i}}}\left( {{\bf{p}}_i^k} \right)} \right)} \right]\\
 &= \mathop \Pi \limits_{{\mathbb{S}_{{2^q}}}} \left[ {{{\mathop{\rm R}\nolimits} _{v{'_i}}}\left( {\frac{1}{{\mu {d_i} - \alpha  + \rho }}\left( {\sum\limits_{\tau  \in {\cal N}\left( i \right)} {\left( {\mu {\bf{z}}_\tau ^{k,\left( i \right)} - {\bf{y}}_\tau ^{k,\left( i \right)}} \right)}  - {{\boldsymbol\lambda}_i} - 0.5\alpha  + \rho {\bf{p}}_i^k} \right)} \right)} \right].
\end{split}
\]
By \emph{Lemma \ref{projection_relative_relation}}, we have
\[ {\bf{v}}_i^{0,k + 1}  = {{\mathop{\rm R}\nolimits} _{v{'_i}}}\left( {{\bf{v}}_i^{k + 1}} \right).\]
Therefore we conclude that
\begin{equation}
{{\bf{v}}^{0,k + 1}} = {{\mathop{\rm R}\nolimits} _{{\bf{v}}'}}\left( {{{\bf{v}}^{k + 1}}} \right).
\end{equation}

By \eqref{z_t_k+1_update}, we have
\[
\begin{split}
{\bf{z}}_\tau ^{0,k + 1} &= \mathop \Pi \limits_{{\mathbb{U}_\tau }} \left[ {\frac{\mu }{{\mu  + \rho }}\left( {{{\bf{Q}}_\tau } \otimes {{\bf{I}}_{{2^q}}}} \right){{\bf{v}}^{0,k + 1}} + \frac{1}{{\mu  + \rho }}{\bf{y}}_\tau ^{0,k} + \frac{\rho }{{\mu  + \rho }}{\bf{q}}_\tau ^{0,k}} \right]\\
 &= \mathop \Pi \limits_{{\mathbb{U}_\tau }} \left[ {\frac{\mu }{{\mu  + \rho }}\left( {{{\bf{Q}}_\tau } \otimes {{\bf{I}}_{{2^q}}}} \right){{\mathop{\rm R}\nolimits} _{{\bf{v}}'}}\left( {{{\bf{v}}^{k + 1}}} \right) + \frac{1}{{\mu  + \rho }}{{\mathop{\rm R}\nolimits} _{{\bf{v}}{'_\tau }}}\left( {{\bf{y}}_\tau ^k} \right) + \frac{\rho }{{\mu  + \rho }}{{\mathop{\rm R}\nolimits} _{{\bf{v}}{'_\tau }}}\left( {{\bf{q}}_\tau ^k} \right)} \right]\\
 &= \mathop \Pi \limits_{{\mathbb{U}_\tau }} \left[ {\frac{\mu }{{\mu  + \rho }}{{\mathop{\rm R}\nolimits} _{{\bf{v}}{'_\tau }}}\left( {\left( {{{\bf{Q}}_\tau } \otimes {{\bf{I}}_{{2^q}}}} \right){{\bf{v}}^{k + 1}}} \right) + \frac{1}{{\mu  + \rho }}{{\mathop{\rm R}\nolimits} _{{\bf{v}}{'_\tau }}}\left( {{\bf{y}}_\tau ^k} \right) + \frac{\rho }{{\mu  + \rho }}{{\mathop{\rm R}\nolimits} _{{\bf{v}}{'_\tau }}}\left( {{\bf{q}}_\tau ^k} \right)} \right]\\
 &= \mathop \Pi \limits_{{\mathbb{U}_\tau }} \left[ {{{\mathop{\rm R}\nolimits} _{{\bf{v}}{'_\tau }}}\left( {\frac{\mu }{{\mu  + \rho }}\left( {\left( {{{\bf{Q}}_\tau } \otimes {{\bf{I}}_{{2^q}}}} \right){{\bf{v}}^{k + 1}}} \right) + \frac{1}{{\mu  + \rho }}{\bf{y}}_\tau ^k + \frac{\rho }{{\mu  + \rho }}{\bf{q}}_\tau ^k} \right)} \right].
\end{split}
\]
By \emph{Lemma \ref{projection_relative_relation}},we conclude that
\begin{equation}
{\bf{z}}_\tau ^{0,k + 1} = {{\mathop{\rm R}\nolimits} _{{\bf{v}}{'_\tau }}}\left( {{\bf{z}}_\tau ^{k + 1}} \right).
\end{equation}

By \eqref{p-proximal-update}-\eqref{yt-proximal-update}, we have
\begin{equation}
\begin{split}
{{\bf{p}}^{0,k + 1}} &= {{\bf{p}}^{0,k}} + \beta \left( {{{\bf{v}}^{0,k + 1}} - {{\bf{p}}^{0,k}}} \right)\\
 &= {{\mathop{\rm R}\nolimits} _{{\bf{v}}'}}\left( {{{\bf{p}}^k}} \right) + \beta \left( {{{\mathop{\rm R}\nolimits} _{{\bf{v}}'}}\left( {{{\bf{v}}^{k + 1}}} \right) - {{\mathop{\rm R}\nolimits} _{{\bf{v}}'}}\left( {{{\bf{p}}^k}} \right)} \right)\\
 &= {{\mathop{\rm R}\nolimits} _{{\bf{v}}'}}\left( {{{\bf{p}}^k} + \beta \left( {{{\bf{v}}^{k + 1}} - {{\bf{p}}^k}} \right)} \right) \\
 &= {{\mathop{\rm R}\nolimits} _{{\bf{v}}'}}\left( {{{\bf{p}}^{k + 1}}} \right),
\end{split}
\end{equation}
\begin{equation}
\begin{split}
{\bf{q}}_\tau ^{0,k + 1} &= {\bf{q}}_\tau ^{0,k} + \beta ({\bf{z}}_\tau ^{0,k + 1} - {\bf{q}}_\tau ^{0,k})\\
 &= {{\mathop{\rm R}\nolimits} _{{\bf{v}}{'_\tau }}}\left( {{\bf{q}}_\tau ^k} \right) + \beta \left( {{{\mathop{\rm R}\nolimits} _{{\bf{v}}{'_\tau }}}\left( {{\bf{z}}_\tau ^{k + 1}} \right) - {{\mathop{\rm R}\nolimits} _{{\bf{v}}{'_\tau }}}\left( {{\bf{q}}_\tau ^k} \right)} \right)\\
 &= {{\mathop{\rm R}\nolimits} _{{\bf{v}}{'_\tau }}}\left( {{\bf{q}}_\tau ^k + \beta \left( {{\bf{z}}_\tau ^{k + 1} - {\bf{q}}_\tau ^k} \right)} \right) \\
 &= {{\mathop{\rm R}\nolimits} _{{\bf{v}}{'_\tau }}}\left( {{\bf{q}}_\tau ^{k + 1}} \right),
\end{split}
\end{equation}
\begin{equation}
\begin{split}
{\bf{y}}_\tau ^{0,k + 1} &= {\bf{y}}_\tau ^{0,k} + \mu \left( {\left( {{{\bf{Q}}_\tau } \otimes {{\bf{I}}_{{2^q}}}} \right){{\bf{v}}^{0,k + 1}} - {\bf{z}}_\tau ^{0,k + 1}} \right)\\
 &= {{\mathop{\rm R}\nolimits} _{{\bf{v}}{'_\tau }}}\left( {{\bf{y}}_\tau ^k} \right) + \mu \left( {\left( {{{\bf{Q}}_\tau } \otimes {{\bf{I}}_{{2^q}}}} \right){{\mathop{\rm R}\nolimits} _{{\bf{v}}'}}\left( {{{\bf{v}}^{k + 1}}} \right) - {{\mathop{\rm R}\nolimits} _{{\bf{v}}{'_\tau }}}\left( {{\bf{z}}_\tau ^{k + 1}} \right)} \right)\\
 &= {{\mathop{\rm R}\nolimits} _{{\bf{v}}{'_\tau }}}\left( {{\bf{y}}_\tau ^k} \right) + \mu \left( {{{\mathop{\rm R}\nolimits} _{{\bf{v}}{'_\tau }}}\left( {\left( {{{\bf{Q}}_\tau } \otimes {{\bf{I}}_{{2^q}}}} \right){{\bf{v}}^{k + 1}}} \right) - {{\mathop{\rm R}\nolimits} _{{\bf{v}}{'_\tau }}}\left( {{\bf{z}}_\tau ^{k + 1}} \right)} \right)\\
 &= {{\mathop{\rm R}\nolimits} _{{\bf{v}}{'_\tau }}}\left( {{\bf{y}}_\tau ^k + \mu \left( {\left( {{{\bf{Q}}_\tau } \otimes {{\bf{I}}_{{2^q}}}} \right){{\bf{v}}^{k + 1}} - {\bf{z}}_\tau ^{k + 1}} \right)} \right) \\
 &= {{\mathop{\rm R}\nolimits} _{{\bf{v}}{'_\tau }}}\left( {{\bf{y}}_\tau ^{k + 1}} \right).
\end{split}
\end{equation}
This completes the proof.
}

We note that we initialize all the variables as all-zeros vectors. Obviously, ${\bf{z}}_\tau ^{0,0} = {{\mathop{\rm R}\nolimits} _{{\bf{v}}{'_\tau }}}\left( {{\bf{z}}_\tau ^0} \right)$, ${{\bf{p}}^{0,0}} = {{\mathop{\rm R}\nolimits} _{{\bf{v}}'}}\left( {{{\bf{p}}^0}} \right)$, ${\bf{q}}_\tau ^{0,0} = {{\mathop{\rm R}\nolimits} _{{\bf{v}}{'_\tau }}}\left( {{\bf{q}}_\tau ^0} \right)$  and ${\bf{y}}_\tau ^{0,0} = {{\mathop{\rm R}\nolimits} _{{\bf{v}}{'_\tau }}}\left( {{\bf{y}}_\tau ^0} \right)$. By induction, we always obtain relative vectors at each iteration. It is easy to verify that both decoding processes stop at the same iteration. Therefore ${{\bf{\hat v}}^0} = {{\mathop{\rm R}\nolimits} _{{\bf{v'}}}}({\bf{\hat v}})$, which means  ${{\bf{r}}^0}$ is decoded unsuccessfully if and only if ${\bf{r}}$  is decoded unsuccessfully, since ${\hat v}_{i,0}^0 = 0$  if and only if ${{\hat v}_{i,v{'_i}}} = 0$. Hence, we can conclude that ${\bf{r}} \in B({\bf{v'}})$  if and only if ${{\bf{r}}^0} \in B({{\bf{0}}^{n + {\Gamma _{\rm{a}}}}})$. Thus, the second statement (b) holds.
}

\section{Proof of \emph{Lemma \ref{equivalent-decoding-problem}}}\label{equivalent-proof}
{\it Proof:} {
We need to prove that \eqref{ML-decoding-all-b-constant}-\eqref{ML-decoding-all-d-constant} is equivalent to $\left( {{{\bf{Q}}_\tau } \otimes {{\bf{I}}_{{2^q}}}} \right){\bf{v}} \in {\mathbb{U}_\tau }$, $\forall \tau  \in \left\{ {1, \ldots, {\Gamma _c}} \right\}$,  i.e., $\left( {{{\bf{Q}}_\tau } \otimes {{\bf{I}}_{{2^q}}}} \right){\bf{v}}$ satisfies condition (a), (b) and (c) in \emph{Definition \ref{code-polytopes}} for all $\tau  \in \left\{ {1, \ldots, {\Gamma _c}} \right\}$.

\eqref{ML-decoding-all-b-constant} can be expressed explicitly as
\begin{equation}
{\hat{\bf{W}}_\tau }\left( {{{\bf{Q}}_\tau } \otimes {{\bf{I}}_{{2^q}}}} \right){\bf{v}}\preceq \hat {\bf{w}},\hspace{0.24cm}  \tau =  {1, \ldots, {\Gamma _c}}.
\end{equation}

For $\forall \tau  \in \left\{ {1, \ldots, {\Gamma _c}} \right\}$,  ${\hat{\bf{W}}_\tau }\left( {{{\bf{Q}}_\tau } \otimes {{\bf{I}}_{{2^q}}}} \right){\bf{v}}\preceq \hat {\bf{w}}$ can be written as
\begin{equation}
{\bf{P}}{\hat {\bf{T}}_\ell }{{\bf{D}}_\tau }\left( {{{\bf{Q}}_\tau } \otimes {{\bf{I}}_{{2^q}}}} \right){\bf{v}} \preceq {\bf{t}},\hspace{0.24cm} \ell =  {1, \ldots, {2^q} - 1}.
\end{equation}

Denote the constant weight embedding function as $f\left(  \cdot  \right)$, then $f\left( {{v{'_i}}} \right) = {{\bf{v}}_i} = \left[ {{v_{i,0}}; \cdots ;{v_{i,{2^q} - 1}}} \right] \in {\{ 0,1\} ^{{2^q}}}$ and  $f\left( {\bf{v'}} \right) = {\bf{v}} = \left[ {{{\bf{v}}_1}; \cdots ;{{\bf{v}}_{n + {\Gamma _a}}}} \right] \in {\{ 0,1\} ^{\left({n + {\Gamma _a}}\right)  {2^q}}}$.

According to the previous content, we have
\begin{equation}
{\bf{D}}({2^q},{h_k})f\left( \alpha  \right) = f\left( {{h_k}\alpha } \right),\hspace{0.24cm} k \in \left\{ {1,2,3} \right\},
\end{equation}
where $\alpha  \in {\mathbb{F}_{{2^q}}}$. Then we have
\[
\left( {\sum\limits_{i \in {{\cal K}_\ell }} {\hat {\bf{b}}_i^T} } \right){\bf{D}}({2^q},{h_k})f\left( \alpha  \right) = \sum\limits_{i \in {{\cal K}_\ell }} {\hat {\bf{b}}_i^T} f\left( {{h_k}\alpha } \right) = \sum\limits_{i \in {{\cal K}_\ell }} {\tilde b{{({h_k}\alpha )}_i}},
\]
where $\hat {\bf{b}}_i^T$ denote the $i$-th row vector of matrix $\bf{B}$, $\tilde{\bf{b}}{({h_k}\alpha )}$ denote the $({h_k}\alpha)$-th column vector of matrix $\bf{B}$, and $\tilde b{{({h_k}\alpha )}_i}$ denotes the  $i$-th entry in $\tilde {\bf{b}}{({h_k}\alpha )}$. We have
\[
\left( {\sum\limits_{i \in {{\cal K}_\ell }} {\hat {\bf{b}}_i^T} } \right){\bf{D}}({2^q},{h_k})f\left( \alpha  \right) = {\left( {\left( {\sum\limits_{i \in {{\cal K}_\ell }} {\hat {\bf{b}}_i^T} } \right){\bf{D}}({2^q},{h_k})} \right)_\alpha },
\]
where ${\left( {\left( {\sum\limits_{i \in {{\cal K}_\ell }} {\hat {\bf{b}}_i^T} } \right){\bf{D}}({2^q},{h_k})} \right)_\alpha }$ denotes the  $\alpha$-th entry of $\left( {\sum\limits_{i \in {{\cal K}_\ell }} {\hat {\bf{b}}_i^T} } \right){\bf{D}}({2^q},{h_k})$. Hence
\[{\left( {\left( {\sum\limits_{i \in {{\cal K}_\ell }} {\hat {\bf{b}}_i^T} } \right){\bf{D}}({2^q},{h_k})} \right)_\alpha } = \sum\limits_{i \in {{\cal K}_\ell }} {\tilde b{{({h_k}\alpha )}_i}}. \]

Therefore we have ${\left( {\left( {\sum\limits_{i \in {{\cal K}_\ell }} {\hat {\bf{b}}_i^T} } \right){\bf{D}}({2^q},{h_k})} \right)_\alpha } = 1$ if and only if $\alpha  \in \tilde {\cal B}({{\cal K}_\ell },{h_k})$, where $\tilde {\cal B}(\cdot)$ is defined by \emph{Definition \ref{code-polytopes}}.

Let $\left( {{{\bf{Q}}_\tau } \otimes {{\bf{I}}_{{2^q}}}} \right){\bf{v}} = {\bf{v}}_{\tau} = [{\bf{v}}_{\tau_1};{{\bf{v}}_{\tau_2}};{{\bf{v}}_{\tau_3}}]$, where ${\bf{v}}_{\tau_1}$, ${\bf{v}}_{\tau_2}$ and ${\bf{v}}_{\tau_3}$ are the $2^q$-length sub-vectors of $\bf{v}$. Then we have
\[\left( {\sum\limits_{i \in {{\cal K}_\ell }} {\hat {\bf{b}}_i^T} } \right){\bf{D}}({2^q},{h_k}){{\bf{v}}_{\tau_k}} = \sum\limits_{\alpha  \in \tilde {\cal B}({{\cal K}_\ell },{h_k})} {{{v}_{\tau_k,\alpha }}}  = g_k^{{\tau, {\cal K}_\ell }},\]
where $k \in \left\{ {1,2,3} \right\}$ and $g_k^{{\tau, {\cal K}_\ell }}$ is defined by \emph{Definition \ref{code-polytopes}}.

Then we have
\[{\hat {\bf{T}}_\ell }{{\bf{D}}_\tau }\left( {{{\bf{Q}}_\tau } \otimes {{\bf{I}}_{{2^q}}}} \right){\bf{v}} = {{\bf{g}}^{{\tau, {\cal K}_\ell }}},\]
where ${{\bf{g}}^{{\tau, {\cal K}_\ell }}} = \left[ {g_1^{{\tau, {\cal K}_\ell }};g_2^{{\tau, {\cal K}_\ell }};g_3^{{\tau, {\cal K}_\ell }}} \right]$, and $g_1^{{\tau, {\cal K}_\ell }}, g_2^{{\tau, {\cal K}_\ell }}, g_3^{{\tau, {\cal K}_\ell }} \in \left[ {0,1} \right]$.

From the reference \cite{minimum-polytope} we have that ${\bf{P}}{{\bf{g}}^{{\tau, {\cal K}_\ell }}} \preceq {\bf{t}}$ is equivalent to ${{\bf{g}}^{{\tau, {\cal K}_\ell }}} \in {\mathbb{P}_3}$. Therefore \eqref{ML-decoding-all-b-constant} is equivalent to that $\left( {{{\bf{Q}}_\tau } \otimes {{\bf{I}}_{{2^q}}}} \right){\bf{v}}$ satisfies condition (c) in \emph{Definition \ref{code-polytopes}} for all $\tau  \in \left\{ {1, \ldots, {\Gamma _c}} \right\}$.


Obviously, \eqref{ML-decoding-all-c-constant} is equivalent to that $\left( {{{\bf{Q}}_\tau } \otimes {{\bf{I}}_{{2^q}}}} \right){\bf{v}}$ satisfies condition (b) in \emph{Definition \ref{code-polytopes}} for all $\tau  \in \left\{ {1, \ldots, {\Gamma _c}} \right\}$. Similarly, \eqref{ML-decoding-all-d-constant} is equivalent to that $\left( {{{\bf{Q}}_\tau } \otimes {{\bf{I}}_{{2^q}}}} \right){\bf{v}}$ satisfies condition (a) in \emph{Definition \ref{code-polytopes}} for all $\tau  \in \left\{ {1, \ldots, {\Gamma _c}} \right\}$. Therefore, \eqref{ML-decoding-all-b-constant}-\eqref{ML-decoding-all-d-constant} is equivalent to $\left( {{{\bf{Q}}_\tau } \otimes {{\bf{I}}_{{2^q}}}} \right){\bf{v}} \in {\mathbb{U}_\tau }$, $\forall \tau  \in \left\{ {1, \ldots, {\Gamma _c}} \right\}$.

Base on the above content, problem \eqref{ML-decoding-all-constant-weight} is equivalent to problem \eqref{LP-decoding}.
}

\section{Proof of \emph{Lemma \ref{relative_in_U}}}\label{relative_in_U_proof}
{\it Proof:} {
We first show that if ${\bf{v}} \in \mathbb{U}$, then ${{\mathop{\rm R}\nolimits} _{\bf{c}}}({\bf{v}}) \in \mathbb{U}$. In other words, we need to verify that  ${{\mathop{\rm R}\nolimits} _{\bf{c}}}({\bf{v}}) $ satisfies all three conditions in \emph{Definition \ref{code-polytopes}}. The first two conditions are obvious. We focus on the third condition. For any ${\cal K} \in \left\{ {{{\cal K}_\ell }|\ell  = 1, \ldots, {2^q} - 1} \right\}$ and any $k \in \left\{ {1,2,3} \right\}$, let $\tilde {\cal B}({\cal K},{h_k})$  be the set defined in \emph{Definition \ref{code-polytopes}}. There are two cases: (i) ${c_k} \notin \tilde {\cal B}({\cal K},{h_k})$  and (ii) ${c_k} \in \tilde {\cal B}({\cal K},{h_k})$.

If ${c_k} \notin \tilde {\cal B}({\cal K},{h_k})$  then for all $\sigma  \in \tilde {\cal B}({\cal K},{h_k})$,
\[\begin{split}
\sum\limits_{i \in {\cal K}} {\tilde b{{\left( {{h_k}\left( {\sigma  + {c_k}} \right)} \right)}_i}}  &= \sum\limits_{i \in {\cal K}} {\tilde b{{\left( {{h_k}\sigma  + {h_k}{c_k}} \right)}_i}} \\
 &= \sum\limits_{i \in {\cal K}} {\left[ {\tilde b{{\left( {{h_k}\sigma } \right)}_i} + \tilde b{{\left( {{h_k}{c_k}} \right)}_i}} \right]} \\
 &= 1 + 0 = 1,
\end{split}\]
where the second equality follows because that the addition in  ${\mathbb{F}_{{2^q}}}$ is equivalent to the vector addition of the corresponding binary vectors. Similarly, for all $\sigma  \notin \tilde {\cal B}({\cal K},{h_k})$, $\sum\limits_{i \in {\cal K}} {\tilde b{{\left( {{h_k}\left( {\sigma  + {c_k}} \right)} \right)}_i} = 0} $. Thus  $\sigma  + {c_k} \in \tilde {\cal B}({\cal K},{h_k})$ if and only if $\sigma  \in \tilde {\cal B}({\cal K},{h_k})$.

By \emph{Definition \ref{relative_matrices}}, ${({{\mathop{\rm R}\nolimits} _{\bf{c}}}({\bf{v}}))_{k,\sigma }} = {{v}_{k,\sigma  + {c_k}}}$. Let $g_k^{\cal K}: = \sum\limits_{\sigma  \in \tilde {\cal B}({\cal K},{h_k})} {{{v}_{k,\sigma }}} $  and $\tilde g_k^{\cal K}: = \sum\limits_{\sigma  \in \tilde {\cal B}({\cal K},{h_k})} {{{({{\mathop{\rm R}\nolimits} _{\bf{c}}}({\bf{v}}))}_{k,\sigma }}} $, then
\[\begin{split}
\sum\limits_{\sigma  \in \tilde {\cal B}({\cal K},{h_k})} {{{({{\mathop{\rm R}\nolimits} _{\bf{c}}}({\bf{v}}))}_{k,\sigma }}}  &= \sum\limits_{\sigma  \in \tilde {\cal B}({\cal K},{h_k})} {{{v}_{k,\sigma  + {c_i}}}} \\
 &= \sum\limits_{\left( {\sigma  + {c_i}} \right) \in \tilde {\cal B}({\cal K},{h_k})} {{{v}_{k,\sigma  + {c_i}}}} \\
 &= \sum\limits_{l \in \tilde {\cal B}({\cal K},{h_k})} {{{v}_{k,l}}}  = g_k^{\cal K}.
\end{split}\]

Therefore we conclude that $\tilde g_k^{\cal K} = g_k^{\cal K}$.

For case (ii), ${c_k} \in \tilde {\cal B}({\cal K},{h_k})$. Using the same argument as above, we can show that $\sigma  + {c_k} \in \tilde {\cal B}({\cal K},{h_k})$ if and only if $\sigma  \notin \tilde {\cal B}({\cal K},{h_k})$. Therefore
\[\begin{split}
\sum\limits_{\sigma  \in \tilde {\cal B}({\cal K},{h_k})} {{{({{\mathop{\rm R}\nolimits} _{\bf{c}}}({\bf{v}}))}_{k,\sigma }}}  &= \sum\limits_{\sigma  \in \tilde {\cal B}({\cal K},{h_k})} {{{v}_{k,\sigma  + {c_i}}}} \\
 &= \sum\limits_{l \notin \tilde {\cal B}({\cal K},{h_k})} {{{v}_{k,l}}} \\
 &= \sum\limits_{l = 0}^{{2^q} - 1} {{{v}_{k,l}}}  - \sum\limits_{l \in \tilde {\cal B}({\cal K},{h_k})} {{{v}_{k,l}}}  = 1 - g_k^{\cal K}.
\end{split}\]

Combining the two cases, we conclude that the vector $\tilde g_k^{\cal K}$  satisfies the following conditions:
\begin{equation}\label{g_tilde_k-g_k}
\tilde g_k^{\cal K} = \left\{ {\begin{array}{*{20}{l}}
{g_k^{\cal K}}&{{\rm{if}}\,{c_k} \notin \tilde {\cal B}({\cal K},{h_k})}\\
{1 - g_k^{\cal K}}&{{\rm{if}}\, {c_k} \in \tilde {\cal B}({\cal K},{h_k})}.
\end{array}} \right.
\end{equation}

We can rephrase this condition by introducing the following notation: Let ${\bf{g}^{{\cal K},{\bf{c}}}}$  be a binary vector for ${\cal K}$  and ${\bf{c}}$  defined by $g_k^{{\cal K},{\bf{c}}}: = \sum\limits_{\sigma  \in \tilde {\cal B}({\cal K},{h_k})} {f{{({\bf{c}})}_{k,\sigma }}} $, where $f({\bf{c}})$  is the Constant-Weight embedding of ${\bf{c}}$, $f{({\bf{c}})_k}$ is the  ${2^q}$-length sub-vector of $f({\bf{c}})$ and ${f{{({\bf{c}})}_{k,\sigma }}}$  denotes the  $\sigma $-th entry in $f{({\bf{c}})_k}$. ${\bf{g}^{{\cal K},{\bf{c}}}}$  is a binary vector with even parity. By its definition, $g_k^{{\cal K},{\bf{c}}} = 1$  if and only if ${c_k} \in \tilde {\cal B}({\cal K},{h_k})$. Thus we can rewrite \eqref{g_tilde_k-g_k} as
\begin{equation}
\tilde g_k^{\cal K} = \left\{ {\begin{array}{*{20}{l}}
{g_k^{\cal K}}&{{\rm{if}}\,g_k^{{\cal K},{\bf{c}}} = 0}\\
{1 - g_k^{\cal K}}&{{\rm{if}}\,g_k^{{\cal K},{\bf{c}}} = 1}.
\end{array}} \right.
\end{equation}

When applying [11, Lemma 17] to the case of binary single parity-check code, we conclude that ${{\bf{\tilde g}}^{\cal K}} \in {\mathbb{P}_3}$  if ${{\bf{g}}^{\cal K}} \in {\mathbb{P}_3}$. This conclude our verification of the third condition of \emph{Definition \ref{code-polytopes}}.	

Next we need to show that if ${{\mathop{\rm R}\nolimits} _{\bf{c}}}({\bf{v}}) \in \mathbb{U}$, then ${\bf{v}} \in \mathbb{U}$. Note that in ${({{\mathop{\rm R}\nolimits} _{\bf{c}}}({\bf{v}}))_{k,\sigma }} = {{v}_{k,\sigma  + {c_k}}}$, $\sigma  + {c_k}$  is equivalent to $\sigma  - {c_k}$  for ${\mathbb{F}_{{2^q}}}$, i.e., ${\bf{v}} = {{\mathop{\rm R}\nolimits} _{\bf{c}}}\left( {{{\mathop{\rm R}\nolimits} _{\bf{c}}}\left( {\bf{v}} \right)} \right)$. Therefore the proof is identical to the previous case.
}

\end{document}